%% file: main.tex
\documentclass[a4paper,11pt]{article}
\pdfoutput=1 

\usepackage[english]{babel}
\usepackage{jheppub}
\AtBeginDocument{%
  \setcounter{tocdepth}{2}%
  \IfFileExists{hyperref.sty}{\hypersetup{bookmarksdepth=2}}{}%
}
\usepackage[T1]{fontenc}
\usepackage{graphicx,calc}
\usepackage{epsfig}
\usepackage{amsmath}
\usepackage{amssymb}
\usepackage{float}
\usepackage{placeins}
\usepackage{braket}
\usepackage{slashed}
\usepackage{mathdots}
\usepackage{lipsum}
\usepackage{feynmf}
\usepackage{bbm}
\usepackage{color}
\usepackage{multicol}
\usepackage{caption}
\usepackage{array}
\usepackage[export]{adjustbox}
\usepackage{subcaption}
\usepackage{multicol}
\usepackage{cleveref}
\usepackage{comment}
\usepackage{colortbl}
\usepackage[most]{tcolorbox}
\definecolor{myred}{RGB}{168,4,4}
\definecolor{myblue}{RGB}{48,53,149}

\makeatletter
\newcommand\xleftrightarrow[2][]{%
  \ext@arrow 9999{\longleftrightarrowfill@}{#1}{#2}}
\newcommand\longleftrightarrowfill@{%
  \arrowfill@\leftarrow\relbar\rightarrow}
\makeatother

\allowdisplaybreaks

\title{All planar three-loop Feynman integrals for the production of two vector bosons at hadron colliders}

\author[]{Dhimiter Canko}
\author[]{and Mattia Pozzoli}
\affiliation[]{Dipartimento di Fisica e Astronomia, Universit\`a di Bologna \\
INFN, Sezione di Bologna, \\
via Irnerio 46, I-40126 Bologna, Italy}
\emailAdd{dhimiter.canko2@unibo.it}
\emailAdd{mattia.pozzoli@unibo.it}

\abstract{We compute all the planar three-loop master integrals relevant for the leading colour N3LO QCD corrections to the production of two massive or off-shell vector bosons at hadron colliders. These integrals are organised into nine four-point integral families with massless internal propagators and two external massive legs. For each family, we construct a basis of pure master integrals and we reconstruct the corresponding canonical differential equations using finite field techniques. We evaluate the master integrals by solving the differential equations using generalised power series expansions.}

\keywords{Feynman integrals, QCD Phenomenology, N3LO Computations}

\begin{document} 
\maketitle
\flushbottom

\input{Introduction}
\input{Notation}
\input{DEs}
\input{Numerical_Evaluation}
\input{Conclusions}

\acknowledgments

We would like to thank Matteo Becchetti, Vsevolod Chestnov, Tiziano Peraro, Lorenzo Tancredi and Simone Zoia for useful discussions and comments on the draft. MP would like to thank the Technical University of Munich and the University of Zurich for hospitality while carrying out this project. This work was supported by the European Research Council (ERC) under the European Union's Horizon Europe research and innovation program grant agreement 101040760, \textit{High-precision multi-leg Higgs and top physics with finite fields} (ERC Starting Grant FFHiggsTop).

\appendix

\input{Ancillary}

\input{New_sectors}

\bibliographystyle{JHEP}
\bibliography{biblio}

\end{document}

%% file: Introduction.tex
\section{Introduction}
\label{Introduction}
Particle physics is entering a precision era. Following the discovery of the Higgs boson, the focus of the Large Hadron Collider (LHC) program has increasingly shifted toward testing the Standard Model of particle physics (SM) with unprecedented accuracy, searching for subtle deviations that could hint at new physics. The upcoming High-Luminosity phase of the LHC will deliver an integrated luminosity roughly an order of magnitude higher than that of Run 3, enabling measurements with sub-percent experimental uncertainties across a wide range of observables. To fully exploit this level of precision, theoretical predictions must reach a comparable degree of accuracy, which for some processes will require next-to-next-to-next-to-leading order (N3LO) QCD corrections~\cite{Gehrmann:2021qex, Caola:2022ayt, Andersen:2024czj, Huss:2025nlt}.

In order to meet these demands, substantial effort has been devoted by the theoretical community to the computation of three-loop scattering amplitudes and the corresponding Feynman integrals that enter the N3LO corrections to these processes. In particular, for processes involving four external particles, all massless Feynman integrals have been evaluated~\cite{Smirnov:2003vi,Henn:2013tua,Henn:2013nsa,Henn:2020lye}, enabling the calculation of three-loop QCD amplitudes for key processes, such as the pair production of photons~\cite{Caola:2020dfu,Bargiela:2021wuy}, light quarks~\cite{Caola:2021rqz,Caola:2022dfa}, and gluons~\cite{Caola:2021izf}, as well as the production of a photon in association with a jet~\cite{Bargiela:2022lxz}. For processes involving massive external particles, all planar~\cite{DiVita:2014pza,Canko:2020gqp,Canko:2021xmn,Syrrakos:2023mor} and non-planar integrals~\cite{Henn:2023vbd,Gehrmann:2024tds,Henn:2024pki,Guan:2025awp} with a single massive particle have been computed. Further progress has been achieved for cases with two massive external particles, with a few of the planar topologies being calculated recently for the equal-mass~\cite{Long:2024bmi} and different-mass~\cite{Canko:2024ara} case. For scattering amplitudes with massive external particles, existing results are so far limited to the generalized leading-colour approximation and to final states involving a vector~\cite{Gehrmann:2023jyv} or a Higgs~\cite{Chen:2025utl} boson produced in association with a jet. In this approximation both the number of colours and the number of light quark flavors are treated as large, while corrections suppressed by inverse powers of these quantities, as well as closed top-quark loop contributions, are neglected.

Some of the processes that occupy a pivotal place within the LHC physics programme are the vector-boson pair production channels, $pp \to V_1 V_2$, where $\{V_1,V_2\}$ denote any possible combination of the electroweak gauge bosons $\{W^{\pm}, Z, \gamma^*\}$~\cite{ATLAS:2019bsc, ATLAS:2019gey, ATLAS:2021mbt, ATLAS:2023dew, CMS:2024hey, CMS:2024ild, ATLAS:2025dhf}. Owing to their relatively large event rates and high-purity multi-leptonic final states, these processes serve as standard candles for calibrating detector performance and the luminosity of the accelerator. They represent one of the most powerful experimental channels to probe the electroweak sector of the SM, sensitive to potential deviations in the trilinear gauge couplings, and allow for detailed tests of QCD dynamics. At the same time, they constitute major backgrounds to Higgs-boson production measurements (e.g.~$pp \to H \to ZZ$ or $pp \to H \to W^+W^-$) as well as to a broad class of searches for physics beyond the Standard Model. Consequently, the precision of theoretical predictions for vector-boson pair production directly impacts both SM measurements and new-physics discovery potential. This, in combination with the future High-Luminosity phase of the LHC, makes imperative the extension of the existing NNLO QCD corrections~\cite{Gehrmann:2013cxs, Gehrmann:2014bfa, Henn:2014lfa, Caola:2014lpa, Caola:2014iua, Caola:2015ila, Gehrmann:2015ora, vonManteuffel:2015msa, Grazzini:2015wpa, Grazzini:2015hta, Caola:2015psa, Grazzini:2016swo, Campbell:2016ivq, Grazzini:2016ctr, Grazzini:2017ckn, Heinrich:2017bvg, Re:2018vac, Kallweit:2018nyv, Grazzini:2018owa, Grazzini:2019jkl, Kallweit:2020gva, Poncelet:2021jmj, Lombardi:2021rvg, Degrassi:2024fye, Pelliccioli:2025com} to N3LO for vector-boson pair production. A step in this direction is the recent computation of the two-loop five-particle Feynman integrals with two external massive particles, which contribute to the real-double-virtual corrections~\cite{Abreu:2024flk}.

In this article we compute the complete set of planar three-loop four-point Feynman integrals with two external massive particles, for both the equal- and different-mass cases. These results pave the way for the computation of the three-loop scattering amplitudes for vector-boson pair production in the generalized leading colour approximation, which contribute to the purely virtual part of the N3LO QCD corrections. Based on their propagator structure, all planar three-loop integrals with two external massive particles can be organized into nine integral families: four of the tennis-court type (\cref{fig:tenniscourts}), three of the ladder-box type (\cref{fig:ladderboxes}), and two of the reducible ladder-box type (\cref{fig:irreducibles}). Results for two ladder-box families in the equal-mass case were obtained in~\cite{Long:2024bmi}, while for the different-mass case we already computed in~\cite{Canko:2024ara} the two reducible families, one tennis-court family, and one ladder-box family. Herein we provide results for all nine families in both mass configurations, re-evaluating the six families previously computed in~\cite{Long:2024bmi,Canko:2024ara} and completing the missing ones.

We compute the integrals using the method of \textit{differential equations} (DEs)~\cite{Kotikov:1990kg,Bern:1993kr,Gehrmann:1999as} and within the framework of dimensional regularisation, shifting the space-time dimension to $d=4 - 2 \varepsilon$ to regulate the ultraviolet and infrared divergences. To this end, we exploit linear relations satisfied by Feynman integrals, called \textit{integration-by-parts relations} (IBPs)~\cite{Tkachov:1981wb,Chetyrkin:1981qh,Laporta:2000dsw}, to find a basis of finitely many \textit{master integrals} (MIs) for each integral family. We evaluate these integrals by solving the linear DEs they satisfy. In order to tackle the significant algebraic complexity associated with the IBPs and the DEs, we employ \textit{finite-field techniques}~\cite{vonManteuffel:2014ixa,Peraro:2016wsq}, as implemented in the \texttt{FiniteFlow} framework~\cite{Peraro:2019svx}. We construct a basis of MIs such that the corresponding DEs are in \textit{canonical form}~\cite{Henn:2013pwa}, which is characterised by the factorisation of the dependence on the dimensional regulator $\varepsilon$ and the presence of logarithmic one-forms only, called \textit{letters}. For the numerical evaluation of the DEs we rely on publicly available packages that implement the method of \textit{generalised power series expansions}~\cite{Pozzorini:2005ff,Moriello:2019yhu,Hidding:2020ytt,Liu:2022chg,Armadillo:2022ugh,Prisco:2025wqs}, using boundary conditions computed through the \texttt{Mathematica} package \texttt{AMFlow}~\cite{Liu:2022chg}, which employs the \textit{auxiliary mass flow} method~\cite{Liu:2017jxz, Liu:2021wks}.

The canonical form of the DEs makes manifest many interesting mathematical properties of the Feynman integrals. For instance, it allows one to express the solution in terms of iterated integrals~\cite{Chen:1977oja} of the logarithmic one-forms. This gives one access to the symbol~\cite{Goncharov:2005sla,Goncharov:2009lql} of the Feynman integrals, enabling several studies of theoretical interest, such as the link between the logarithmic one-forms and cluster algebras~\cite{Drummond:2017ssj,Drummond:2018dfd,Chicherin:2020umh,Aliaj:2024zgp} and (extended) Steinman relations~\cite{Steinmann1,Steinmann2,Cahill:1973px}, as well as opening the door to bootstrap approaches. Moreover, the construction of canonical DEs sheds light on the stability of the \textit{alphabet}, the set of all letters, with the loop order. For the planar families with one massive external leg~\cite{Canko:2021xmn} and for the already computed families with two external masses~\cite{Long:2024bmi,Canko:2024ara}, it was observed that the alphabet of the  was identical with the two-loop one. On the contrary, new letters appeared in the non-planar case~\cite{Henn:2023vbd,Gehrmann:2024tds}.

This article is organized as follows. In \cref{Notation} we define the kinematics relevant for vector-boson pair production and we introduce the Feynman integral families under consideration. In \cref{DEs} we describe the techniques employed for the construction of the canonical DEs and the relevant alphabet of \textit{dlog}-forms. In \cref{Numerical_Evaluation} we discuss the numerical evaluation of the DEs and the validation of the results. Finally, in \cref{Conclusions} we conclude and give an outlook on future work.

%% file: Notation.tex
\section{Notation}
\label{Notation}

\subsection{Kinematics}
\label{Kinematics}

The kinematics related to vector-boson pair production in hadron collisions correspond to the scattering of two massless particles in the initial state into two massive particles in the final state. We denote the momenta of the massless particles by $\{p_1,p_2\}$ and those of the massive particles by $\{p_3,p_4\}$. In the different-mass case we have $p_3^2 = m_3^2$ and $p_4^2 = m_4^2$, while in the equal-mass case we impose $p_3^2 = p_4^2 = m^2$. All external momenta are treated as incoming, so that momentum conservation implies
\begin{equation}
p_4 = -p_1 - p_2 - p_3 \>.
\end{equation}

In the general case the process is fully described by four independent Mandelstam invariants. A convenient choice is
\begin{equation}
s_{12} = (p_1 + p_2)^2, \qquad
s_{23} = (p_2 + p_3)^2, \qquad
m_3^2 = p_3^2, \qquad
m_4^2 = p_4^2 \>.
\end{equation}
All other scalar products can be expressed in terms of these variables. For instance,
\begin{equation}
\label{s13}
s_{13} = (p_1 + p_3)^2 = m_3^2 + m_4^2 - s_{12} - s_{23} \>.
\end{equation}
In the equal-mass case the number of independent invariants reduces to three, since $m_3^2 = m_4^2 = m^2$, and \cref{s13} becomes
$s_{13} = 2m^2 - s_{12} - s_{23}$.

The physical region relevant for the production of two vector bosons $(V,V')$ from the scattering of two partons ($P = q,g$) corresponds in our notation to the $s_{12}$-channel
\begin{equation}
g/q(p_1) + g/\bar{q}(p_2) \;\to\; V(-p_3) \, V'(-p_4) \>.
\end{equation}
Imposing the reality of the scattering angle, the positivity of the external energies, and the Gram-determinant conditions of~\cite{Byers:1964ryc}, we find that the physical region for the different-mass case is described by the following set of inequalities:
\begin{equation}
\label{PhysRegion}
\begin{split}
\mathcal{P} = \{ &
 m_3^2 (m_4^2 - s_{23}) \le s_{23} (m_4^2 - s_{12} - s_{23}), \; s_{12} \geq (m_3 + m_4)^2, \; m_3^2 > 0,\\
& m_4^2 > 0, \;  m_3^2+m_4^2 \leq s_{12}+s_{23}, \; s_{12}^2 + (m_4^2 - m_3^2)^2 \ge 2 s_{12} (m_4^2 + m_3^2) \} \>.
\end{split}
\end{equation}
As expected, the form of the physical region simplifies substantially when the masses degenerate. In this case we obtain
\begin{equation}
\label{PhysRegionSameMass}
\mathcal{P}^{(\mathrm{EM})}
=\{s_{12} \ge 4 m^2,\; 2m^2 \leq s_{12}+s_{23}, \; (m^2 - s_{23})^2 \le -\, s_{12} s_{23}, \; m^2 > 0 \} \>.
\end{equation}

\subsection{Feynman integral families}
\label{Families}

\begin{figure}[t]
  \centering  
  \begin{subfigure}[b]{0.496\textwidth}
    \includegraphics[width=\textwidth]{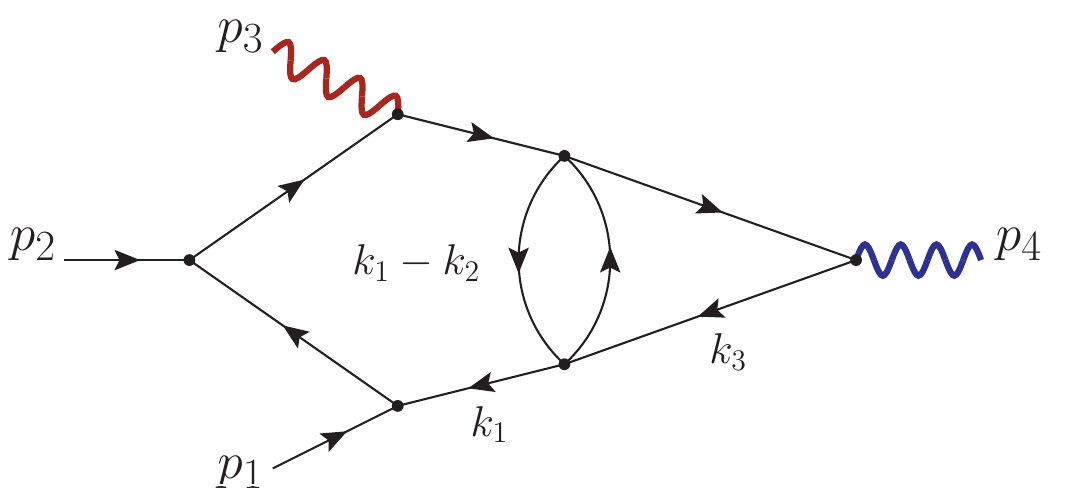}
    \caption{$\mathrm{RL}_1={G}^{(F_{123})}_{\{1,1,1,1,0,0,0,0,1,0,0,1,1,0,1\}}$}
    \label{fig:RL1}
  \end{subfigure}
  \hfill
  \begin{subfigure}[b]{0.496\textwidth}
    \includegraphics[width=\textwidth]{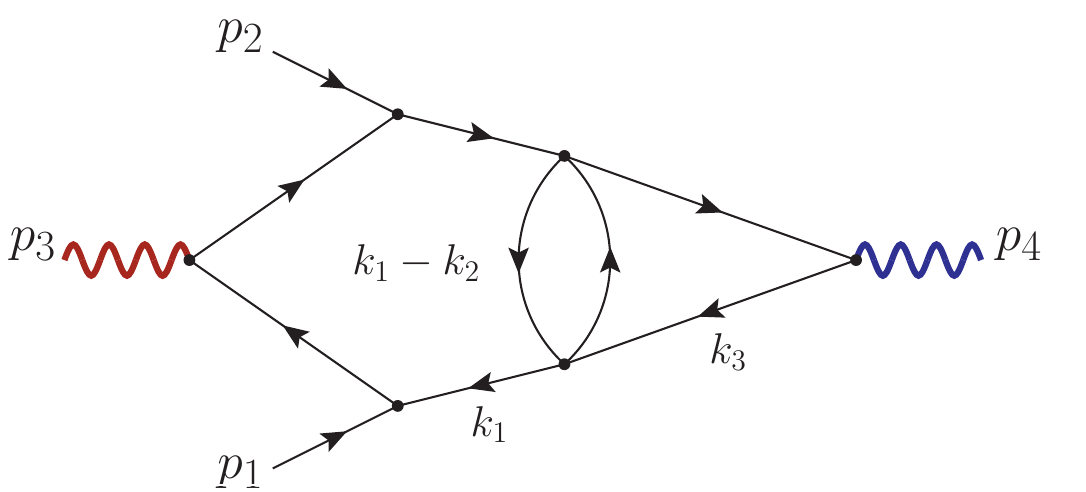}
    \caption{$\text{RL}_2={G}^{(F_{132})}_{\{1,1,1,1,0,0,0,0,1,0,0,1,1,0,1\}}$}
    \label{fig:RL2}
  \end{subfigure}  
  \caption{The top-sectors of the two reducible integral families. Black lines represent massless particles, curled coloured lines correspond to the vector bosons. The particle with mass $m_4$ ($m_3$) is depicted in blue (red), respectively.}
    \label{fig:irreducibles}
\end{figure}

\begin{figure}[t]
  \centering  
  \begin{subfigure}[b]{0.496\textwidth}
    \includegraphics[width=\textwidth]{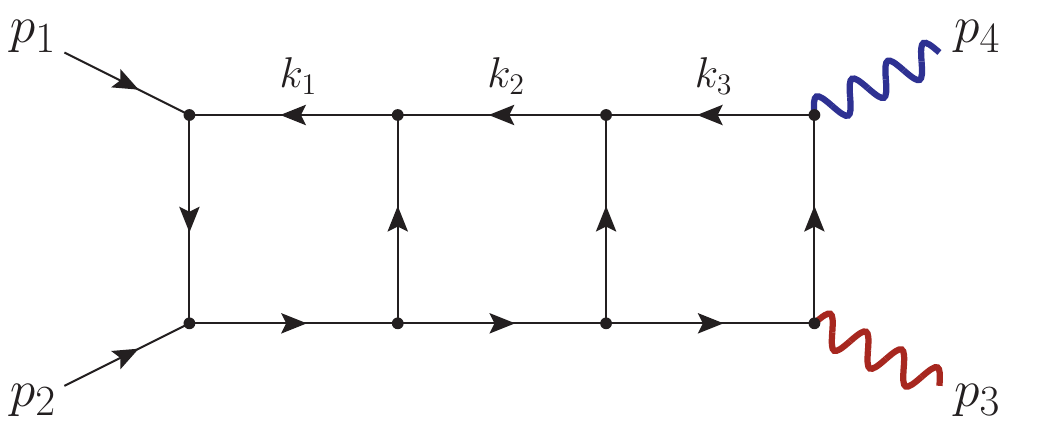}
    \caption{$\text{PL}_1={ G}^{(F_{123})}_{\{1,1,1,0,1,0,1,0,1,0,1,1,1,0,1\}}$}
    \label{fig:PL1}
  \end{subfigure}
  \hfill
  \begin{subfigure}[b]{0.496\textwidth}
    \includegraphics[width=\textwidth]{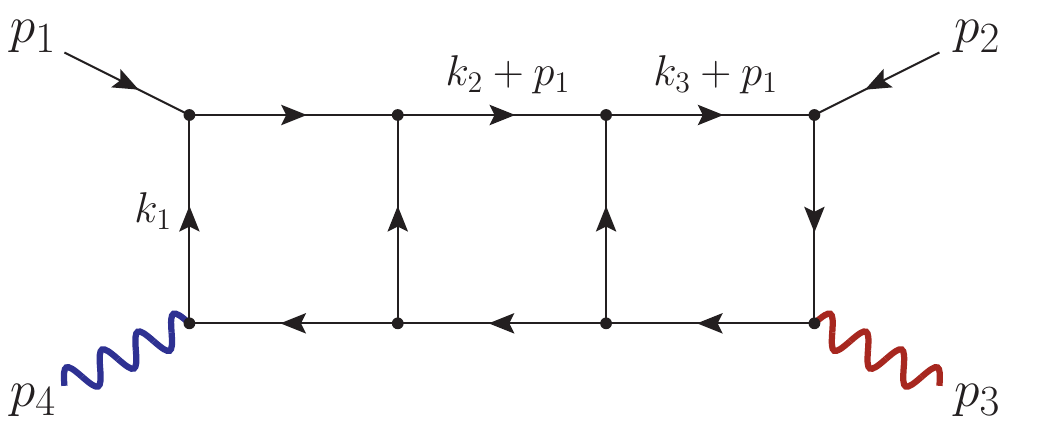}
    \caption{$\text{PL}_2={G}^{(F_{123})}_{\{1,1,0,1,0,1,0,1,0,1,1,1,1,0,1\}}$}
    \label{fig:PL2}
  \end{subfigure} \\
  \vspace{0.2cm}
  \begin{subfigure}[b]{0.496\textwidth}
    \includegraphics[width=\textwidth]{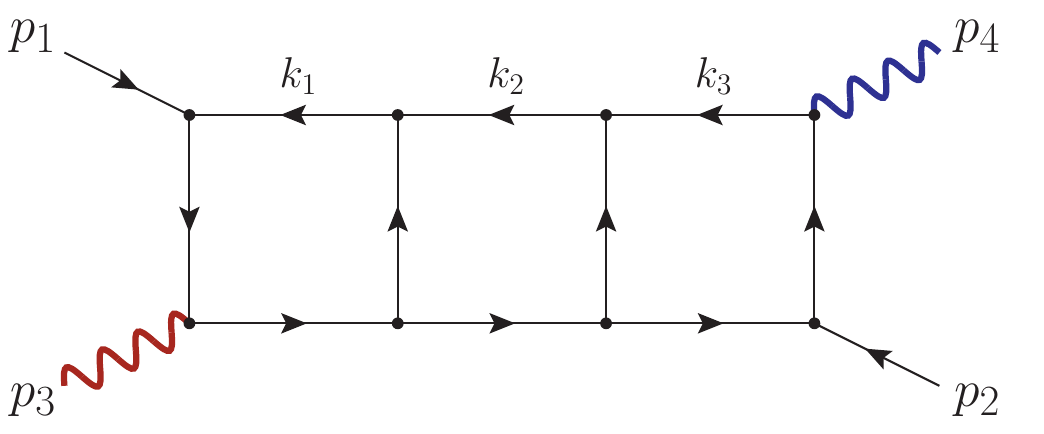}
    \caption{$\text{PL}_3={G}^{(F_{132})}_{\{1,1,1,0,1,0,1,0,1,0,1,1,1,0,1\}}$}
    \label{fig:PL3}
  \end{subfigure}
  \hfill  
  \caption{The top-sectors of the three ladder-box integral families. The same conventions as in \cref{fig:irreducibles} for line styles and colours are used.}
    \label{fig:ladderboxes}
\end{figure}

\begin{figure}[t]
  \centering  
  \begin{subfigure}[b]{0.4\textwidth}
    \includegraphics[width=\textwidth]{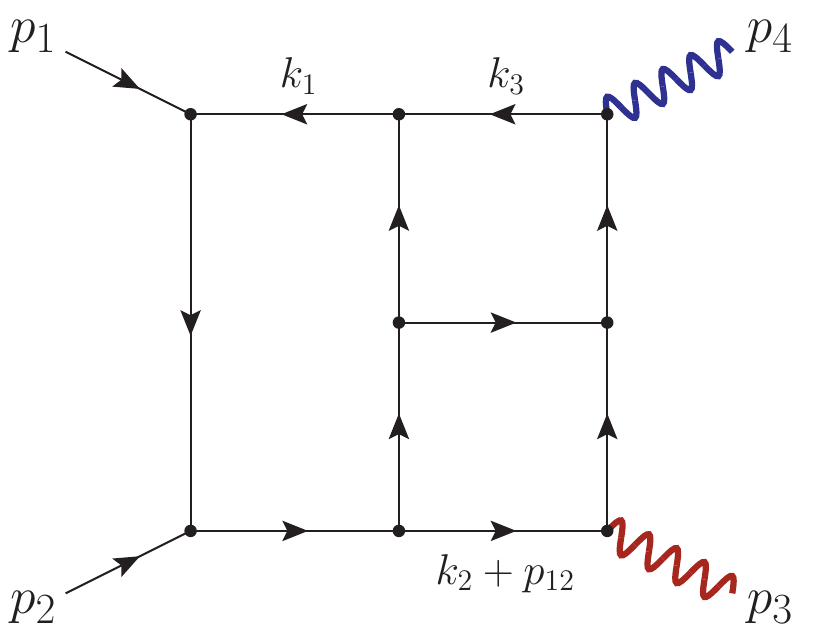}
    \caption{$\text{PT}_1={G}^{(F_{123})}_{\{1,1,1,0,0,0,1,1,1,0,0,1,1,1,1\}}$}
    \label{fig:PT1}
  \end{subfigure}
  \hspace{0.5cm}
  \begin{subfigure}[b]{0.4\textwidth}
    \includegraphics[width=\textwidth]{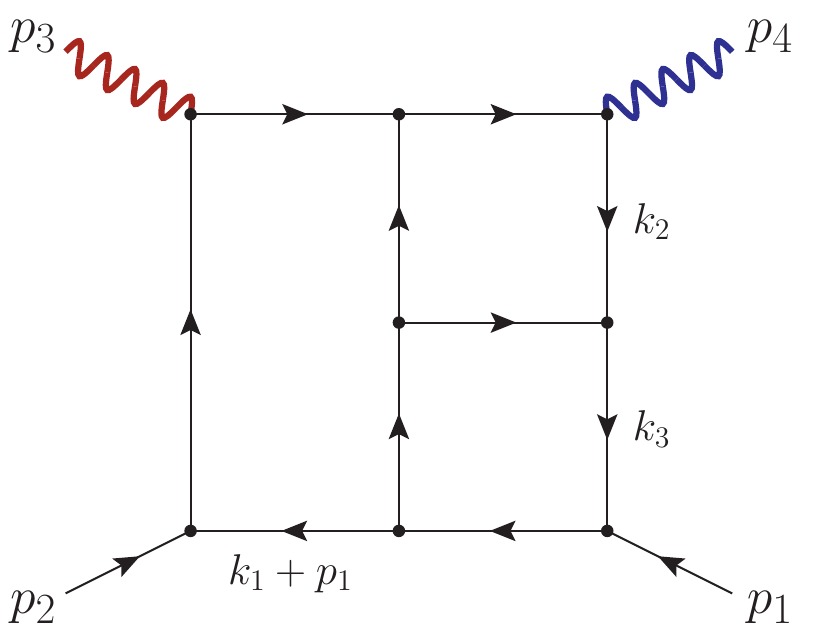}
    \caption{$\text{PT}_2={G}^{(F_{123})}_{\{0,1,1,1,1,0,0,1,1,1,0,0,1,1,1\}}$}
    \label{fig:PT2}
  \end{subfigure}
      \hfill \\
  \vspace{0.2cm}
  \begin{subfigure}[b]{0.4\textwidth}
    \includegraphics[width=\textwidth]{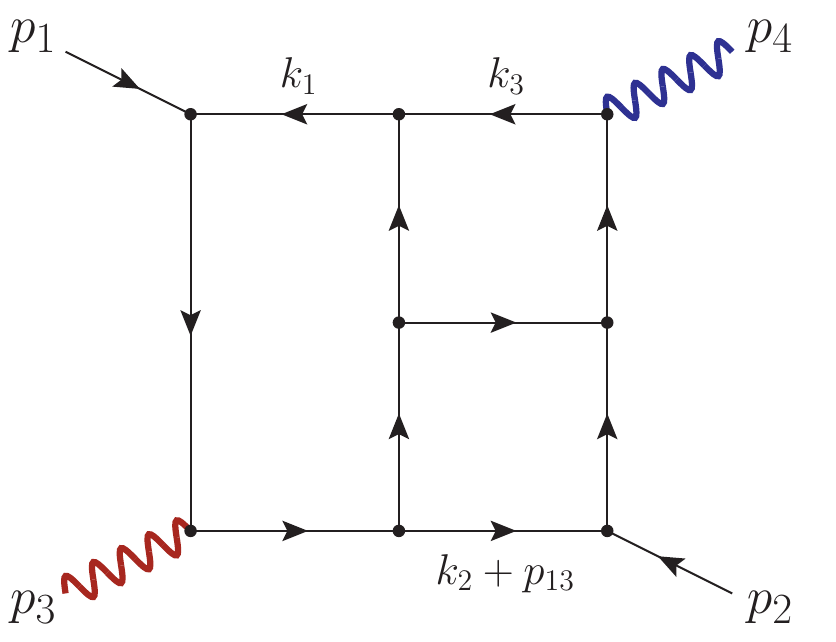}
    \caption{$\text{PT}_3={G}^{(F_{132})}_{\{1,1,1,0,0,0,1,1,1,0,0,1,1,1,1\}}$}
    \label{fig:PT3}
  \end{subfigure}
  \hspace{0.5cm}
  \begin{subfigure}[b]{0.4\textwidth}
    \includegraphics[width=\textwidth]{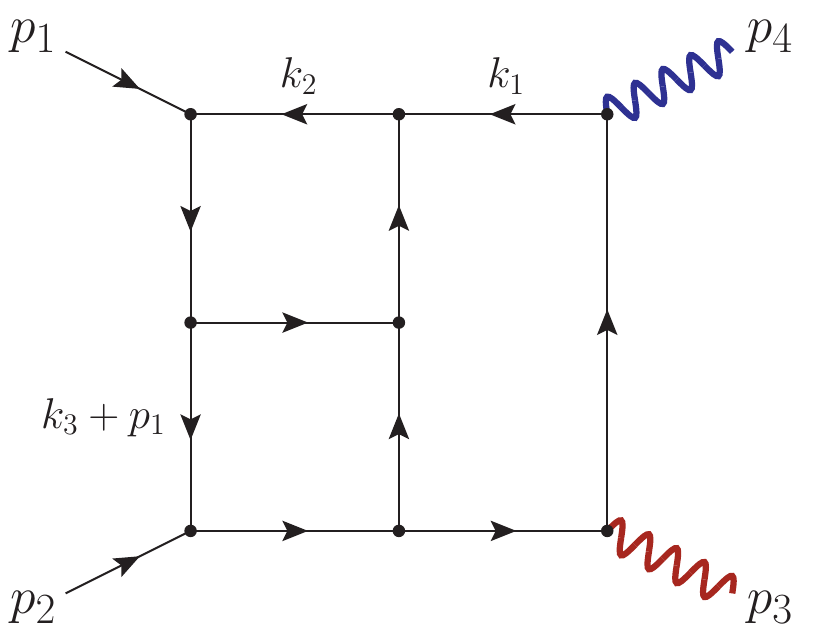}
    \caption{$\text{PT}_4={G}^{(F_{123})}_{\{1,0,1,1,1,1,0,0,0,1,1,0,1,1,1\}}$}
    \label{fig:PT4}
  \end{subfigure}
  \hfill  
  \caption{The top-sectors of the four tennis-court integral families. The same colour and line conventions as in \cref{fig:irreducibles} are followed.}
    \label{fig:tenniscourts}
\end{figure}

Neglecting loop-contributions from massive particles, all planar three-loop Feynman integrals relevant to the production of two vector bosons belong to the nine integral families depicted in \cref{fig:irreducibles,fig:ladderboxes,fig:tenniscourts}, for the different-mass case. These include: two reducible families, $\{\text{RL}_1,\text{RL}_2\}$ (\cref{fig:irreducibles}), three ladder-box families, $\{\text{PL}_1,\text{PL}_2,\text{PL}_3\}$ (\cref{fig:ladderboxes}) and four tennis-court families, $\{\text{PT}_1,\text{PT}_2,\text{PT}_3,\text{PT}_4\}$ (\cref{fig:tenniscourts}). We find it convenient to group all the families into two superfamilies, meaning that the different families can be described by the same set of propagators and they are distinguished by their set of \textit{irreducible scalar products} (ISPs).

The two superfamilies are related by the transformation $p_2 \leftrightarrow p_3$, and we shall denote them by $F_{123}$ and $F_{132}$. The scalar integrals belonging to superfamily $F_x$ can be written in the standard form
\begin{equation}
{G}^{(F_x)}_{\vec{\nu}} =\displaystyle \int \prod_{l=1}^3\frac{\text{d}^d \, k_l\,e^{\epsilon\gamma_E}}{\text{i} \pi^{\frac{d}{2}}} \, \prod_{j=1}^{15}\frac{1}{D^{\nu_j}_{(F_x),\, j}} \>,
\end{equation}
where $\vec{\nu} = \{\nu_1, \ldots, \nu_{15}\} \in \mathbb{Z}^{15}$ and $d = 4 - 2 \varepsilon$, as we work in dimensional regularization. For the superfamily $F_{123}$, we choose the following set of propagators:
\begin{equation}
\begin{aligned}
\label{propagators_F123}
D_{(F_{123}), \, 1} &= k_1^2, & D_{(F_{123}),\, 2}  &= (k_1 + p_1)^2,          & D_{(F_{123}),\, 3}  &= (k_1 + p_{12})^2, \\
D_{(F_{123}),\, 4} &= (k_1 + p_{123})^2,  & D_{(F_{123}),\, 5}  &= k_2^2,                  & D_{(F_{123}), \, 6}  &= (k_2 + p_1)^2,  \\
D_{(F_{123}), \, 7} &= (k_2 + p_{12})^2,   & D_{(F_{123}), \, 8}  &= (k_2 + p_{123})^2, & D_{(F_{123}),\, 9}  &= k_3^2,  \\
D_{(F_{123}), \, 10} &= (k_3 + p_1)^2,     & D_{(F_{123}), \, 11} &= (k_3 + p_{12})^2, & D_{(F_{123}), \, 12} &= (k_3 + p_{123})^2,  \\
D_{(F_{123}), \, 13} &= (k_1 - k_2)^2,     & D_{(F_{123}), \, 14} &= (k_1 - k_3)^2, & D_{(F_{123}), \, 15} &= (k_2 - k_3)^2 \>.
\end{aligned}
\end{equation}
The propagators of the superfamily $F_{132}$ are obtained by applying the substitution $p_2 \leftrightarrow p_3$ to those of $F_{123}$. Although only the three propagators $\{D_{(F_{123}), \, 3},D_{(F_{123}), \, 7},D_{(F_{123}), \, 11} \}$ are affected by this substitution, we list the complete set for completeness:
\begin{equation}
\begin{aligned}
\label{propagators_F132}
D_{F_{(132)}, \, 1} &= k_1^2, & D_{F_{(132)}, \, 2}  &= (k_1 + p_1)^2,          & D_{F_{(132)}, \, 3}  &= (k_1 + p_{13})^2, \\
D_{F_{(132)}, \, 4} &= (k_1 + p_{123})^2,  & D_{F_{(132)}, \, 5}  &= k_2^2,                  & D_{F_{(132)}, \, 6}  &= (k_2 + p_1)^2, \\
D_{F_{(132)}, \, 7} &= (k_2 + p_{13})^2,   & D_{F_{(132)}, \, 8}  &= (k_2 + p_{123})^2,      & D_{F_{(132)}, \, 9}  &= k_3^2,  \\
D_{F_{(132)}, \, 10} &= (k_3 + p_1)^2,     & D_{F_{(132)}, \, 11} &= (k_3 + p_{13})^2,       & D_{F_{(132)}, \, 12} &= (k_3 + p_{123})^2, \\
D_{F_{(132)}, \, 13} &= (k_1 - k_2)^2,     & D_{F_{(132)}, \, 14} &= (k_1 - k_3)^2, & D_{F_{(132)}, \, 15} &= (k_2 - k_3)^2 \>.
\end{aligned}
\end{equation}
Above and in what follows, we adopt the shorthand notation $p_{i \ldots j} = p_i + \ldots + p_j$ for the external momenta. These definitions are equally valid in the general and in the mass-degenerate case.

In \cref{fig:irreducibles,fig:ladderboxes,fig:tenniscourts} we depict the graphs associated with the top sectors of the families, indicating the defining set of exponents and the superfamily. To determine the MIs of each family, we generate the IBP relations using \texttt{FFIntRed}\footnote{An in-house \texttt{Mathematica} package by Tiziano Peraro.} and we solve the resulting systems with \texttt{FiniteFlow}~\cite{Peraro:2019svx}. The number of MIs obtained for each family, in both the equal-mass and different-mass case, is summarized in \cref{tab:MIs}. Across all families, we find 823 independent master integrals in the general case, and 523 in the equal-mass case. The families in the different-mass case contain noticeably more MIs than those in the equal-mass case, with the maximum numbers being 344 and 196, respectively. For comparison, the planar three-loop four-point one-mass families~\cite{DiVita:2014pza,Canko:2020gqp,Canko:2021xmn,Syrrakos:2023mor} range from 83 to 166 MIs, while the non-planar case~\cite{Henn:2023vbd,Gehrmann:2024tds,Henn:2024pki,Guan:2025awp} is comparable to our calculation, with families having from 114 up to 371 MIs, although the largest top-sector contains in that case 19 MIs. Interestingly, while $\text{PT}_{1}$ is the family with the highest number of MIs in the different-mass case, this is no longer true for the equal-mass configuration, where $\text{PT}_{2}$ becomes the largest one. This behavior is expected, as $\text{PT}_{1}$ becomes topologically more symmetric in the equal-mass limit, leading to a reduction in the number of independent MIs. Finally, we note that the number of top-sector MIs is generally identical between the two mass configurations, except for $\text{PT}_{1}$, which contains one fewer MI in the equal-mass case.

We verified, by means of IBP reductions, that all remaining planar three-loop topologies with two off-shell legs do not give rise to independent master integrals. Instead, they reduce either to master integrals belonging to the families considered in this work, or to products of lower-loop integrals, namely one-loop times two-loop integrals or products of three one-loop integrals. This confirms that the set of integral families studied here is complete for describing all the three-loop planar Feynman integrals with two off-shell legs and massless internal propagators.

\renewcommand{\arraystretch}{1.2}
\begin{table}[t!]
\centering
\begin{tabular}{|c|c|c|c|c|}
\hline
 & \multicolumn{2}{c|}{Different masses} & \multicolumn{2}{c|}{Equal masses} \\
\hline
Family & MIs & Top-Sector & MIs  & Top-Sector \\
\hline
$\mathrm{RL}_1$ & 27 & 1 & 24 & 1 \\
$\mathrm{RL}_2$ & 25 & 1 & 22 & 1 \\
$\mathrm{PL}_1$ & 150 & 5 & 94 & 5 \\
$\mathrm{PL}_2$ & 143 & 3 & 84 & 3 \\
$\mathrm{PL}_3$ & 142 & 3 & 81 & 3 \\
$\mathrm{PT}_1$ & 344 & 6 & 188 & 5 \\
$\mathrm{PT}_2$ & 252 & 4 & 196 & 4 \\
$\mathrm{PT}_3$ & 240 & 4 & 181 & 4 \\
$\mathrm{PT}_4$ & 189 & 5 & 109 & 5 \\
\hline
\end{tabular}
\caption{Summary of the MI content for all planar three-loop four-point families with two external massive legs. The table lists, for both the equal- and different-mass cases, the total number of MIs and the subset corresponding to the top sector of each family.}
\label{tab:MIs}
\end{table}
\renewcommand{\arraystretch}{1.0}

%% file: DEs.tex
\section{Canonical differential equations}
\label{DEs}

In this section, we outline the strategies employed to obtain canonical DEs for all the families under consideration, with particular emphasis on the most challenging sector we encountered. We also discuss the procedure used to determine the alphabet of DEs and comment on the emergence of new letters compared to the two-loop case.

\subsection{Construction of the pure bases}
Combining the IBP relations with the fact that each integral family is closed under differentiation with respect to the kinematic invariants, we obtain a system of linear DEs for the vector of MIs, $\vec{G}$:
\begin{equation}
\partial_\xi \Vec{G} (\Vec{s}) = B_\xi(\Vec{s};\epsilon)\cdot \Vec{G}(\Vec{s})  \qquad \forall \ \xi \in \Vec{s}:= \begin{cases}
\left( s_{12},s_{23},m_3^2,m_4^2\right), \quad &\text{different-mass case}\\ \left( s_{12},s_{23},m^2\right), \quad &\text{equal-mass case} \end{cases} \>, \\
\label{eq:DEs_general}
\end{equation}
where we used the shorthand notation $\partial_{\xi} X \equiv \partial X /\partial \xi$. We use \texttt{FFIntRed} to compute the derivatives of the MIs and \texttt{FiniteFlow}~\cite{Peraro:2019svx} to reduce the resulting expressions to MIs and reconstruct the DEs over finite-fields.

In general, the matrices $B_\xi$ can be very complicated and mix the dependence on the kinematics and on the dimensional regulator $\varepsilon$. However, it was observed that in many cases one can find a basis of MIs $\Vec{I}$ such that the DEs take a canonical form~\cite{Henn:2013pwa}:
\begin{equation}
\label{eq:canonical_DE}
\mathrm{d} \Vec{I}(\Vec{s}) = \varepsilon \, \mathrm{d}\tilde{A}(\Vec{s}) \cdot \Vec{I}(\Vec{s}) \>,
\end{equation}
where the connection matrix $\mathrm{d}\tilde{A}$ contains differential one-forms with at most simple poles near their singular loci. The $\varepsilon$-factorised structure of \cref{eq:canonical_DE} allows one to obtain the solution of the MIs iteratively from their Laurent expansion around $\varepsilon$,
\begin{equation}
    \Vec{I}(\Vec{s};\varepsilon) = \sum_{k=0}^{\infty} \varepsilon^{k} \, \Vec{I}^{(k)}(\Vec{s}) \>,
    \label{eq:MIs_eps_expansion_iterative}
\end{equation}
where at each order $k$, the coefficient $\Vec{I}^{(k)}$ is completely determined by a boundary condition, $\vec{I}^{(k)}(\vec{s}_0)$, and the coefficient of the previous order $\Vec{I}^{(k-1)}$ through
\begin{equation}
    \Vec{I}^{(k)}(\vec{s}) = \vec{I}^{(k)}(\vec{s}_0) + \int_{\vec{s}_0}^{\vec{s}} \, \mathrm{d}\tilde{A}(\vec{s} \,') \cdot \vec{I}^{(k-1)}(\vec{s} \,') \>.
\end{equation}

The simplest and best-understood case of canonical DEs for Feynman integrals is the \textit{dlog}-form, where the connection matrix takes the form
\begin{equation}
\label{eq:dlog Form}
\mathrm{d} \tilde{A}(\Vec{s}) = \sum_i a_i \ \mathrm{d} \log w_i (\Vec{s}),
\end{equation}
with the functions $w_i(\vec{s})$ referred to as \textit{letters}. Using the canonical DEs, the coefficients $\vec{I}^{(k)}$ can be straightforwardly expressed in terms of iterated integrals~\cite{Chen:1977oja} over the logarithmic one-forms of \cref{eq:dlog Form}. As mentioned in \cref{Introduction}, this enables formal studies of the mathematical properties of the solution.

For the families at hand, we construct pure bases of MIs following the same strategy used in~\cite{Canko:2024ara}, which we briefly summarise here. Some sectors are topologically identical to sectors from families already known in the literature~\cite{DiVita:2014pza,Henn:2020lye,Canko:2021xmn,Canko:2024ara}, in which case we directly adopt the pure integrals introduced in those works. For the genuinely new sectors, we employ a bottom-up approach, working sector by sector, starting from those with the fewest number of propagators. We first search for candidate integrals whose DEs are $\varepsilon$-factorised on the maximal cut, i.e.~after setting to zero all integrals belonging to lower sectors. To this end, we use the method of Magnus exponential~\cite{Argeri:2014qva,Gehrmann:2014bfa}. This method requires the identification of a pre-canonical basis of MIs that renders the DEs linear in $\varepsilon$. The canonical form is then obtained by rotating this basis through a transformation matrix determined by solving a simple system of DEs. An explicit example is presented in \cref{mushroom sector}. 

A crucial step is the selection of the pre-canonical candidates, for which we employed several complementary techniques. For sectors with only one or a few MIs, educated guesses involving squared propagators in lower sectors and ISPs in higher sectors were typically sufficient. For more complicated cases, we applied the methods of building blocks~\cite{Wasser:2018qvj} and integrand analysis~\cite{Arkani-Hamed:2010pyv,Henn:2020lye,Wasser:2022kwg}. For the latter, we worked both in four dimensions using the package \texttt{DlogBasis}~\cite{Henn:2020lye}, including some tricks to handle massive external momenta~\cite{Papadopoulos:2014lla,Canko:2021hvh}, and in $d$ dimensions using the Baikov representation~\cite{Baikov:1996iu,Baikov:1996rk}, for which we used the package \texttt{Baikov.m}~\cite{Frellesvig:2017aai}. After obtaining an $\varepsilon$-factorised form of the DEs on the maximal cut, we release the cuts and we bring the DEs into canonical form also in the sub-sectors. In practice, we found that most entries were already either $\varepsilon$-factorised or linear in $\varepsilon$. In the latter case, we achieved the canonical form by rotating the basis, analogously to what we did on the maximal cut. In a few cases, notably the ten- and nine-propagators sectors of $\mathrm{PT}_1$ and $\mathrm{PT}_2$, the couplings to some of the lower sectors still involved factorised poles in $\varepsilon$ after releasing the cuts. This was a consequence of missing contributions to the canonical integrals coming from some reducible sub-sector, which we managed to find using \texttt{DlogBasis}.

It is known that the computational cost of reconstructing the DEs depends on their complexity, for example the degree of the rational functions involved. As the reconstruction of the canonical DEs\footnote{Here we mean canonical up to a rotation by a diagonal matrix involving the square roots of \cref{eq:sqrts_general,eq:sqrts_degenerate}.} for $\mathrm{PT}_1$ takes $\mathcal{O}$(hours) on a cluster, it is clear that reconstructing the DEs in a non-canonical form would be time-consuming and inefficient. Fortunately, most of the steps in the procedure only require the knowledge of the $\varepsilon$-dependence of the differential equation, while the full kinematic dependence is only needed to determine the rotation from the pre-canonical candidates to the canonical basis. Therefore, in order to be more efficient and to be able to test many candidates, we resorted to numerical slices, reconstructing only the dependence of the DEs on $\varepsilon$.  

In our study, we first derived the canonical bases of the families in the general kinematic configuration. We then obtained the bases for the equal-mass case by setting $m_3 = m_4 = m$ in the normalisation of our pure integrals and reducing them via IBPs in the equal-mass kinematics. The resulting bases were already in canonical form, requiring no further manipulation. For both mass configurations, the pure bases and the canonical DEs are provided in the ancillary files~\cite{canko_2025_17727482}. Moreover, we refer to \cref{newsectors} for the graphs of the genuinely new integrals appearing in our calculation.

\subsubsection{The sector $\{0,1,0,0,0,0,1,1,1,0,0,1,1,1,0\}$ of $F_{123}$}
\label{mushroom sector}

\begin{figure}[t]
  \centering  
  \begin{subfigure}[b]{0.4\textwidth}
    \includegraphics[width=\textwidth]{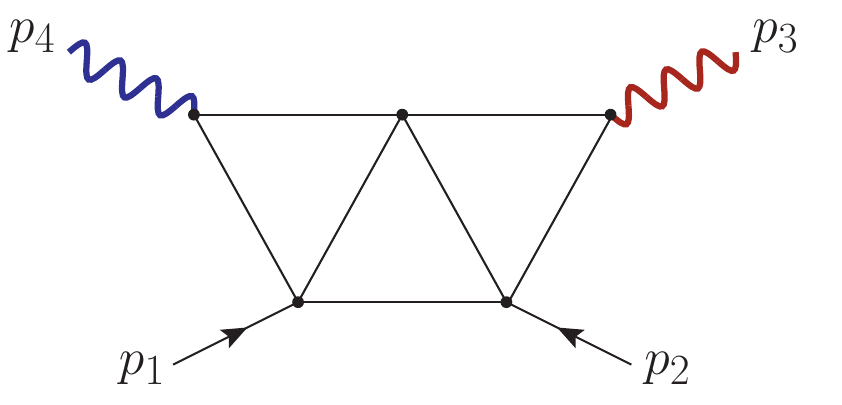}
  \end{subfigure}
  \hfill  
  \caption{The sector $\{0,1,0,0,0,0,1,1,1,0,0,1,1,1,0\}$ of superfamily $F_{123}$.}
    \label{fig:mushroom}
\end{figure}

Here we provide additional details on the construction of the canonical DEs for the sector $\{0,1,0,0,0,0,1,1,1,0,0,1,1,1,0\}$ of superfamily $F_{123}$. This sector appears in the families $\mathrm{PL}_2$, $\mathrm{PT}_1$ and $\mathrm{PT}_2$, and is the most intricate sector we encountered, as it contains 15 MIs in the $m_3 \neq m_4$ case. To construct a pre-canonical basis, we took inspiration from the corresponding sector in the $m_3 = m_4$ case, which was studied in~\cite{Long:2024bmi}. As can be seen from \cref{fig:mushroom}, in this limit the graph of the sector is symmetric with respect to the exchange $(p_1 \leftrightarrow p_2, \, p_3 \leftrightarrow p_4)$. Because of this symmetry, the number of MIs reduces from 15 to 10. These 10 integrals turn out to be suitable pre-canonical candidates also in the different-mass case. The remaining 5 candidates can be obtained by applying the transformation $(p_1 \leftrightarrow p_2, \, p_3 \leftrightarrow p_4)$ to these 10 integrals. Among the 6 integrals that are not mapped to themselves\footnote{The integrals $\mathcal{T}_{59}$, $\mathcal{T}_{61}$, $\mathcal{T}_{62}$, $\mathcal{T}_{63}$, $\mathcal{T}_{66}$ and $\mathcal{T}_{67}$ in~\cite{Long:2024bmi}.}, we select 5 that are linearly independent in order to complete the basis.

\begin{figure}[t]
  \centering  
  \begin{subfigure}[b]{1.0\textwidth}
    \includegraphics[width=\textwidth]{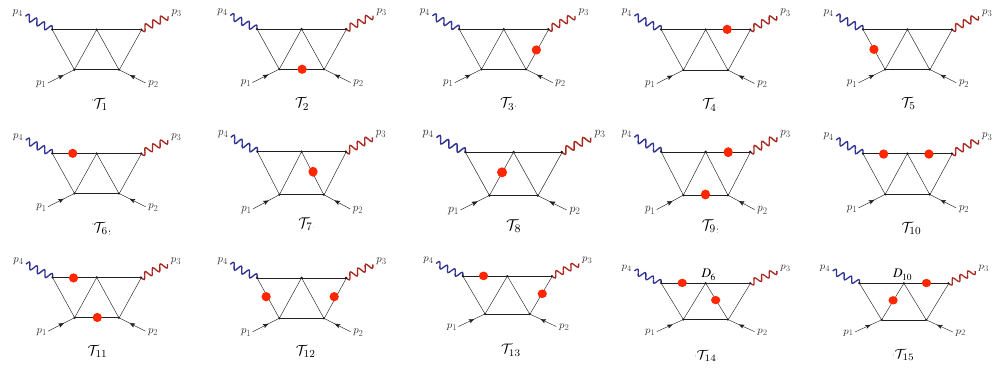}
  \end{subfigure}
  \caption{The starting basis of integrals for the sector $\{0,1,0,0,0,0,1,1,1,0,0,1,1,1,0\}$ of superfamily $F_{123}$. Red dots indicate that the corresponding propagator is squared. The insertion of $D_i$ indicates the presence of an ISP.}
    \label{fig:basis_sector}
\end{figure}

Our starting set of scalar integrals is shown in \cref{fig:basis_sector}. We then fix the polynomial normalisation in $\varepsilon$ for the integrals as follows:
\begin{align}
\label{pre_canonical_mushroom}
f_1 &= \varepsilon^6 \mathcal{T}_1, 
& f_6 &= \varepsilon^5 \mathcal{T}_6, 
& f_{11} &= \varepsilon^4 \mathcal{T}_{11}, \nonumber\\
f_2 &= \varepsilon^5 \mathcal{T}_2, 
& f_7 &= \varepsilon^5 \mathcal{T}_7, 
& f_{12} &= \varepsilon^4 \mathcal{T}_{12}, \nonumber \\
f_3 &= \varepsilon^5 \mathcal{T}_3, 
& f_8 &= \varepsilon^5 \mathcal{T}_8, 
& f_{13} &= \varepsilon^4 \mathcal{T}_{13}, \\
f_4 &= \varepsilon^5 \mathcal{T}_4, 
& f_9 &= \varepsilon^4 \mathcal{T}_9, 
& f_{14} &= \varepsilon^3 (1 + 2 \varepsilon)\mathcal{T}_{14}, \nonumber \\
f_5 &= \varepsilon^5 \mathcal{T}_5, 
& f_{10} &= \varepsilon^3 (1 + 2 \varepsilon)\mathcal{T}_{10}, 
& f_{15} &= \varepsilon^3 (1 + 2 \varepsilon)\mathcal{T}_{15}. \nonumber
\end{align}
The vector $\vec{f} = (f_1, \dots, f_{15})^\mathrm{T}$ satisfies a system of DEs of the form
\begin{equation}
\mathrm{d} \vec{f} = (A_0 + \varepsilon A_1 + \frac{1}{\varepsilon}A_{-1}) \cdot \vec{f}
\end{equation}
with
\begin{equation}
(A_{-1})_{ij} = \begin{cases} a_{-1,i,j} & \text{ if } i \in \{14,15\}, \, j \in \{4,6\}\\
0 & \text{ else}
\end{cases}.
\end{equation}
Analogously to the equal-mass case, the matrix $A_{-1}$ can be removed by an $\varepsilon$-dependent transformation $\vec{f} \rightarrow (\mathbf{1}_{15} + T^\varepsilon) \cdot \vec{f}$ with 
\begin{equation}
(T^\varepsilon)_{ij} = \begin{cases}
t^{(0)}_{ij} + \varepsilon t^{(1)}_{ij} + \frac{t^{(-1)}_{ij}}{\varepsilon} & \text{ if } i \in \{14,15\}, \, j \in \{4,6\}\\
0 & \text{ else}
\end{cases},
\end{equation}
where the coefficients $t_{ij}^{(k)}$ could in principle depend on the kinematics. Imposing that $A_{-1}$ is removed by the transformation and that the linearity in $\varepsilon$ of the other entries is not spoilt, we obtain that $t_{ij}^{(1)} = t_{14,4}^{(-1)} = t_{15,6}^{(-1)} = 0$ and $t^{(-1)}_{14,6} = t_{15,4}^{(-1)} = -2$. 
Therefore, the replacement
\begin{equation}
\begin{split}
f_{14} \rightarrow f_{14} - \frac{2}{\varepsilon} f_6,\\
f_{15} \rightarrow f_{15} - \frac{2}{\varepsilon} f_4,
\end{split}
\end{equation}
leads to DEs linear in $\varepsilon$.

The rotation from the pre-canonical to the canonical basis is straightforward for the first 13 integrals, not so for the last two. The block of the DEs involving these two integrals is fully coupled at $\varepsilon^0$, implying that the canonical basis involves linear combinations of the two integrals. As a consequence, in order to remove the $\varepsilon^0$-term with the usual method we would need to solve a system of coupled differential equations.

The transformation that decouples the last two integrals can, in principle, be an arbitrarily complicated matrix of algebraic functions. The only constraint is that, since the two integrals have the same mass dimension, the coefficients of this transformation must be dimensionless. The problem is then more easily tackled by studying the differential equation on a numerical slice
\begin{equation}
\xi (\lambda) = \alpha_\xi \lambda + \beta_\xi, \quad \forall \ \xi  \in \vec{s},
\end{equation}
thereby reducing the number of variables from four to one. On this slice, one can consider increasingly complicated \textit{ans\"atze} for the transformation coefficients. A convenient initial attempt is
\begin{equation}
T^{\text{decp}} = \begin{pmatrix}
\mathbf{1}_{13} & 0\\
0 & T^\lambda
\end{pmatrix}, \qquad T^\lambda = \begin{pmatrix}
1 & \frac{a_1 \lambda + a_2}{b_1 \lambda + b_2}\\
\frac{c_1 \lambda + c_2}{d_1 \lambda + d_2} & 1
\end{pmatrix}.
\end{equation}
In this case, this simple \textit{ansatz} already proved sufficient to decouple the two integrals on the slice. The full dependence on the kinematic invariants can then be straightforwardly reconstructed.

Performing this rotation and computing the normalisation of the last two integrals from the DEs, we find that the integrals
\begin{equation}
\begin{split}
\tilde{f}_{14} = \frac{m_3^2 m_4^2}{\sqrt{s_{12}^2 + (m_4^2 - m_3^2)^2 - 2 s_{12} (m_4^2 + m_3^2)}} \big( m_4^2 f_{14} +\frac{1}{2} (m_3^2 + m_4^2 -s_{12}) f_{15} \big),\\
\tilde{f}_{15} = \frac{m_3^2 m_4^2}{\sqrt{s_{12}^2 + (m_4^2 - m_3^2)^2 - 2 s_{12} (m_4^2 + m_3^2)}} \big( m_3^2 f_{15} +\frac{1}{2} (m_3^2 + m_4^2 -s_{12}) f_{14}   \big),
\end{split}
\end{equation}
are decoupled at $\varepsilon^0$. The resulting DEs are still linear in $\varepsilon$ in the entries coupling these two integrals to the rest of the basis, but these $\varepsilon^0$-terms can simply be removed with the techniques of~\cite{Gehrmann:2014bfa}, rendering the DEs canonical on the maximal cut.

\subsection{Construction of the alphabet}

To determine explicitly the form of the connection matrix in \cref{eq:dlog Form}, we begin by making an \textit{ansatz} for the full set of letters, i.e.~the \textit{alphabet}. The entries of the partial-derivative matrices are then expressed as linear combinations of the logarithmic derivatives of these letters. The alphabet splits into a rational and algebraic component.

The rational letters are straightforward to identify, as they correspond to the denominators appearing in the DEs. The algebraic letters, on the other hand, take the form
\begin{equation}
w_j^{r_i} (\vec{s}) = \frac{\pi_j(\vec{s})-q_j(\vec{s}) \cdot r_i}{\pi_j(\vec{s})+q_j(\vec{s}) \cdot r_i},
\label{eq:algebraic_letters}
\end{equation}
where $r_i$ is one of the square roots appearing in the normalisation of the MIs. Contrary to what we observed for the families we studied in~\cite{Canko:2024ara}, in the families $\mathrm{PL}_3$, $\mathrm{PT}_1$ and $\mathrm{PT}_2$ we find four additional square roots that were absent in the two-loop families~\cite{Henn:2014lfa, Caola:2014lpa}:
\begin{equation}
\begin{aligned}
r_1&=\sqrt{s_{12}^2 + (m_4^2 - m_3^2)^2 - 2 s_{12} (m_4^2 + m_3^2)} \\
r_2&=\sqrt{(m_3^2 m_4^2 + s_{13} (s_{12} + s_{13}))^2 -4 m_3^2 m_4^2 s_{13}^2} \\
r_3&=\sqrt{s_{23} ((m_3^2 - m_4^2)^2 s_{23} -4 m_3^2 m_4^2 s_{12})} \\
r_4&=\sqrt{(m_4^2 - s_{23})((m_4^2 - s_{12})^2 (m_4^2 - s_{23})-4 m_3^2 m_4^2 s_{12})} \\
r_5&=\sqrt{(m_3^2 - s_{23})((m_3^2 - s_{12})^2 (m_3^2 - s_{23})-4 m_3^2 m_4^2 s_{12})}
\end{aligned} \> ,
\label{eq:sqrts_general} 
\end{equation}
where above we used $s_{13}=m_3^2+m_4^2-s_{12}-s_{23}$ to simplify the expression of $r_4$. We also remark that we renamed the square root $R$ from~\cite{Canko:2024ara} to $r_1$. In the equal-mass case, the number of square roots reduces to four, as $r_4$ and $r_5$ degenerate in this limit:
\begin{equation}
\begin{aligned}
r_1^{\text{(EM)}}&=\sqrt{s_{12}(s_{12}-4 m^2)} \\
r_2^{\text{(EM)}}&=\sqrt{(m^4 + s_{13} (s_{12} + s_{13}))^2 -4 m^4 s_{13}^2}\\
r_3^{\text{(EM)}}&=\sqrt{-s_{12} s_{23}}\\
r_4^{\text{(EM)}}&=\sqrt{(m^2 - s_{23})((m^2 - s_{12})^2 (m^2 - s_{23})-4 m^4 s_{12})}
\end{aligned} \> ,
\label{eq:sqrts_degenerate}
\end{equation}
Similarly to the different-mass case, the square roots $r_2^{\text{(EM)}}, r_3^{\text{(EM)}}$ and $r_4^{\text{(EM)}}$ did not appear in the two-loop families~\cite{Gehrmann:2013cxs,Gehrmann:2014bfa}.

There are several packages that implement algorithms to generate candidates for algebraic letters, such as \texttt{BaikovLetter}~\cite{Jiang:2024eaj}, which we employed in~\cite{Canko:2024ara} to identify the letters involving $r_1$. However, using \texttt{BaikovLetter} for the letters depending on $r_2$, $r_4$ and $r_5$ proved to be too time-consuming. For these cases, we instead used the package \texttt{SOFIA}~\cite{Correia:2025yao}, which automates the detection of the singularities of the Feynman integrals using the method proposed in~\cite{Caron-Huot:2024brh} and constructs candidate letters utilizing \texttt{Effortless}. With the resulting \textit{ansatz} for the alphabet, we were able to determine all entries of the connection matrices except for a few, involving $r_1$ and one of the new square roots. We constructed the missing letters following the same strategy of \texttt{Effortless}, described in~\cite{Matijasic:2024too}. Assuming that $q_j(\vec{s})=1$ in \cref{eq:algebraic_letters}, we know that the product of numerator and denominator of each letter must factorise in the alphabet~\cite{Heller:2019gkq,Zoia:2021zmb}:
\begin{equation}
\pi_j^2 - r_i^2 = c \prod_{w_i \text{even}} w_i^{e_i},
\label{eq:algebraic_letter_factorisation}
\end{equation}
where $c, e_i \in \mathbb{Q}$. In~\cite{Matijasic:2024too} it was observed that the set of letters that are allowed to appear on the right-hand side of \cref{eq:algebraic_letter_factorisation} is restricted by the condition
\begin{equation}
\forall \ \vec{s}_0 \in \mathbb{Q}^4 \vert w_i (\vec{s}_0) = 0 \Longrightarrow r_i (\vec{s}_0) = \pi_j (\vec{s}_0) \in \mathbb{Q}.
\end{equation}
Moreover, it was also observed that solutions of \cref{eq:algebraic_letter_factorisation} are typically found only for $c = \pm 4$. We can use this knowledge to construct a reduced set of candidates for $p_j^2$ and verify whether they can be written as perfect squares. Following this strategy, we were able to identify all remaining letters and fully determine the form of the connection matrices.

The full alphabet for both mass configurations is provided in the ancillary files~\cite{canko_2025_17727482}. It consists of 37 (22) letters in the general (mass-degenerate) case, of which 22 (13) are rational and 15 (9) are algebraic. The appearance of new algebraic letters compared to the two-loop case~\cite{Gehrmann:2013cxs,Gehrmann:2014bfa,Henn:2014lfa,Caola:2014lpa} is an inevitable consequence of the additional square roots in \cref{eq:sqrts_general,eq:sqrts_degenerate}. We remark that none of these new letters involves only $r_1$ or $r_1^{\text{(EM)}}$, which however appear in combination with some other square root. It is interesting to note that also the rational component of the alphabet includes additional elements, four in the mass-degenerate case and nine in the general case. These new rational letters exhibit an increase in the degree compared to the two-loop case: in the mass-degenerate case we found a letter with mass dimension three, while in the general case polynomials of degree three and four appear. In the case of only one massive leg, the study of the planar families~\cite{Canko:2021xmn} suggested stability of the alphabet going from two to three loops, while new letters and square roots were observed in the non-planar case~\cite{Henn:2023vbd,Gehrmann:2024tds}. Our observation of new analytic structures and differential forms already in the planar case indicates a more generalised lack of stability of the alphabet with the loop order.

%% file: Numerical_Evaluation.tex
\section{Numerical evaluation}
\label{Numerical_Evaluation}
Since we did not identify a change of variables capable of simultaneously rationalising all five square roots appearing in \cref{eq:sqrts_general,eq:sqrts_degenerate}, we do not pursue a fully analytic solution of the DEs. Instead, we rely on semi-numerical methods to solve them. In particular, we use the method of generalised power series expansions~\cite{Pozzorini:2005ff,Moriello:2019yhu,Hidding:2020ytt,Liu:2022chg,Armadillo:2022ugh,Prisco:2025wqs}. Given the numerical value of the MIs at a boundary point $\vec{s}_0$ and the system of DEs they satisfy, the method allows us to propagate the solution to a target point in the phase-space using truncated analytic expansions of the solution, whose coefficients are determined numerically from the DEs. 

Among the various publicly available tools that implement the method of generalised power series expansions~\cite{Hidding:2020ytt,Liu:2022chg,Armadillo:2022ugh,Prisco:2025wqs}, we make use of the \texttt{Mathematica} package \texttt{DiffExp}~\cite{Hidding:2020ytt} and of the DE solver provided by the \texttt{AMFlow}~\cite{Liu:2022chg} package. Although in principle both implementation do not put any restriction on the location of the boundary- and target-point in the phase-space, we find it convenient to work directly in the physical scattering region \cref{PhysRegion,PhysRegionSameMass}. This guarantees that no physical singularities are crossed during the evaluation, which would require us to perform an analytic continuation of the solution.

We computed the boundary values at a physical phase-space point using \texttt{AMFlow}~\cite{Liu:2022chg}, which implements  the \textit{auxiliary mass flow method}~\cite{Liu:2017jxz}, requesting 60-digit precision. For the different-mass case, we use the boundary point
\begin{equation}
\label{eq:base_point_general}
\vec{s}_0:=\left(s_{12},s_{23},m_3^2,m_4^2\right) = \left(\frac{13}{4},-\frac{3}{4},1,\frac{9}{20}\right),
\end{equation}
while for the equal-mass case we use
\begin{equation}
\label{eq:base_point_degenerate}
\vec{s}_0^{\, \text{(EM)}} := \left(s_{12},s_{23},m^2\right)= \left(\frac{23}{4},-\frac{3}{4},1\right).
\end{equation}
Using our setup, one can evaluate the integrals at any point that is connected to these boundary points by a path in the physical region. The most convenient choice of parametrisations is to use straight lines. Therefore, we define a point in the physical region to be connected to our boundary points in \cref{eq:base_point_general,eq:base_point_degenerate} if it can be reached by a straight line lying entirely in the physical region. We investigated whether every physical point satisfies this property using the \texttt{Mathematica} function \texttt{Reduce}. For the degenerate mass case, we found that every point with $m^2 = 1$ is connected to the one in \cref{eq:base_point_degenerate}. Since values for $m^2 \neq 1$ can be recovered by exploiting the scaling properties of the integrals, the entirety of the physical region is covered by this boundary point. For the general case, we could not verify whether every point of the physical region is connected to the one in \cref{eq:base_point_general}. However, we observed that for fixed values of the two masses, setting $m_3 = 1$ for convenience and keeping $m_4$ generic, the physical region is star-shaped. In other words, one can use the point
\begin{equation}
\label{eq:base_point_qq}
\vec{s}_{0,m_4^2}:=\left(s_{12},s_{23},m_3^2,m_4^2\right) = \left(\frac{13}{4},-\frac{3}{4},1,m_4^2 \right),
\end{equation}
to evaluate the integrals for any physical value of the Mandelstam invariants and fixed masses. We verified that any point in \cref{eq:base_point_qq} is connected to our boundary point. Therefore, one can use our setup to first evolve the mass $m_4$ to the desired value and then evaluate the integrals at the target point. We also remark that there is a letter, $m_3^2-m_4^2$, which can vanish in the physical region, possibly requiring analytic continuation. This issue is however easily circumvented by choosing $m_3$ as the largest mass.

As a validation of the values obtained for the MIs with our setup, we compared them against independent evaluations with \texttt{AMFlow}. The extent of this comparison is constrained by the substantial computational cost of three-loop calculations in \texttt{AMFlow}, which limits us to a small set of phase-space points whose numerators and denominators are chosen to be small primes. Specifically, we performed the comparison at the following three points
\begin{equation}
    \begin{split}
        \left(s_{12},s_{23},m_3^2,m_4^2\right) &= \left(\frac{304}{33},-\frac{5}{7},3,\frac{25}{33}\right),\\
        \left(s_{12},s_{23},m_3^2,m_4^2\right) &= \left(\frac{88}{5},-\frac{35}{6},5,\frac{49}{15}\right),\\
        \left(s_{12},s_{23},m_3^2,m_4^2\right) &= \left(25,-\frac{36}{5},9,\frac{16}{7}\right),
    \end{split}
    \label{eq:checkpoints_general}
\end{equation}
for the different-mass case, and at the points
\begin{equation}
    \begin{split}
        \left(s_{12},s_{23},m^2\right) &= \left(\frac{704}{33},-\frac{5}{7},3\right),\\
        \left(s_{12},s_{23},m^2\right) &= \left(\frac{288}{5},-\frac{35}{6},5\right),\\
        \left(s_{12},s_{23},m^2\right) &= \left(75,-\frac{36}{5},8\right),
    \end{split}
    \label{eq:checkpoints_degenerate}
\end{equation}
for the equal-mass case. In all cases, we found agreement up to 16 digits, matching the target precision imposed on both computational setups. In addition, for the equal-mass families $\mathrm{PL_1}$ and $\mathrm{PL_2}$, we have compared our results with the numerical values obtained from the analytic solution of Ref.~\cite{Long:2024bmi}, finding prefect agreement.

In the ancillary files~\cite{canko_2025_17727482} we include two scripts that evaluate the MIs of each family for either mass configuration at any target point lying within the physical regions defined in \cref{PhysRegion,PhysRegionSameMass}. We tested these codes by evaluating the integrals twice at 20 random physical phase space points, increasing the precision for the second run. We found agreement between the two results up to 16 digits. Our setup allows for the evaluation of the integrals of one family in the order of a few minutes for generic phase-space points and a precision of 16 digits. The exact evaluation time per point depends on the number of segments in which the integration path is split, which is determined by the singularity structure of the DEs along the path.

%% file: Conclusions.tex
\section{Conclusions}
\label{Conclusions}
In this paper we studied all the planar families of Feynman integrals relevant for the N3LO QCD corrections to the production of two massive vector bosons, complementing the results of~\cite{Long:2024bmi,Canko:2024ara}. For each family we constructed a basis of MIs that satisfy canonical DEs, which we solved for physical kinematics using the method of generalised power series expansions and public tools. We validated our results through independent numerical cross-checks with \texttt{AMFlow}. Beyond providing numerical evaluations, our results allow one to express the MIs in terms of iterated integrals, immediately enabling theoretical studies of their mathematical properties.

An interesting structural feature emerged in our study: the alphabet of all planar families for the general mass (equal-mass) configuration contains four new square roots and 21 (11) additional letters beyond those present in the complete two-loop alphabet. This contrasts with known results from the planar three-loop four-point families with one massive leg, and illustrates the additional analytic complexity introduced by having multiple external massive legs at three loops. The enlarged alphabet also comes with an increased number of independent square roots, which we were unable to rationalise simultaneously. For this reason, we did not pursue an analytic solution in terms of multiple polylogarithms~\cite{Goncharov:1998kja, Goncharov:2010jf}. Nevertheless, it would be interesting to further investigate this possibility. In particular, following the approach of \cite{Papathanasiou:2025stn}, it may be feasible to obtain analytic expressions by employing different rationalisations in different parts of the computation, provided that not all square roots appear simultaneously along a given integration path.

The efficiency of our implementation is sufficient for the purposes of this work and therefore we have not put any effort into the optimisation of the numerical evaluation of the solution. For phenomenological applications, the performance can be further improved with a more aggressive setting, tailored to the stability of the amplitude evaluation. To this end, several promising avenues are already available in the literature. One option is to replace the \texttt{Mathematica}-based implementations of the generalised power series expansion method with the C++ code \texttt{LINE}~\cite{Prisco:2025wqs}. A similar strategy is to investigate fully numerical integration methods, which have already been successfully employed for $2 \to 2$ and $2 \to 3$ scattering processes~\cite{Boughezal:2007ny,Czakon:2008zk,Czakon:2020vql,Czakon:2021yub,Calisto:2023vmm,PetitRosas:2025xhm,Badger:2025ljy}. Lastly, one could remove the redundancies in the representation of the integrals by constructing a basis of special functions~\cite{Gehrmann:2015bfy,Gehrmann:2018yef,Chicherin:2020oor,Chicherin:2021dyp,Abreu:2023rco,Gehrmann:2024tds,Becchetti:2025osw} and investigating if further relation can be obtained for them using the reduction approach of~\cite{Ma:2025mog}. We postpone the investigation of these optimisations to forthcoming studies.

Taken together, our results lay the groundwork for the computation of the three-loop scattering amplitudes required for diboson production at N3LO QCD in the leading-colour approximation, a task we plan to pursue in future work.

%% file: Ancillary.tex
\section{Ancillary files}
\label{Ancillary}
The ancillary files can be downloaded from~\cite{canko_2025_17727482}. The correspondence with the notation of the paper is:
\begin{table}[ht!]
    \centering
    \begin{tabular}{ll}
        \texttt{sij} = $s_{ij}$, \quad & \texttt{FFG[Fam,\{a1,...,a15\}]} = $G^{Fam}_{\{a_1,\dots,a_{15}\}}$,\\
        \texttt{mi2} = $m_i^2$, \texttt{mm2} = $m^2$, \quad & \texttt{W[i]},\texttt{W[r[j],i]} = $w_i$,\\
        \texttt{eps} = $\epsilon$, \quad & \texttt{dlog[W[i]]} = $\mathrm{d}\log(w_i)$,\\
         \texttt{r[i]} = $r_i$, \quad & \texttt{rEM[i]} = $r^{(EM)}_i$.
    \end{tabular}
    \label{tab:ancillary_notation}
\end{table}

The folder \texttt{Alphabets} contains the definitions of the square roots \cref{eq:sqrts_general,eq:sqrts_degenerate} and of the alphabets in the format
\begin{equation*}
    \texttt{\{\dots, W[i] -> expression, \dots\}}.
\end{equation*}
We remark that for the different-mass case the naming of the letters is consistent with~\cite{canko_2024_14284044}.

For each family, we provide four folders:
\begin{itemize}
    \item \texttt{MIs\_DEs};
    \item \texttt{MIs\_DEs\_same\_mass};
    \item \texttt{values};
    \item \texttt{values\_same\_mass}.
\end{itemize}
The first two contain three files each:
\begin{itemize}
    \item \texttt{masters.m}: the basis of pure MIs, in terms of the dimensional regulator \texttt{eps}, the scalar integrals \texttt{FFG[Fam,\{a1,...,a15\}]} and the kinematic invariants;
    \item \texttt{sqrt\_normalisation.m}: the square root normalisation of each MI.
    \item \texttt{dAtilde.m}: the matrix $\mathrm{d} \Tilde{A}$ from \cref{eq:dlog Form}.
\end{itemize}
The last two folders contain the boundary values for both mass configurations.

In the folder \texttt{info}, we provide the following files:
\begin{itemize}
\item \texttt{MI\_relations.m} (\texttt{MI\_relations\_same\_mass.m}): the relations between the canonical master integrals of different families;
\item \texttt{independent\_integrals.m} (\texttt{independent\_integrals\_same\_mass.m}): the list of 823 (523) independent master integrals across all families in the different-mass (equal-mass) case.
\end{itemize}
In the above files, the notation \texttt{ut[fam,j]} refers to the $j$-th entry in the list of master integrals of family \texttt{fam}. Moreover, the folder also contains the file \texttt{iterated\_integrals.wl}, which generates the expressions of the MIs in terms of Chen iterated integrals~\cite{Chen:1977oja}. These are defined as
\begin{equation}
    I_{\gamma}\left(\omega_1,...,\omega_n;\lambda\right) = 
    \int\limits_a^{\lambda} d\lambda_1 \mathfrak{f}_1\left(\lambda_1\right)
    \int\limits_a^{\lambda_1} d\lambda_2 \mathfrak{f}_2\left(\lambda_2\right)
    \dots
    \int\limits_a^{\lambda_{n-1}} d\lambda_n \mathfrak{f}_n\left(\lambda_n\right),
    \label{eq:def_iterated_integral}
\end{equation}
where $\mathfrak{f}_j\left(\lambda\right) d\lambda = \gamma^\ast \omega_j$ is the pull-back of the one-form $\omega_j$ along a path $\gamma$. In the \textit{dlog} case $\omega_i = \mathrm{d}\log \left(w_{k_i}\right)$, hence we use the short-hand notation
\begin{equation}
    \mathtt{II}[\mathtt{W[k_1]},\dots,\mathtt{W[k_n]}] := I_{\gamma}\left(\omega_1,...,\omega_n;\lambda\right),
    \label{eq:Chen_iterated_integrals}
\end{equation}
where $k_i$ is the letter index.

Lastly, we provide the codes \texttt{DiffExpRun.wl} and \texttt{AMFSolver.wl}, which compute the master integrals using \texttt{DiffExp} and the solver of \texttt{AMFlow}, respectively. In order to use it, one should first modify the files \texttt{PathToDiffExp.txt} and \texttt{PathToAMFlow.txt} appropriately. Then the only input needed is a file with the physical point where one wants to evaluate the integrals, in the same format as \texttt{testpoint.m} (respectively \texttt{testpoint\_same\_mass.m}). The computation can be ran with the commands
\begin{align*}
    \texttt{math -script DiffExpRun.wl } &\texttt{-family 'fam' -equalMass False }\\
    &\texttt{-filein "point.m" -fileout "output.m"},\\
    \texttt{math -script AMFSolver.wl } &\texttt{-family 'fam' -equalMass False }\\
    &\texttt{-filein "point.m" -fileout "output.m"},
\end{align*}
where \texttt{'fam'} is the integral family one wants to compute, \texttt{'equalMass'} is a boolean that selects the mass configuration and the results are stored as a list in \texttt{output.m}.

%% file: New_sectors.tex
\section{New topologies}
\label{newsectors}

In this appendix, we collect the pure candidates for the genuinely new sectors in the general-mass case, referring to their position within the basis of the corresponding integral family (see the ancillary files~\cite{canko_2025_17727482}). We exclude the sectors presented in Appendix B of Ref.~\cite{Canko:2024ara}. Regarding the sector definition, we use the notation
\begin{equation*}
\text{SF}[\nu_1,\ldots,\nu_{15}],
\end{equation*}
where $\text{SF}=\{\text{$F_{123}$}, \text{$F_{132}$}\}$ denotes the superfamily, and $\nu_i \in \{0,1\}$ are the exponents of the propagators of the corresponding superfamily. Pure candidates are denoted by
\begin{equation*}
I_{\text{pos}}^{\text{Fam}},
\end{equation*}
where $\text{Fam}=\{\text{PL2}, \text{PL3}, \text{PT1}, \text{PT2}, \text{PT3}\}$ labels the integral family, and $\text{pos}\in \{1,\ldots,344\}$ specifies the position of the candidate within the corresponding basis.
In total, we identified 76 new sectors, comprising 212 MIs. In the following, we classify the sectors according to the number of propagators they contain, starting with those with six propagators and ending with those with ten.

\vspace{0.15cm}
\begin{center}
\textbf{\textit{Six-Propagator Pure Candidates}}\\
\end{center}
\vspace{0.2cm}

\textbf{Sector $\mathbf{F_{132}}$[0,1,0,0,1,0,0,0,0,0,1,1,1,0,1]}

\begin{multicols}{2}
\includegraphics[scale=0.20]{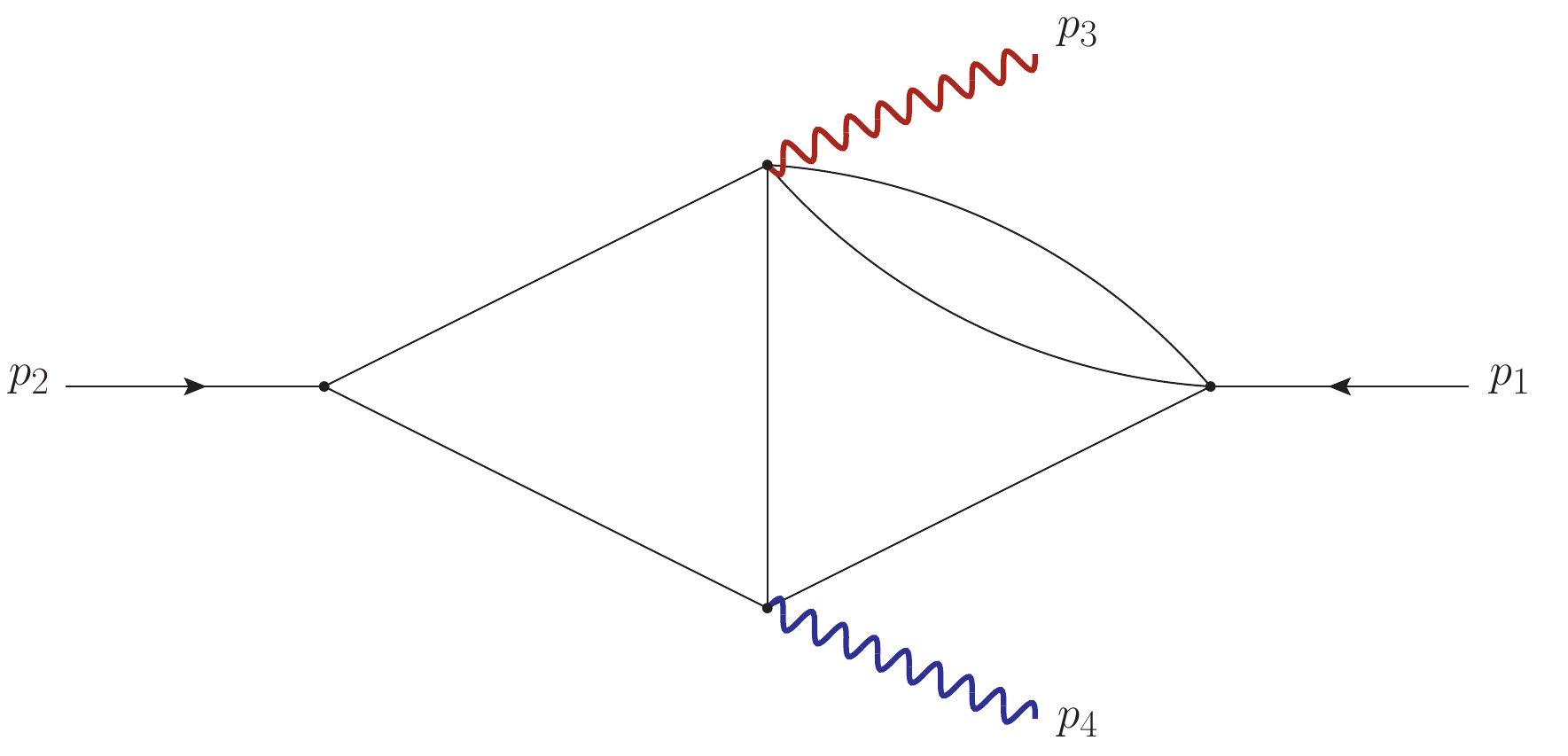}\\
\newline
\begin{equation*}
\{I_{113}^{\text{PL3}}\}
\end{equation*}
\end{multicols}

\textbf{Sector $\mathbf{F_{132}}$[0,1,0,0,0,0,1,0,1,0,0,1,1,0,1]}

\begin{multicols}{2}
\includegraphics[scale=0.20]{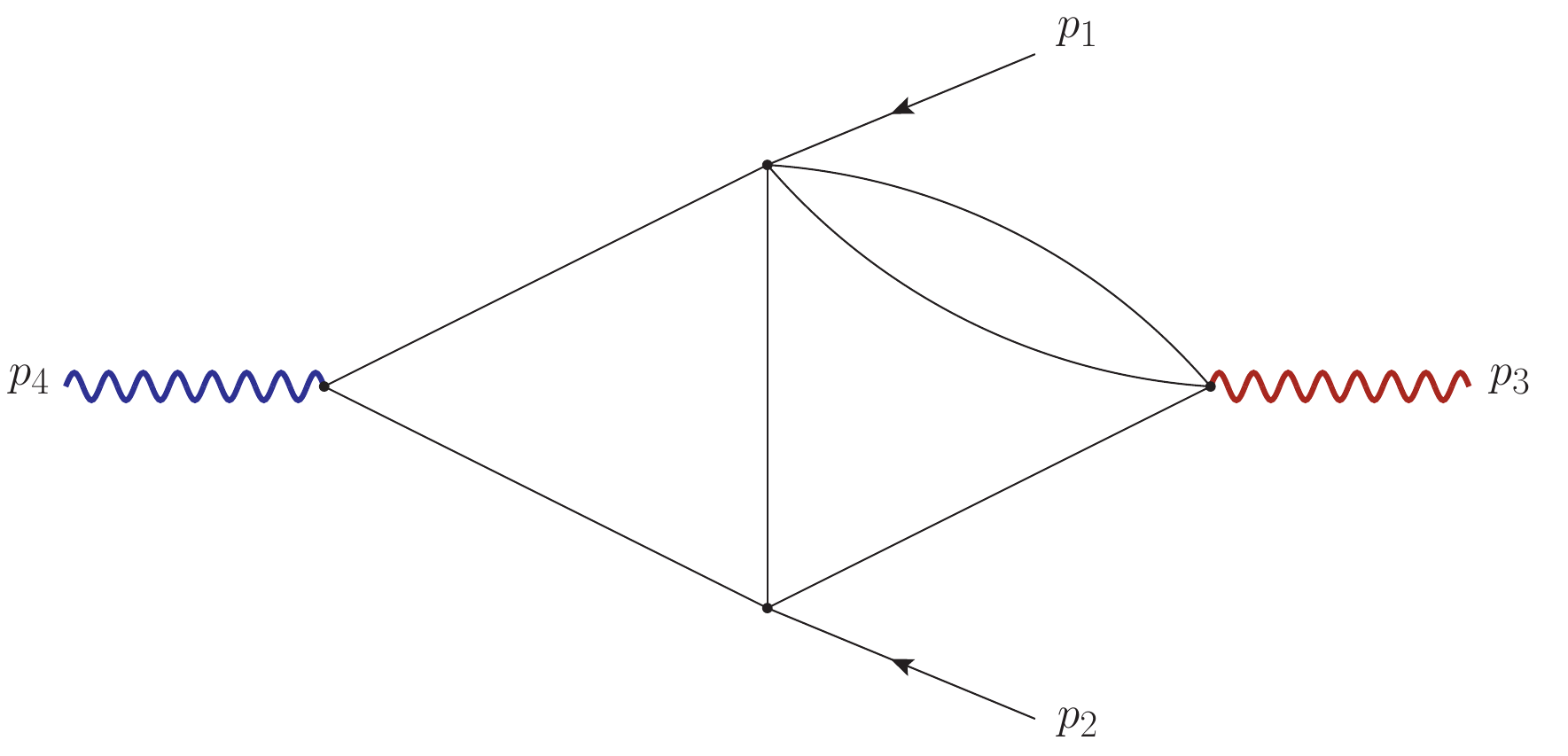}\\
\newline
\begin{equation*}
\{I_{79}^{\text{PL3}},I_{80}^{\text{PL3}},I_{81}^{\text{PL3}},I_{82}^{\text{PL3}},I_{83}^{\text{PL3}},I_{84}^{\text{PL3}}\}
\end{equation*}
\end{multicols}

\textbf{Sector $\mathbf{F_{132}}$[1,1,0,0,0,0,0,0,0,0,1,1,1,0,1]}

\begin{multicols}{2}
\includegraphics[scale=0.20]{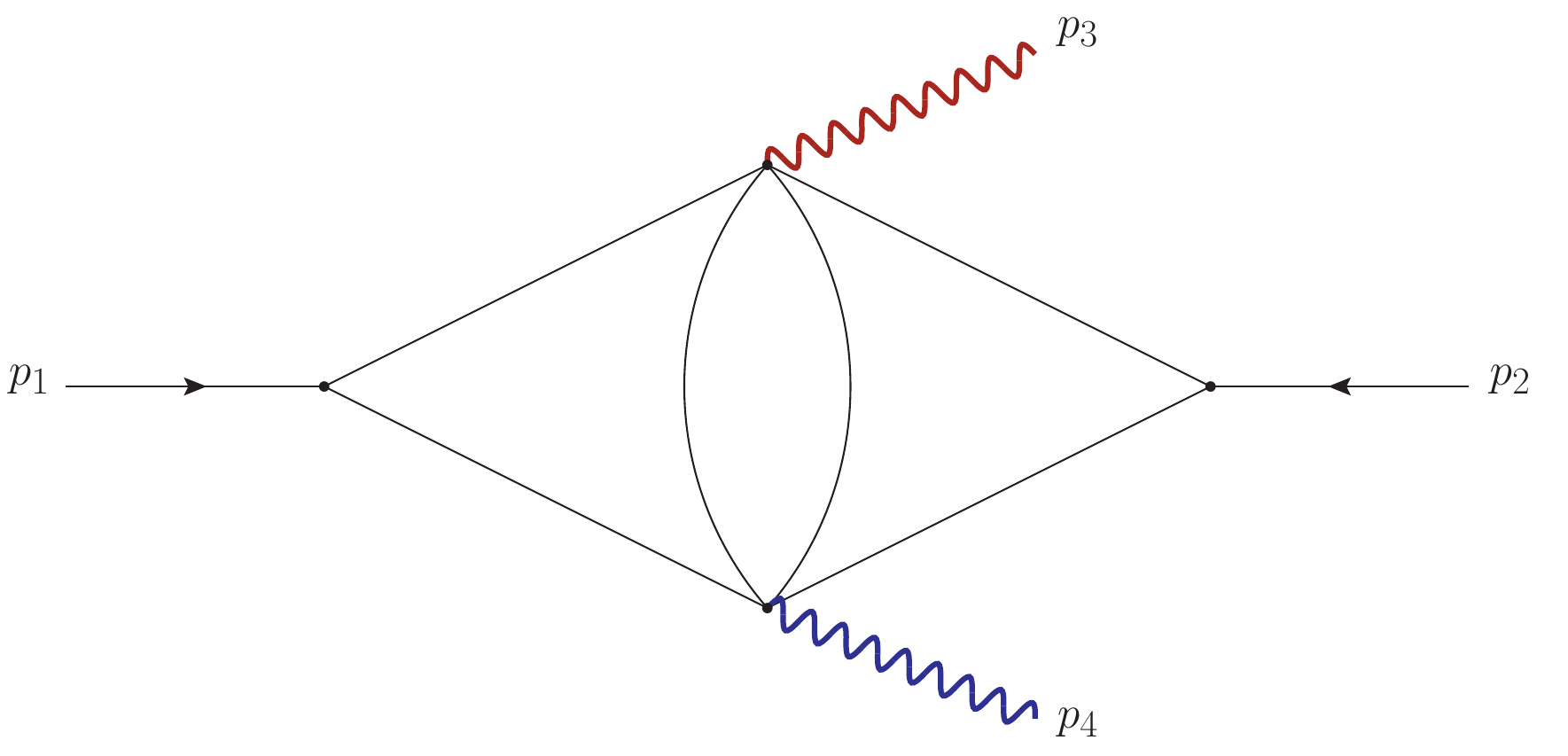}\\
\newline
\begin{equation*}
\{I_{97}^{\text{PL3}}\}
\end{equation*}
\end{multicols}

\textbf{Sector $\mathbf{F_{132}}$[0,1,0,0,0,0,1,1,1,0,0,0,1,0,1]}

\begin{multicols}{2}
\includegraphics[scale=0.20]{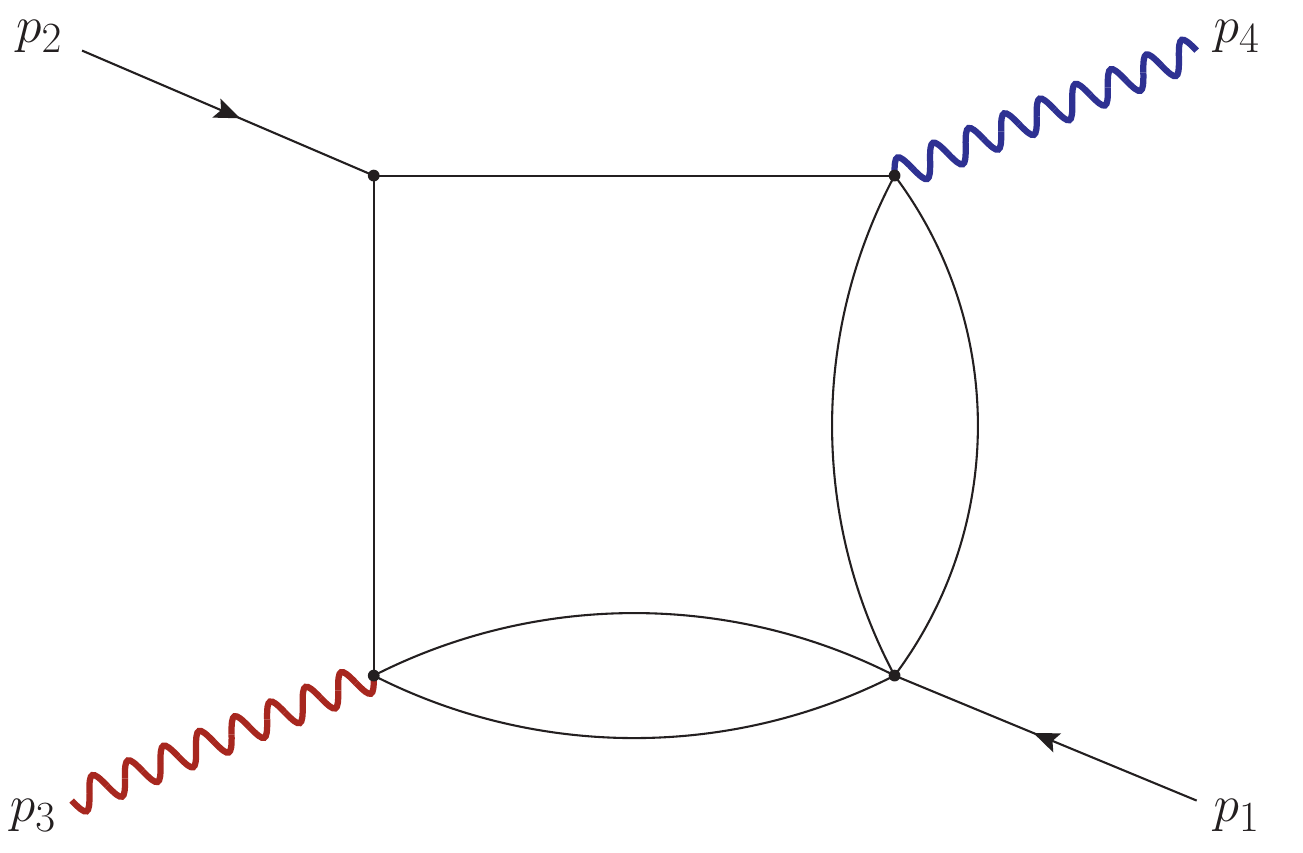}\\
\newline
\begin{equation*}
\{I_{200}^{\text{PT3}}\}
\end{equation*}
\end{multicols}

\textbf{Sector $\mathbf{F_{132}}$[0,1,0,0,0,0,1,0,1,0,0,1,0,1,1]}

\begin{multicols}{2}
\includegraphics[scale=0.20]{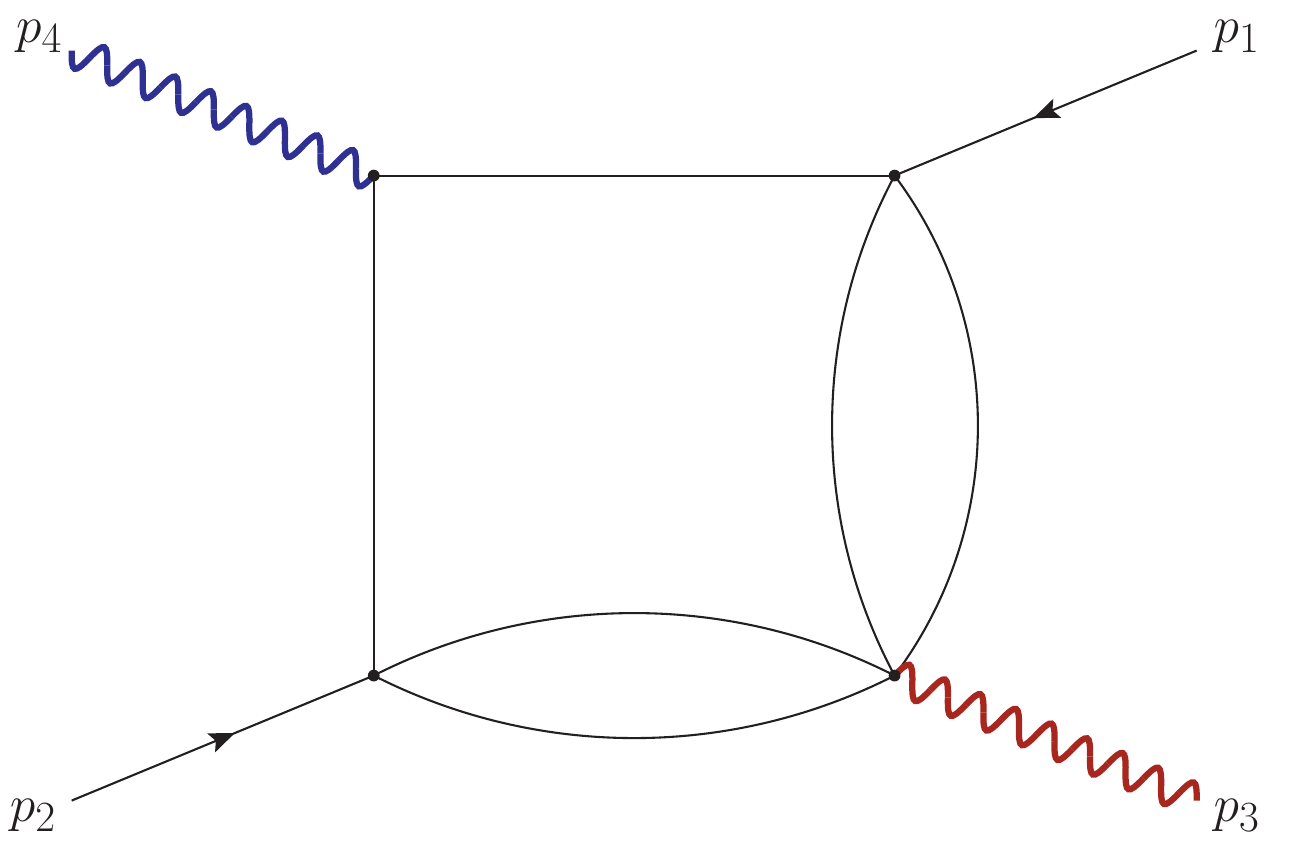}\\
\newline
\begin{equation*}
\{I_{171}^{\text{PT3}},I_{172}^{\text{PT3}}\}
\end{equation*}
\end{multicols}

\textbf{Sector $\mathbf{F_{123}}$[0,0,1,0,0,0,1,0,1,0,0,1,0,1,1]}

\begin{multicols}{2}
\includegraphics[scale=0.20]{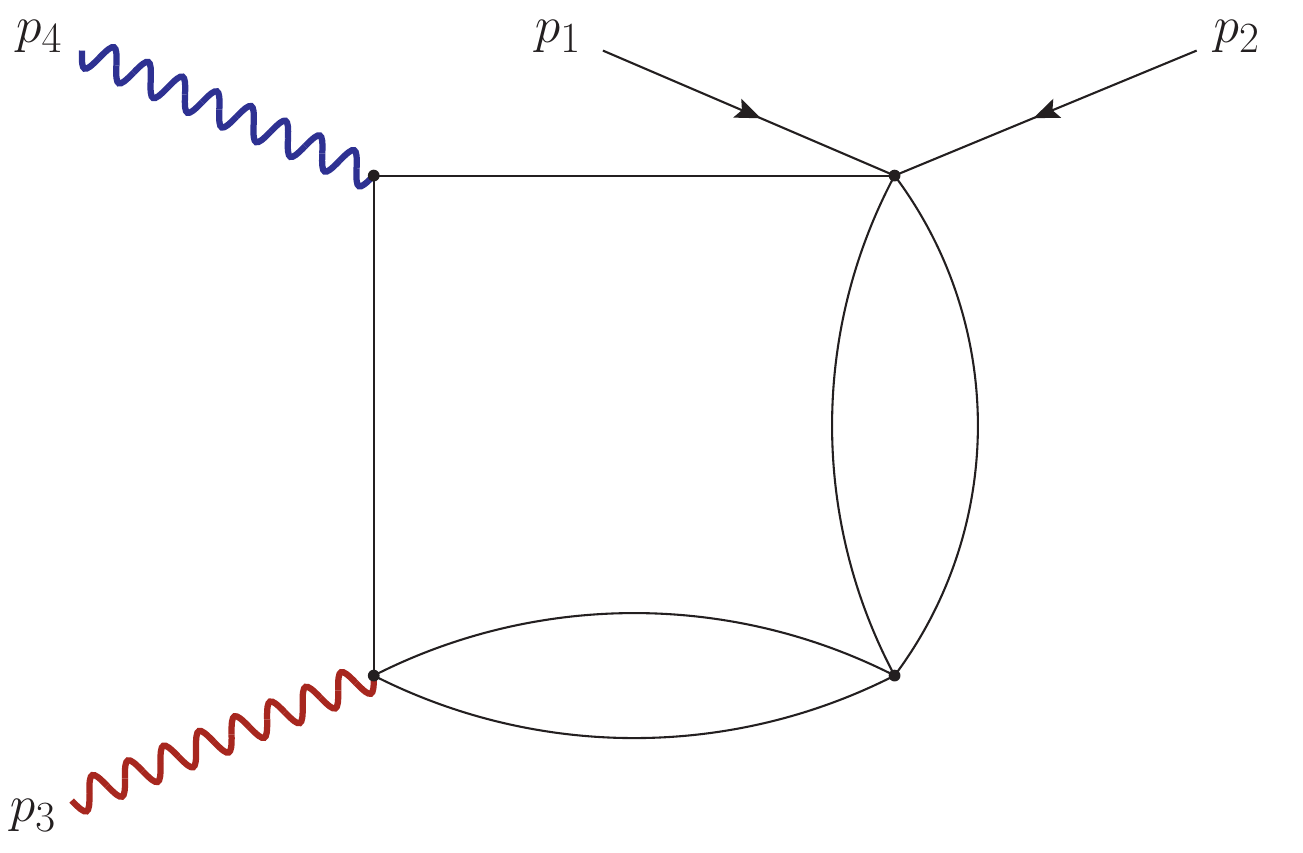}\\
\newline
\begin{equation*}
\{I_{281}^{\text{PT1}},I_{282}^{\text{PT1}}\}
\end{equation*}
\end{multicols}

\textbf{Sector $\mathbf{F_{123}}$[1,0,0,0,0,1,0,1,0,0,1,0,1,0,1]}

\begin{multicols}{2}
\includegraphics[scale=0.20]{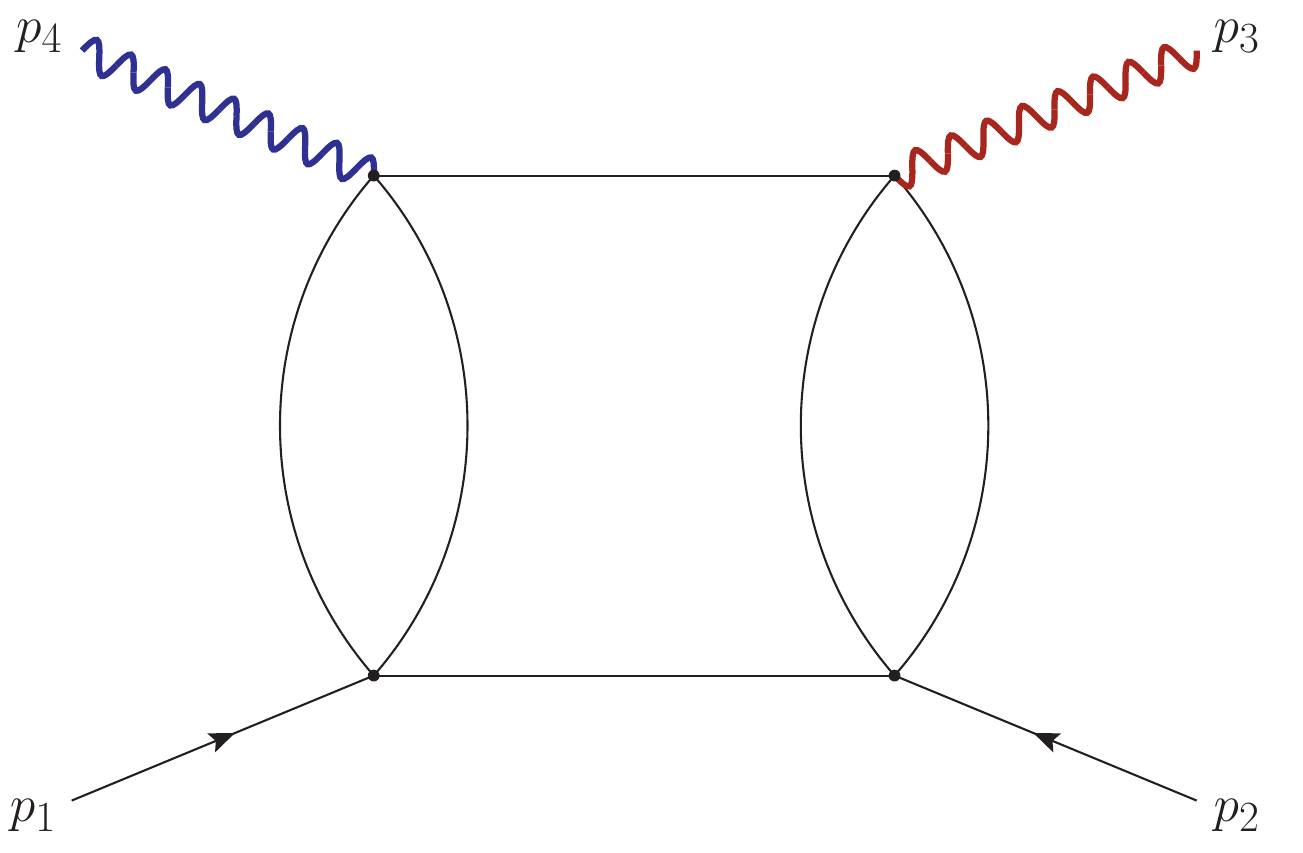}\\
\newline
\begin{equation*}
\{I_{86}^{\text{PL2}},I_{87}^{\text{PL2}}\}
\end{equation*}
\end{multicols}

\textbf{Sector $\mathbf{F_{132}}$[0,1,0,0,1,0,1,0,0,0,0,1,1,0,1]}

\begin{multicols}{2}
\includegraphics[scale=0.20]{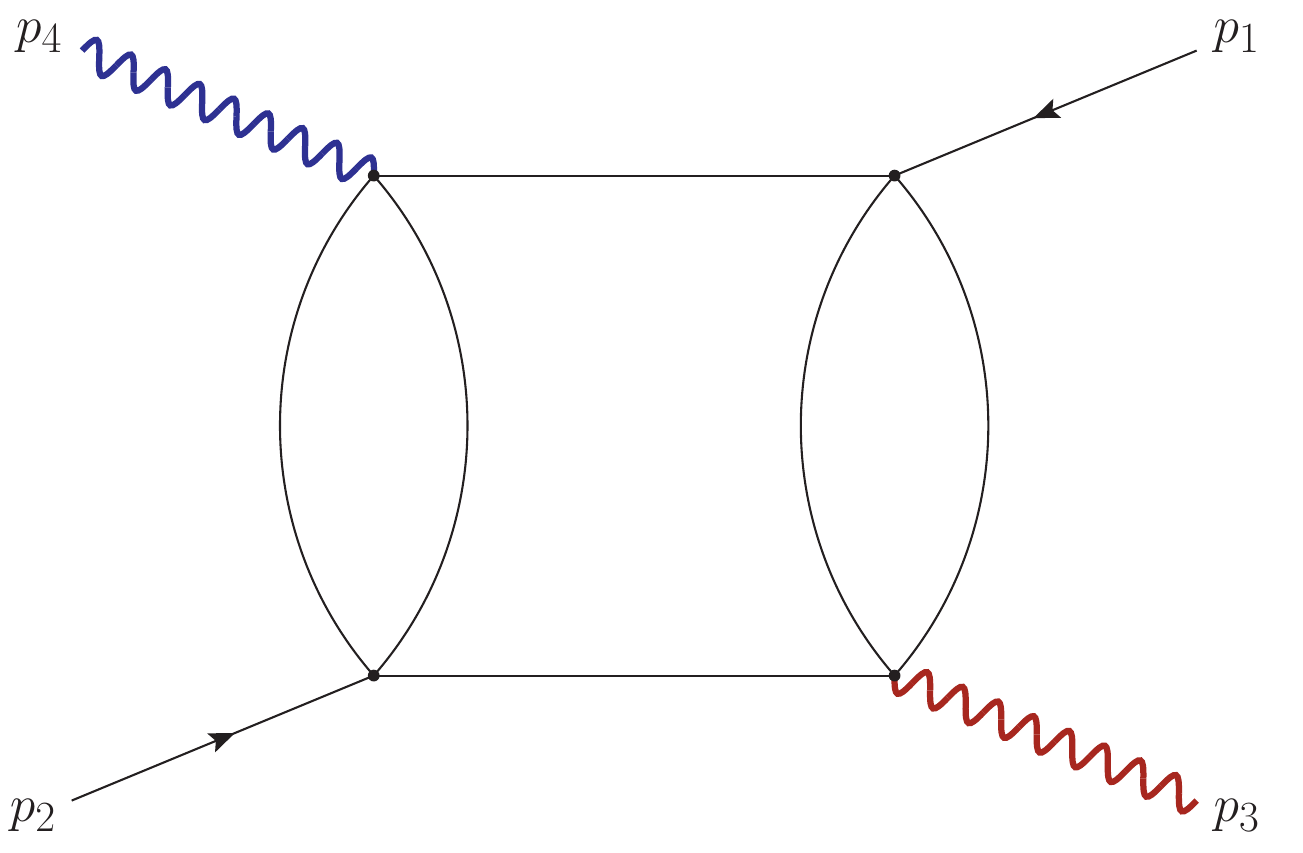}\\
\newline
\begin{equation*}
\{I_{94}^{\text{PL3}},I_{93}^{\text{PL3}}\}
\end{equation*}
\end{multicols}

\vspace{0.15cm}
\begin{center}
\textbf{\textit{Seven-Propagator Pure Candidates}}\\
\end{center}
\vspace{0.2cm}

\textbf{Sector $\mathbf{F_{123}}$[1,0,0,0,0,0,0,1,0,1,1,1,1,0,1]}

\begin{multicols}{2}
\includegraphics[scale=0.20]{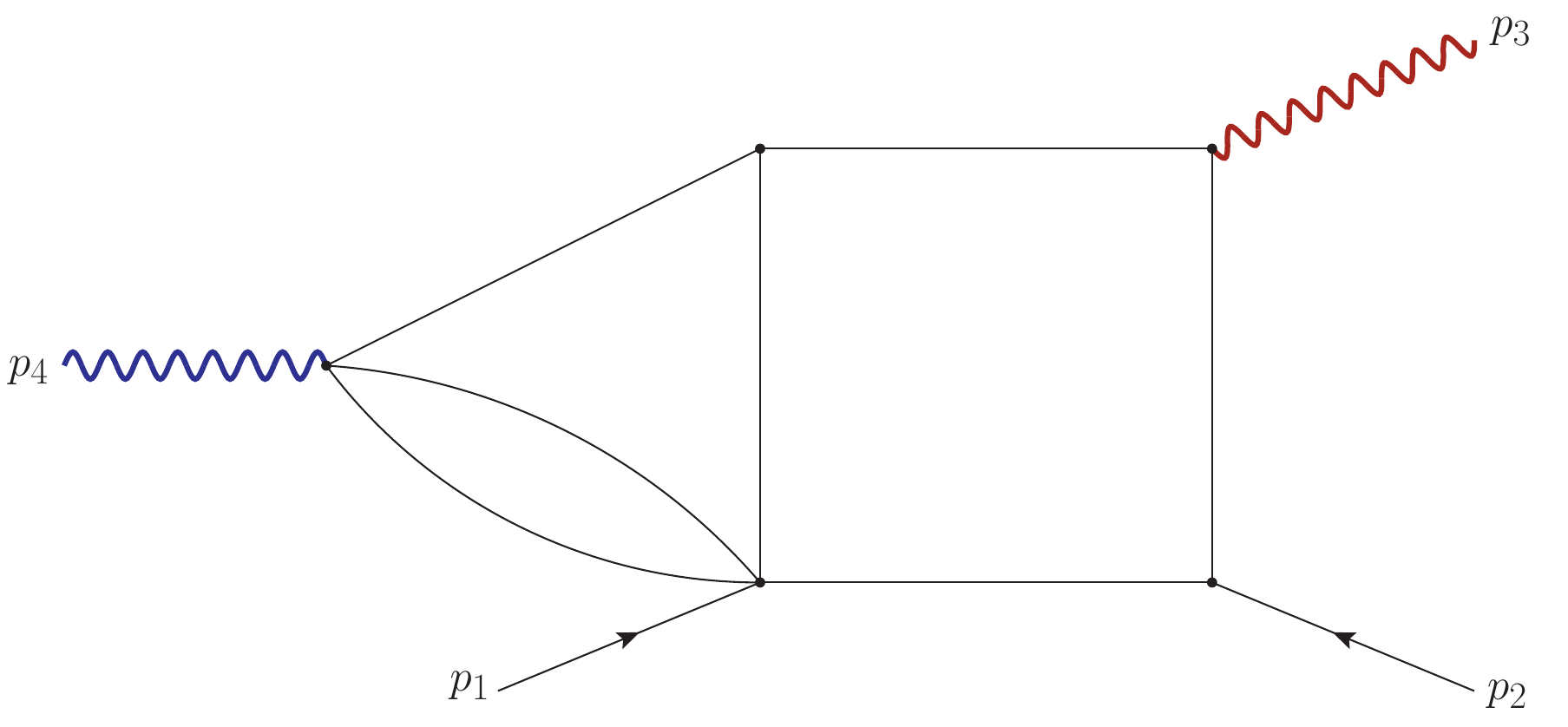}\\
\newline
\begin{equation*}
\{I_{65}^{\text{PL2}}\}
\end{equation*}
\end{multicols}

\textbf{Sector $\mathbf{F_{132}}$[0,1,0,0,0,0,1,0,1,0,1,1,1,0,1]}

\begin{multicols}{2}
\includegraphics[scale=0.20]{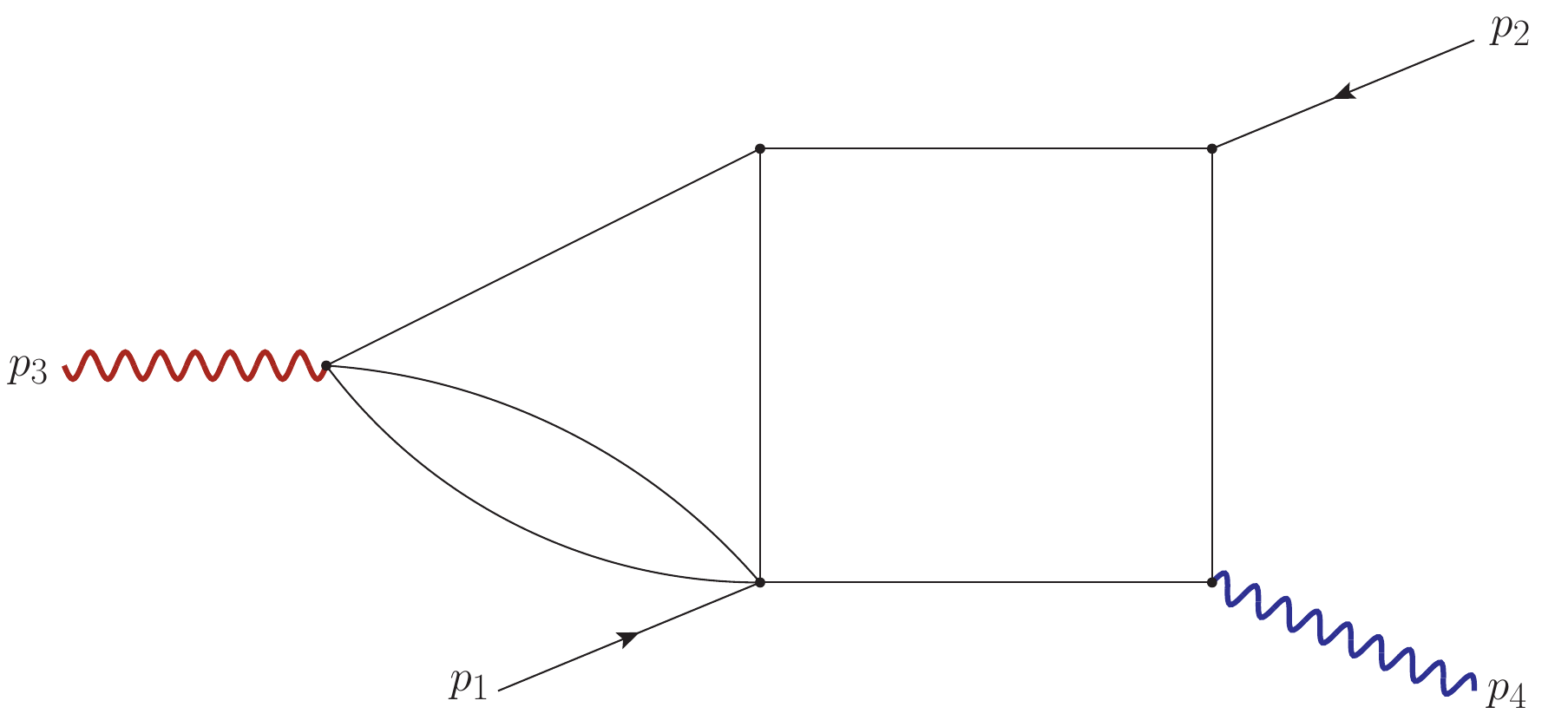}\\
\newline
\begin{equation*}
\{I_{66}^{\text{PL3}}\}
\end{equation*}
\end{multicols}

\textbf{Sector $\mathbf{F_{123}}$[0,1,1,1,1,0,0,0,0,1,0,0,1,0,1]}

\begin{multicols}{2}
\includegraphics[scale=0.20]{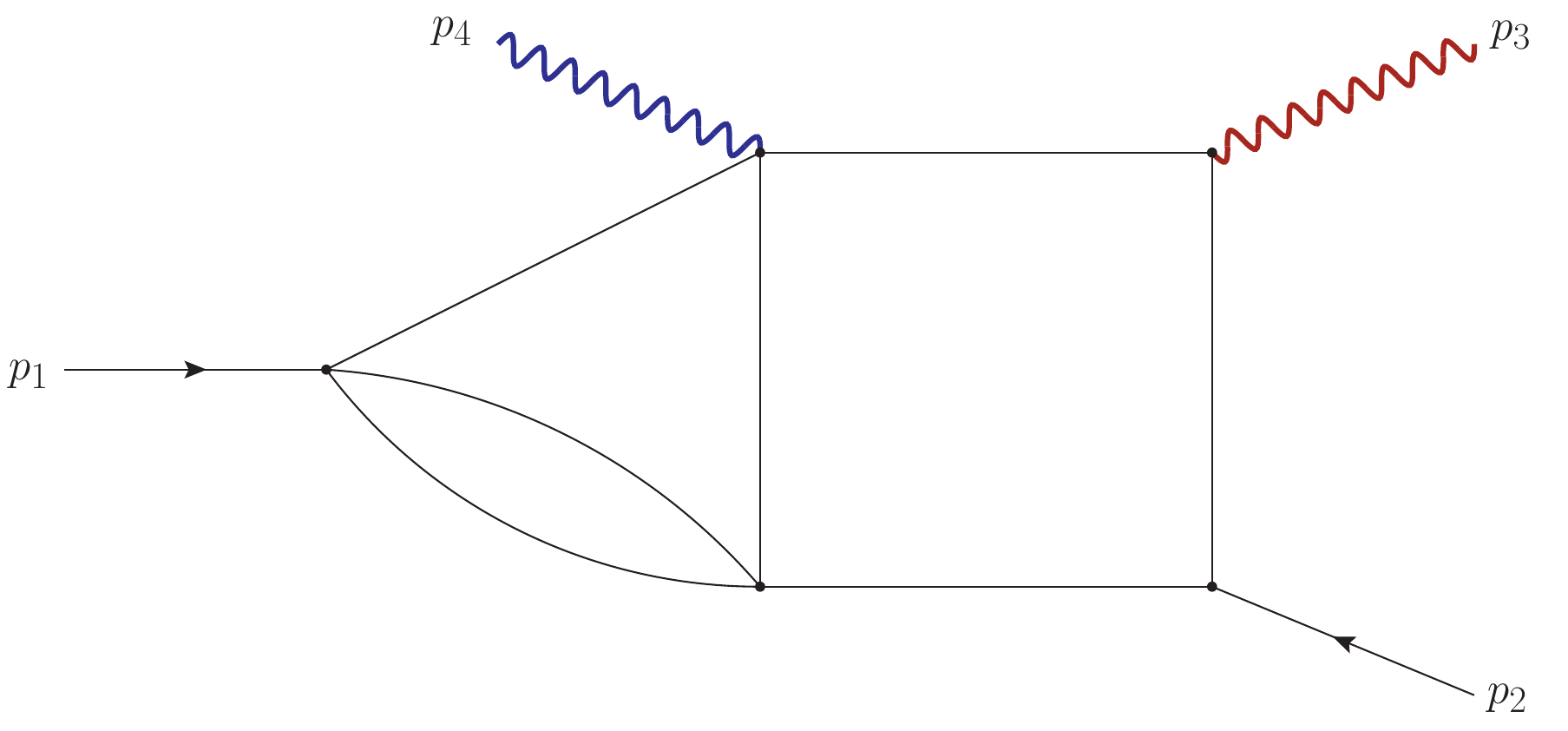}\\
\newline
\begin{equation*}
\{I_{129}^{\text{PT2}}\}
\end{equation*}
\end{multicols}

\textbf{Sector $\mathbf{F_{132}}$[1,1,1,0,0,0,1,0,0,0,0,1,0,1,1]}

\begin{multicols}{2}
\includegraphics[scale=0.20]{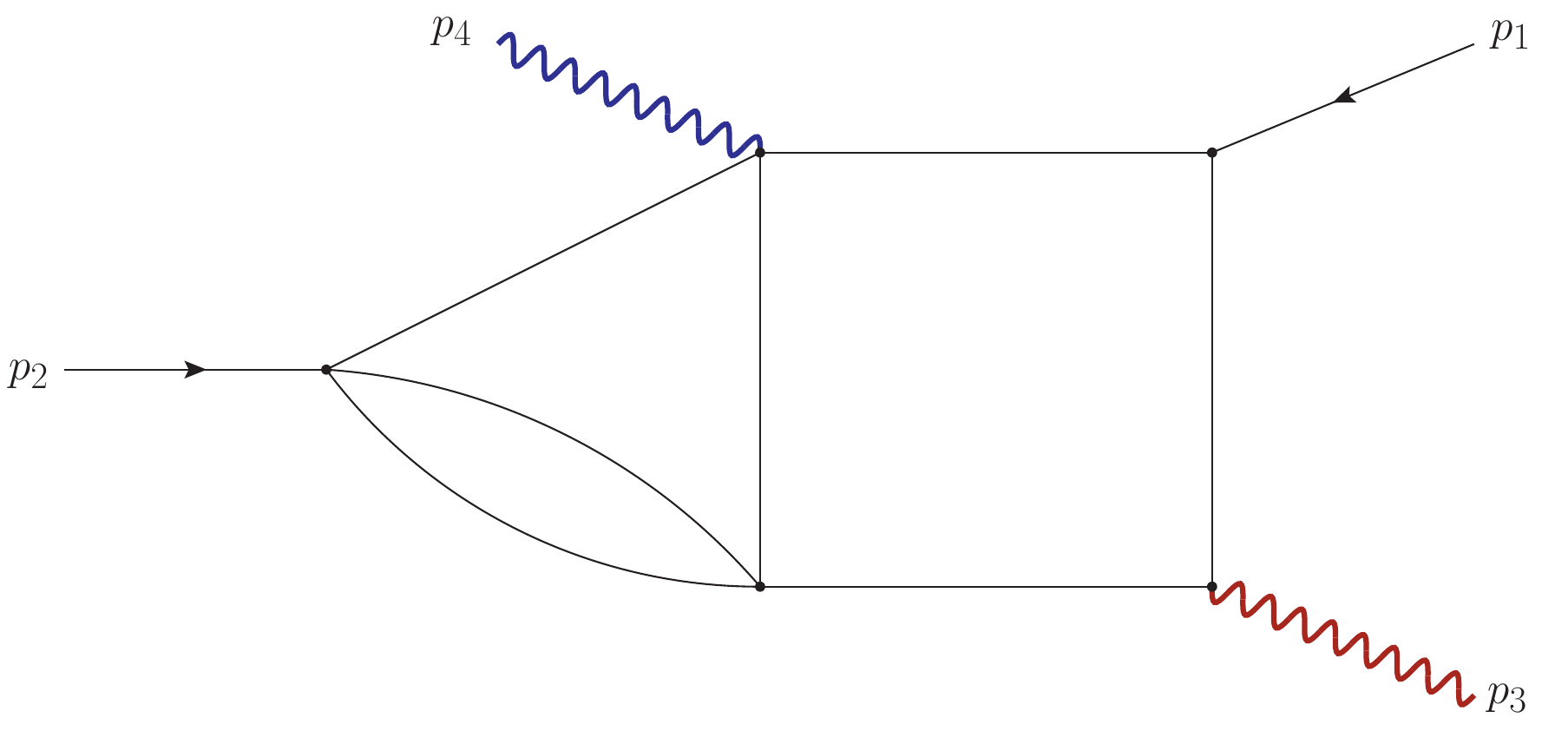}\\
\newline
\begin{equation*}
\{I_{111}^{\text{PT3}}\}
\end{equation*}
\end{multicols}

\textbf{Sector $\mathbf{F_{123}}$[1,1,1,0,0,0,0,1,1,0,0,0,1,0,1]}

\begin{multicols}{2}
\includegraphics[scale=0.20]{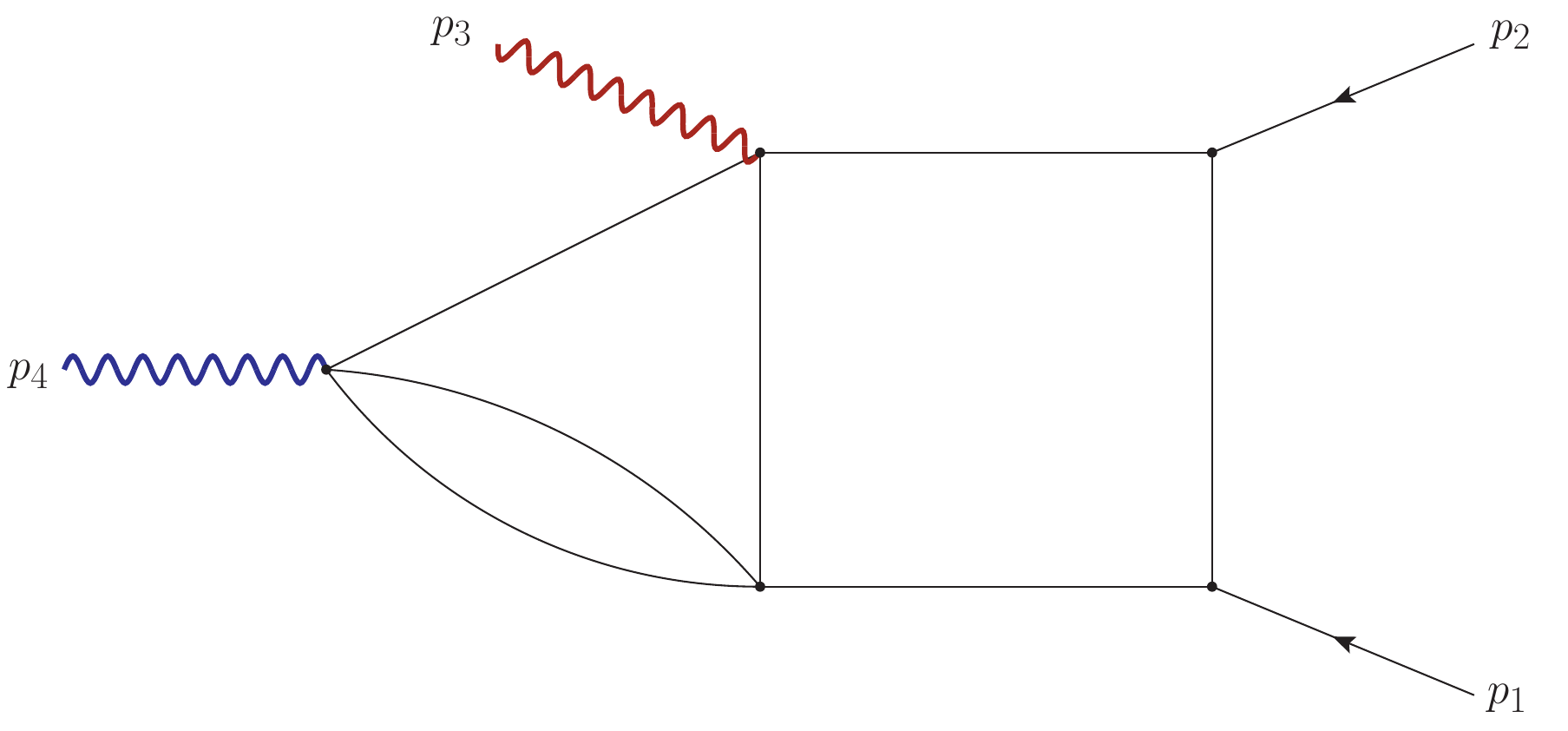}\\
\newline
\begin{equation*}
\{I_{161}^{\text{PT1}},I_{162}^{\text{PT1}}\}
\end{equation*}
\end{multicols}

\textbf{Sector $\mathbf{F_{123}}$[0,1,1,1,0,0,0,1,1,0,0,0,0,1,1]}

\begin{multicols}{2}
\includegraphics[scale=0.20]{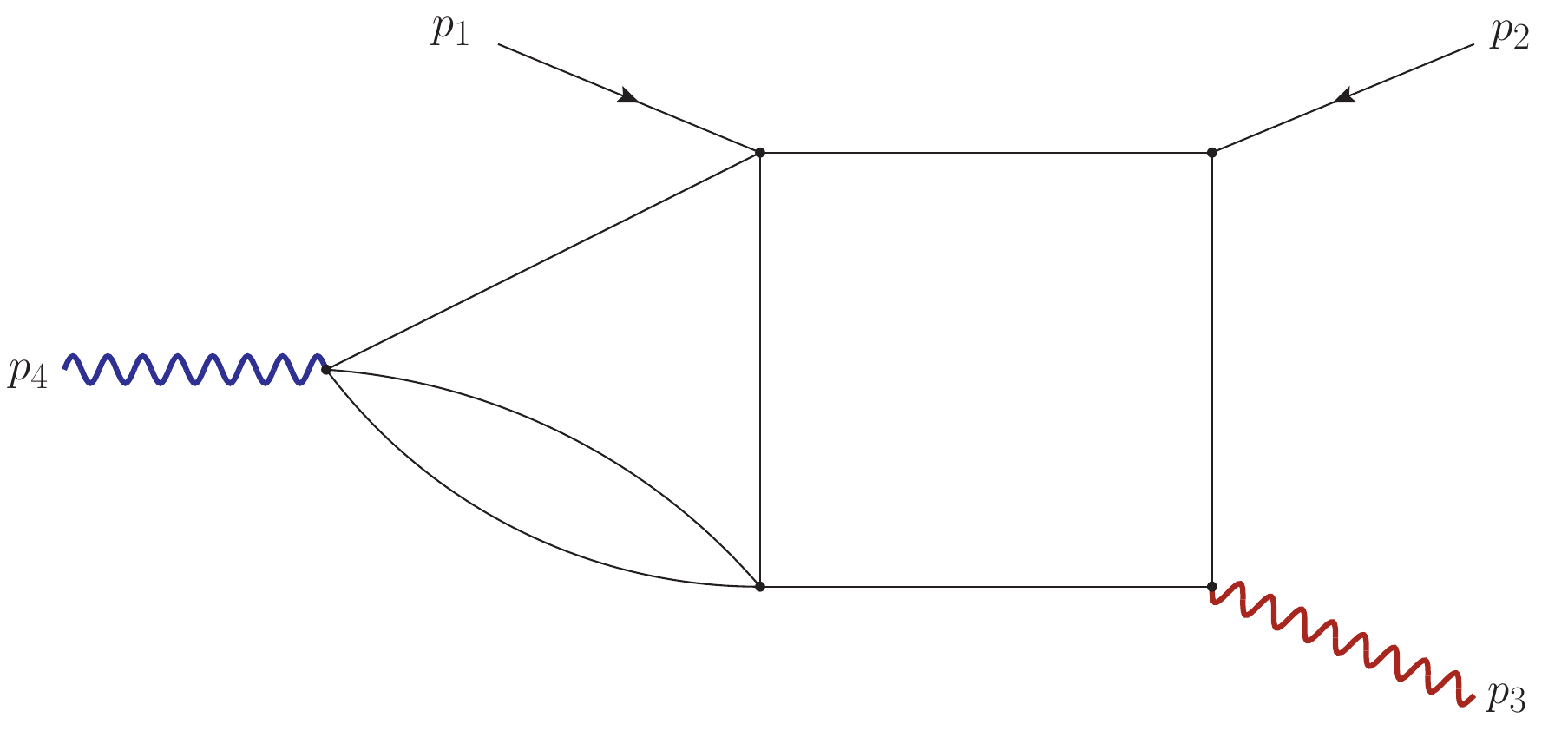}\\
\newline
\begin{equation*}
\{I_{108}^{\text{PT2}},I_{109}^{\text{PT2}}\}
\end{equation*}
\end{multicols}

\textbf{Sector $\mathbf{F_{132}}$[1,1,1,0,0,0,0,1,1,0,0,0,1,0,1]}

\begin{multicols}{2}
\includegraphics[scale=0.20]{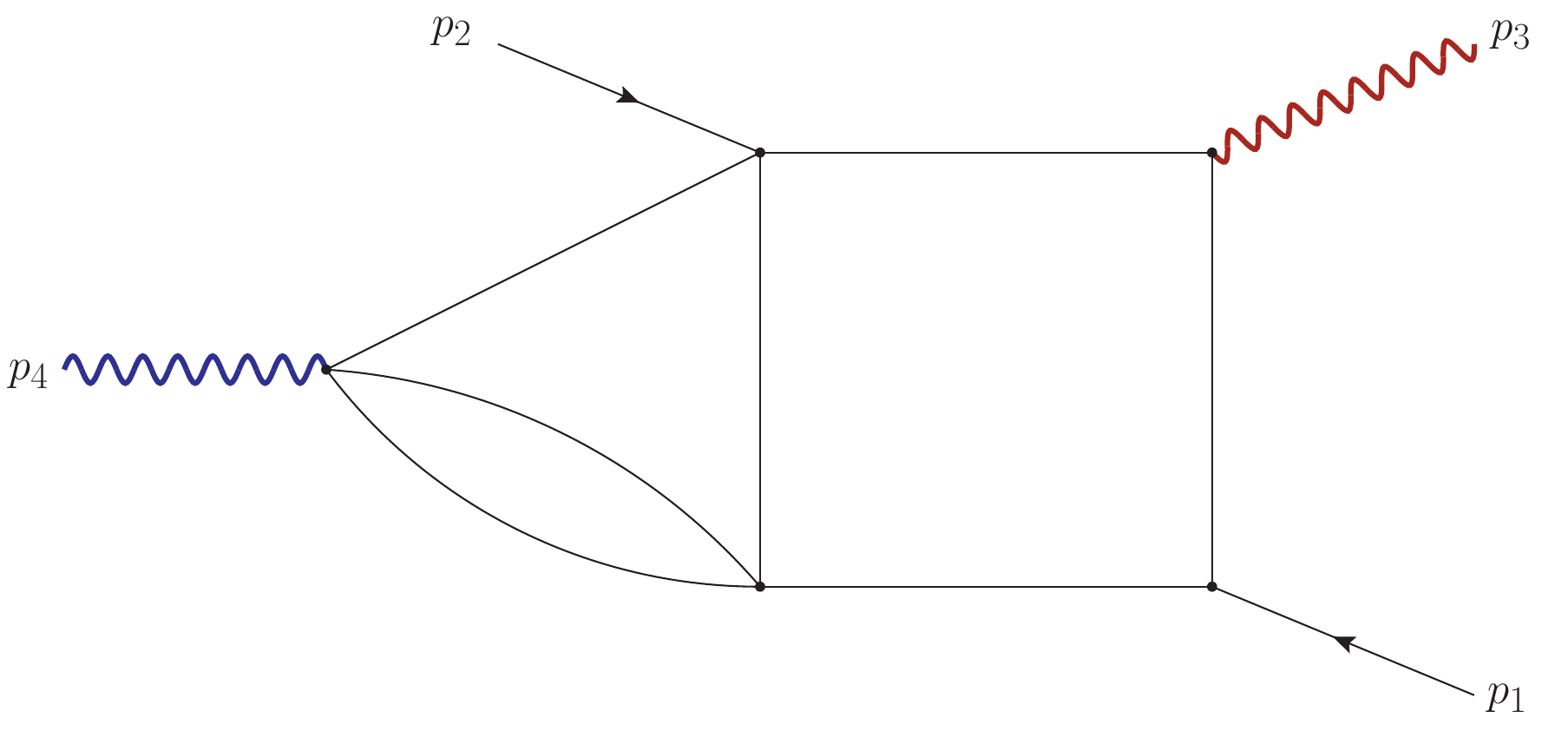}\\
\newline
\begin{equation*}
\{I_{105}^{\text{PT3}},I_{106}^{\text{PT3}}\}
\end{equation*}
\end{multicols}

\textbf{Sector $\mathbf{F_{123}}$[0,1,0,1,1,0,0,1,0,1,0,0,1,0,1]}

\begin{multicols}{2}
\includegraphics[scale=0.20]{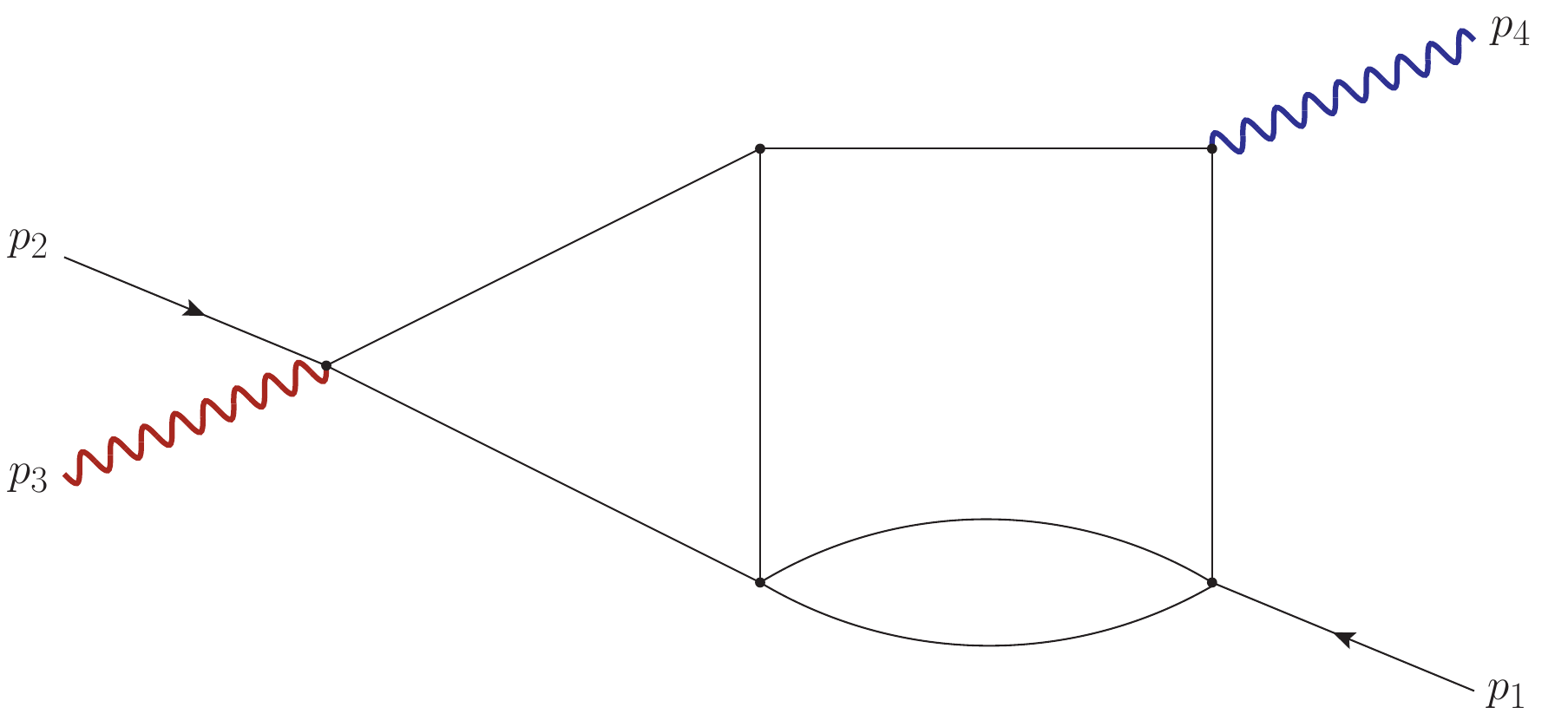}\\
\newline
\begin{equation*}
\{I_{132}^{\text{PT2}}\}
\end{equation*}
\end{multicols}

\textbf{Sector $\mathbf{F_{123}}$[0,0,1,0,0,0,1,1,1,0,0,1,0,1,1]}

\begin{multicols}{2}
\includegraphics[scale=0.20]{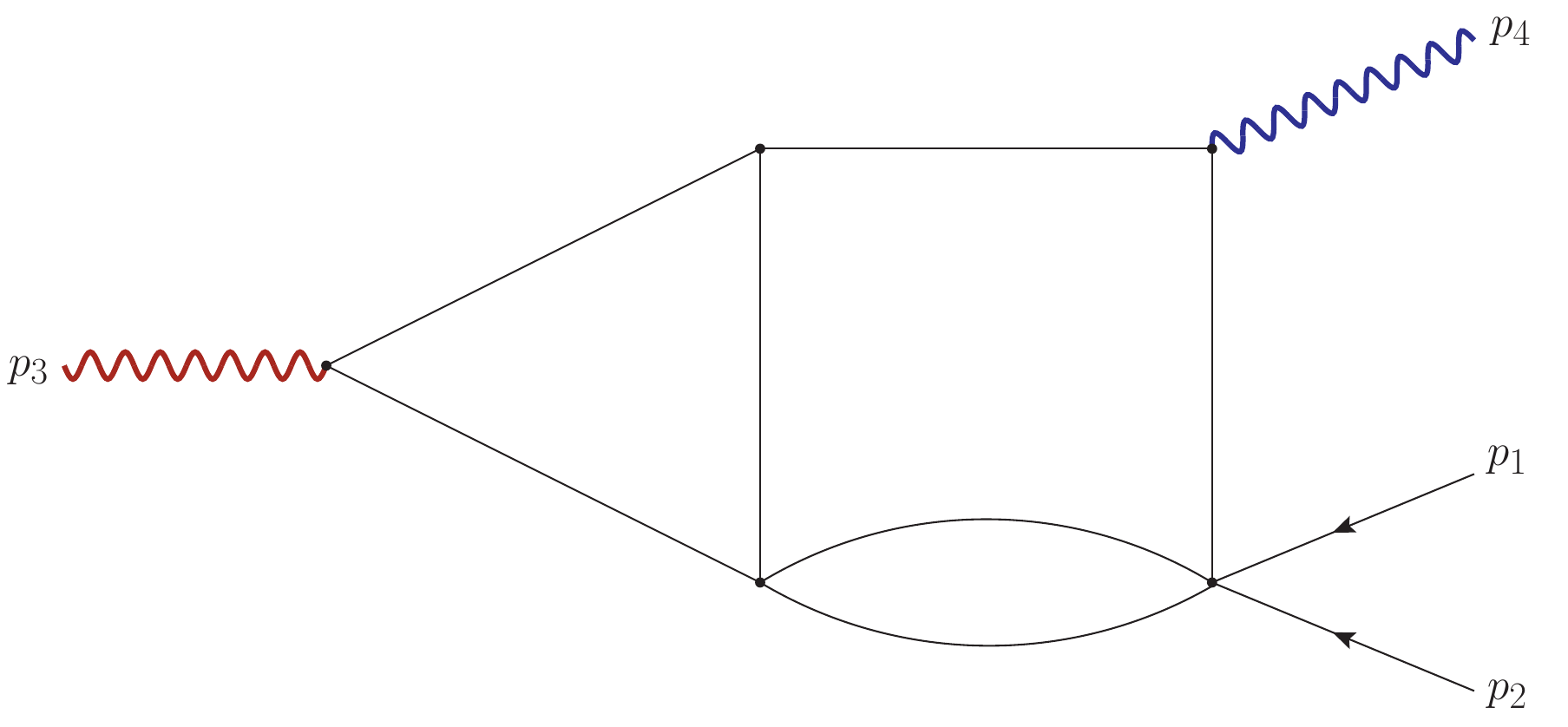}\\
\newline
\begin{equation*}
\{I_{198}^{\text{PT1}}\}
\end{equation*}
\end{multicols}

\textbf{Sector $\mathbf{F_{132}}$[0,1,0,0,0,0,1,1,1,0,0,1,1,0,1]}

\begin{multicols}{2}
\includegraphics[scale=0.20]{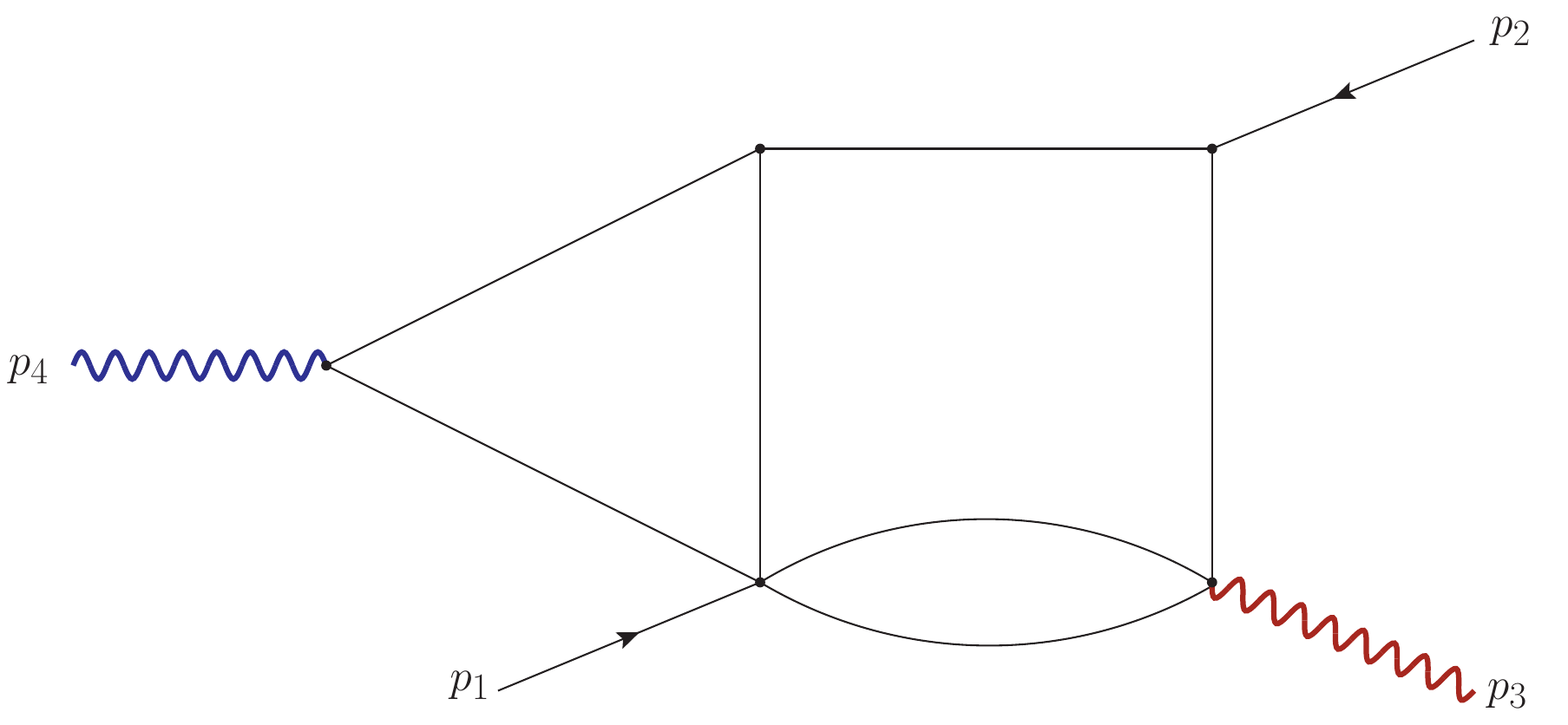}\\
\newline
\begin{equation*}
\{I_{130}^{\text{PT3}}\}
\end{equation*}
\end{multicols}

\textbf{Sector $\mathbf{F_{132}}$[1,1,0,0,0,0,1,1,0,0,0,1,1,1,0]}

\begin{multicols}{2}
\includegraphics[scale=0.20]{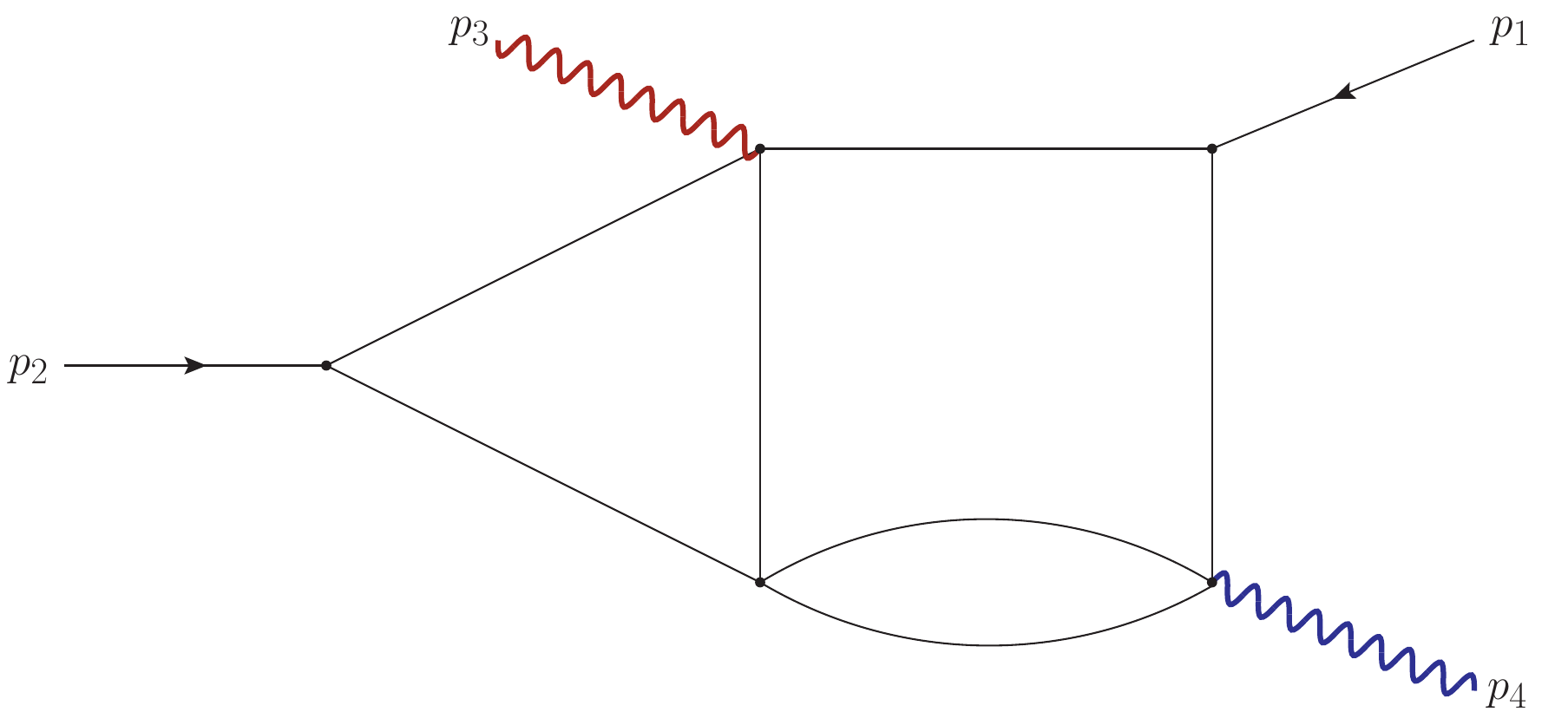}\\
\newline
\begin{equation*}
\{I_{122}^{\text{PT3}}\}
\end{equation*}
\end{multicols}

\textbf{Sector $\mathbf{F_{123}}$[0,1,0,0,0,0,1,1,1,0,0,1,0,1,1]}

\begin{multicols}{2}
\includegraphics[scale=0.20]{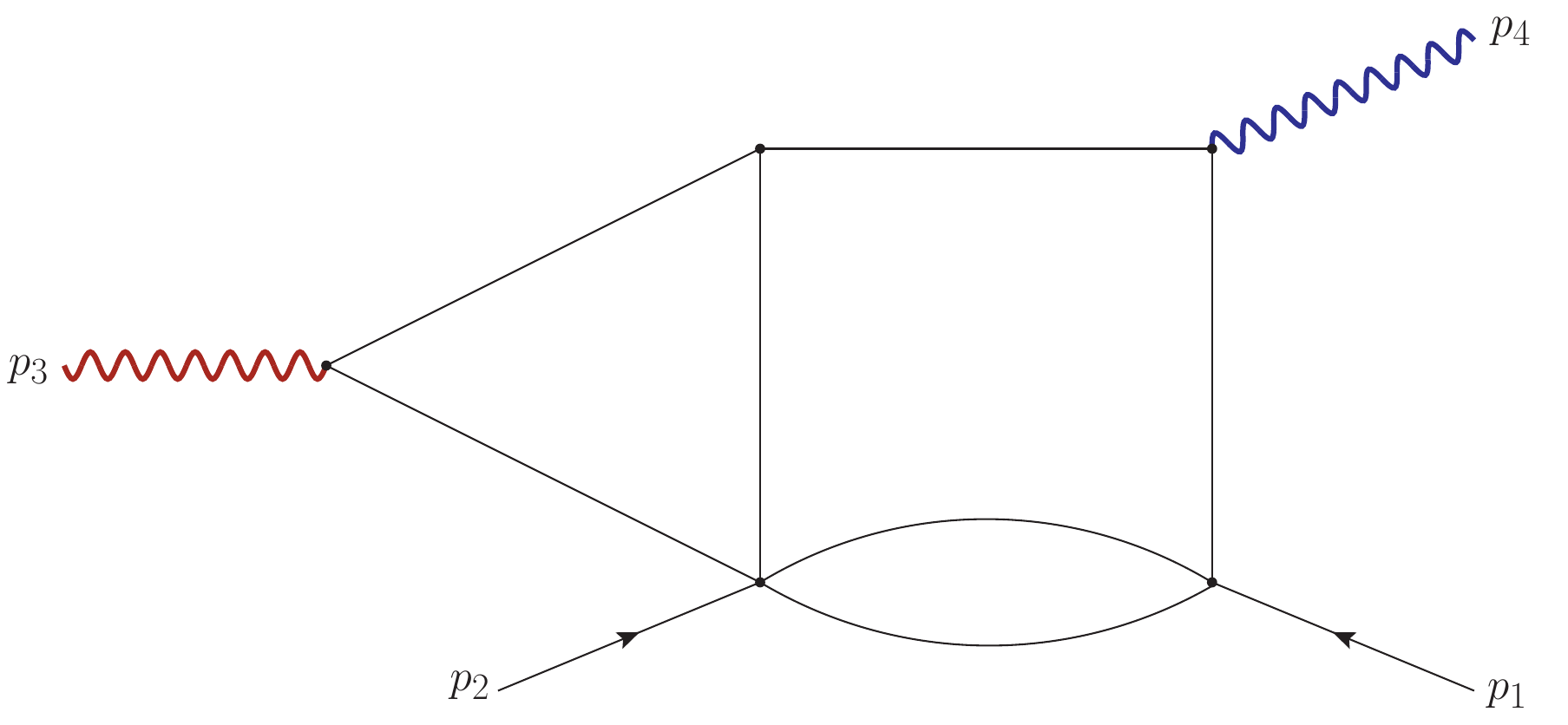}\\
\newline
\begin{equation*}
\{I_{177}^{\text{PT1}},I_{178}^{\text{PT1}}\}
\end{equation*}
\end{multicols}

\textbf{Sector $\mathbf{F_{123}}$[0,1,1,0,0,0,1,0,1,0,0,1,0,1,1]}

\begin{multicols}{2}
\includegraphics[scale=0.20]{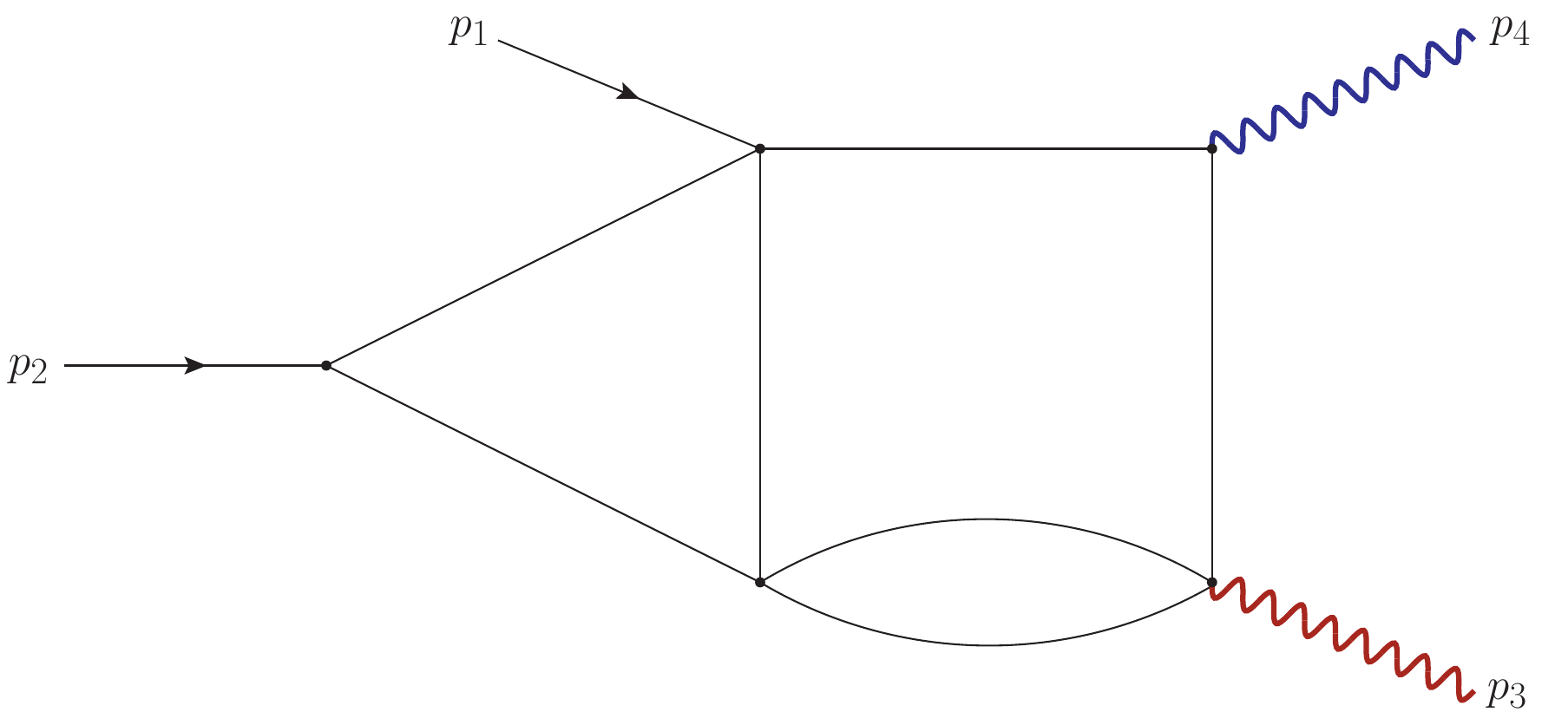}\\
\newline
\begin{equation*}
\{I_{171}^{\text{PT1}},I_{172}^{\text{PT1}}\}
\end{equation*}
\end{multicols}

\textbf{Sector $\mathbf{F_{123}}$[0,1,1,0,0,0,0,1,1,0,0,1,1,1,0]}

\begin{multicols}{2}
\includegraphics[scale=0.20]{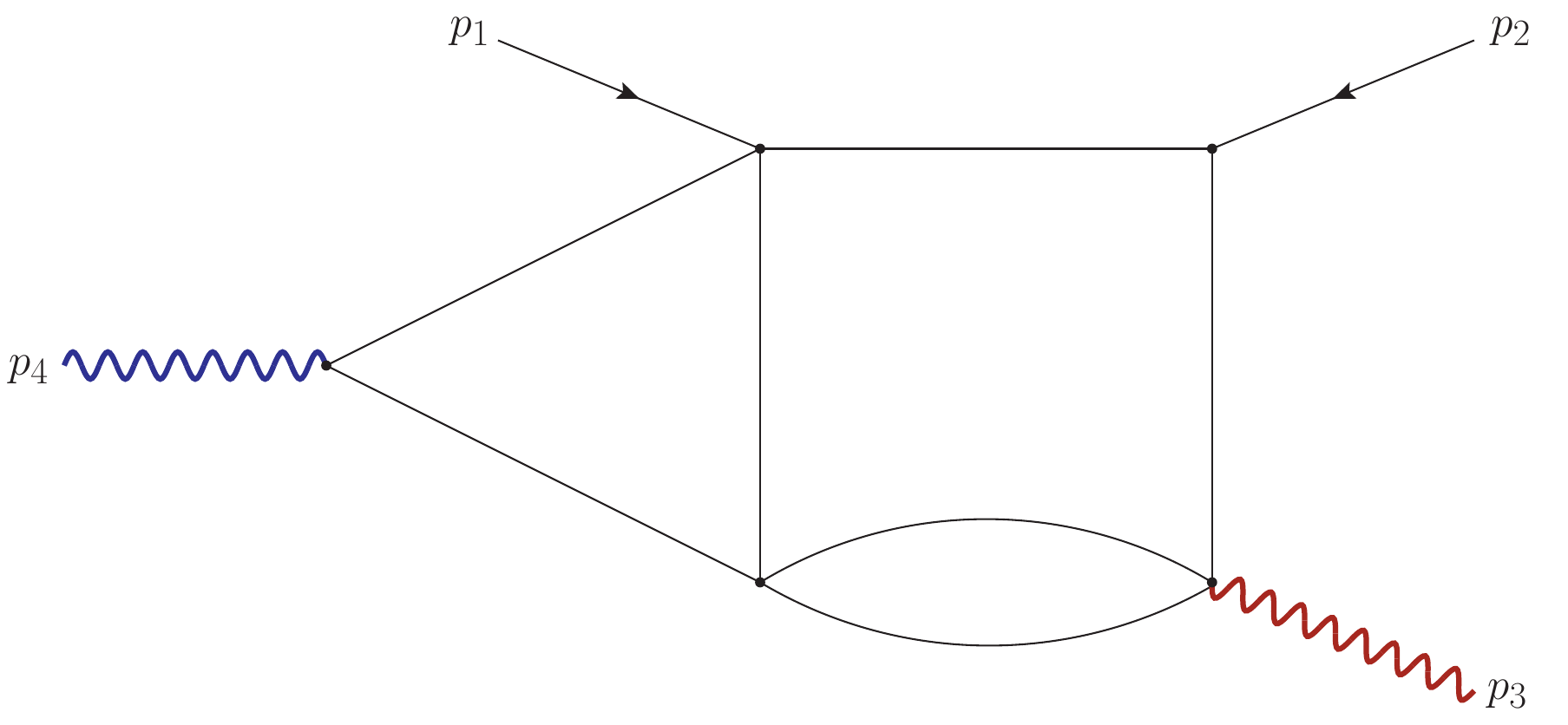}\\
\newline
\begin{equation*}
\{I_{153}^{\text{PT1}},I_{154}^{\text{PT1}},I_{155}^{\text{PT1}}\}
\end{equation*}
\end{multicols}

\textbf{Sector $\mathbf{F_{132}}$[0,1,1,0,0,0,0,1,1,0,0,1,1,1,0]}

\begin{multicols}{2}
\includegraphics[scale=0.20]{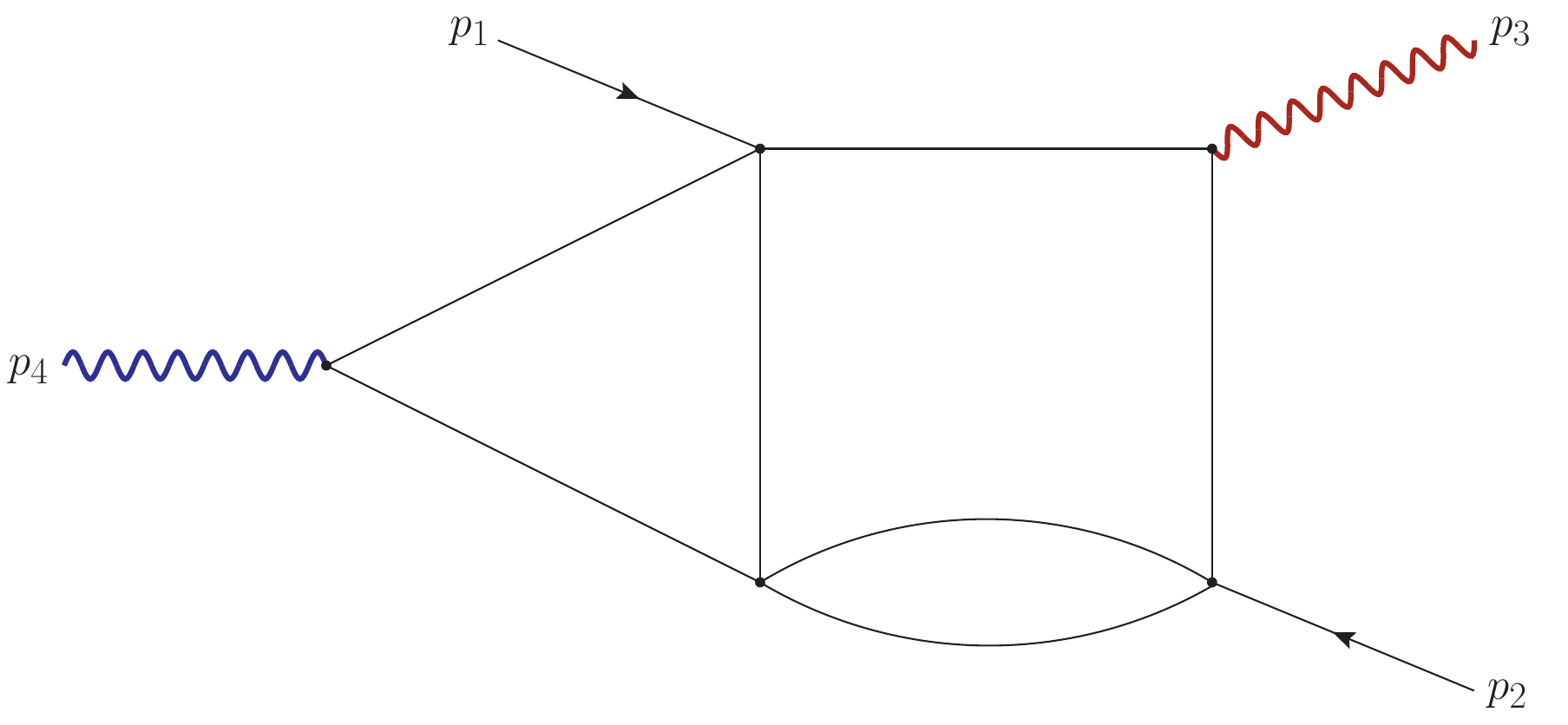}\\
\newline
\begin{equation*}
\{I_{74}^{\text{PT3}},I_{75}^{\text{PT3}},I_{76}^{\text{PT3}},I_{77}^{\text{PT3}},I_{78}^{\text{PT3}},I_{79}^{\text{PT3}}\}
\end{equation*}
\end{multicols}

\textbf{Sector $\mathbf{F_{123}}$[1,0,0,0,0,1,0,1,0,0,1,1,1,0,1]}

\begin{multicols}{2}
\includegraphics[scale=0.20]{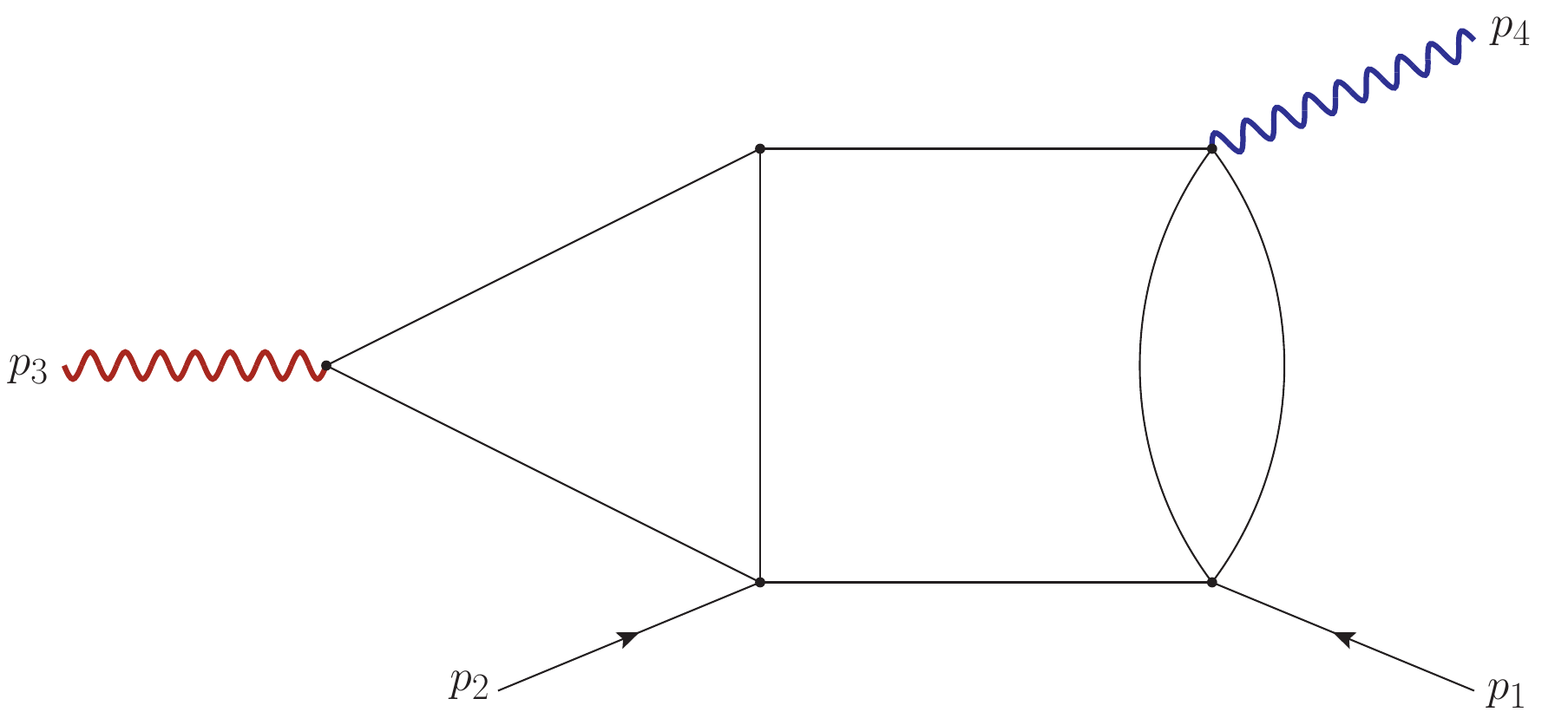}\\
\newline
\begin{equation*}
\{I_{54}^{\text{PL2}},I_{55}^{\text{PL2}}\}
\end{equation*}
\end{multicols}

\textbf{Sector $\mathbf{F_{132}}$[0,1,0,0,1,0,1,0,1,0,0,1,1,0,1]}

\begin{multicols}{2}
\includegraphics[scale=0.20]{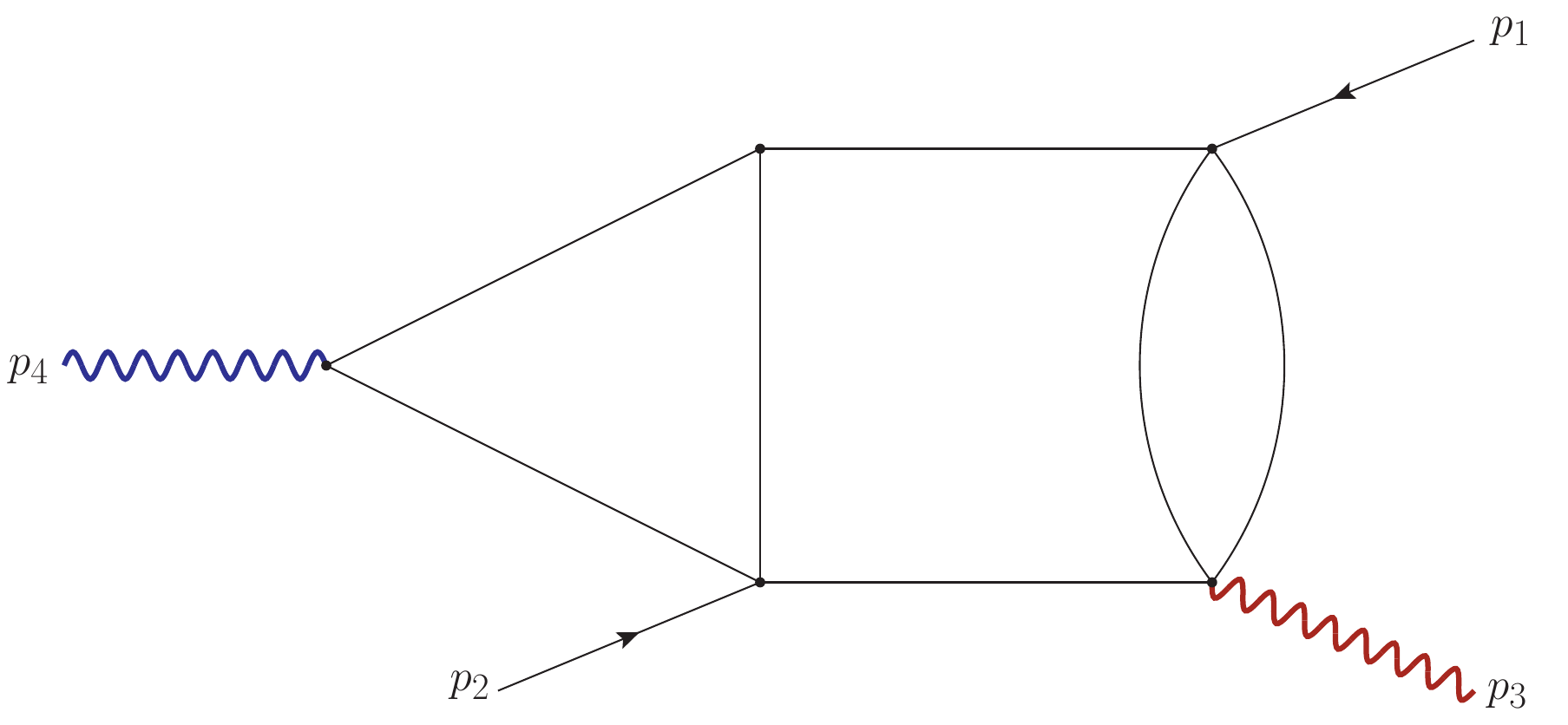}\\
\newline
\begin{equation*}
\{I_{55}^{\text{PL3}},I_{56}^{\text{PL3}}\}
\end{equation*}
\end{multicols}

\textbf{Sector $\mathbf{F_{123}}$[1,1,1,0,0,0,0,1,0,0,0,1,1,1,0]}

\begin{multicols}{2}
\includegraphics[scale=0.20]{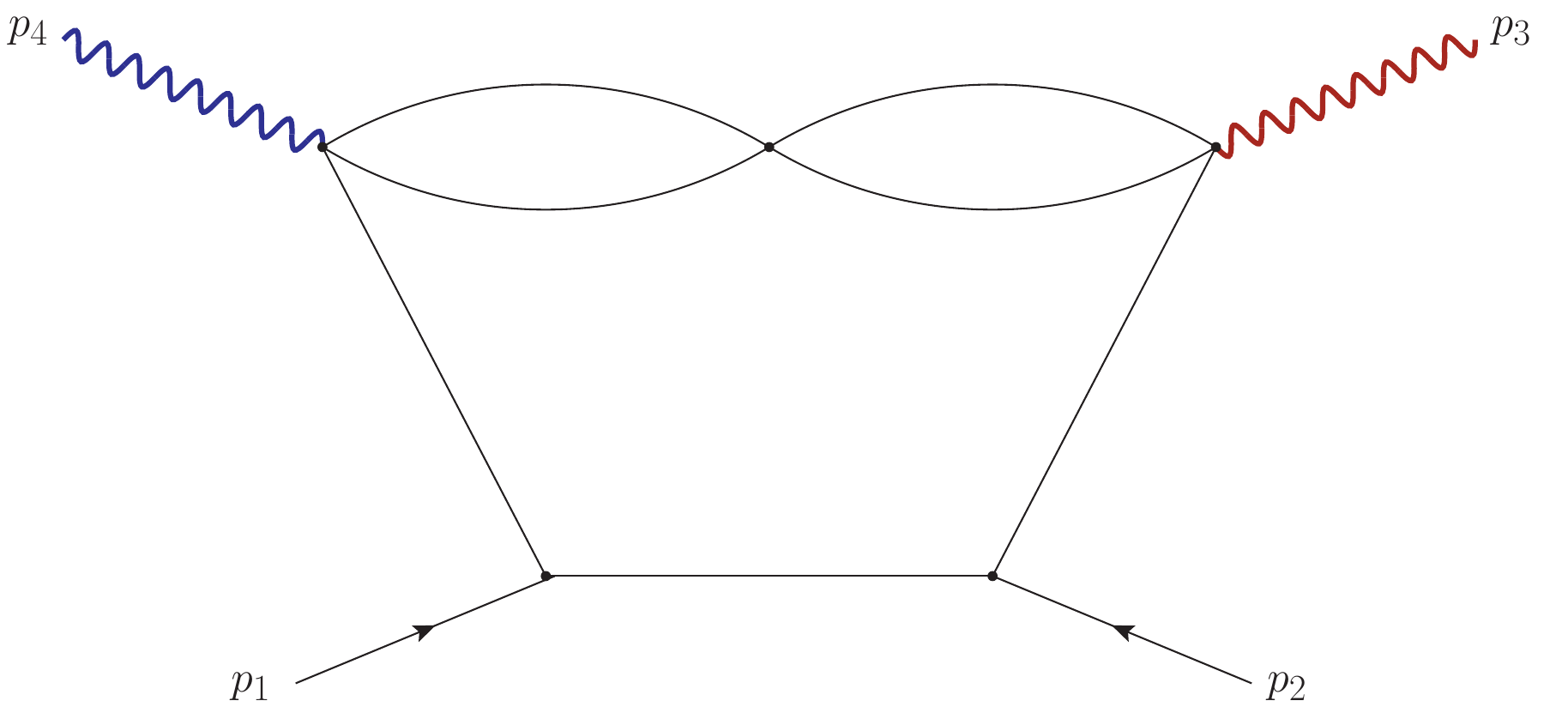}\\
\newline
\begin{equation*}
\{I_{184}^{\text{PT1}}\}
\end{equation*}
\end{multicols}

\textbf{Sector $\mathbf{F_{123}}$[0,1,1,1,1,0,0,0,1,0,0,0,1,1,0]}

\begin{multicols}{2}
\includegraphics[scale=0.20]{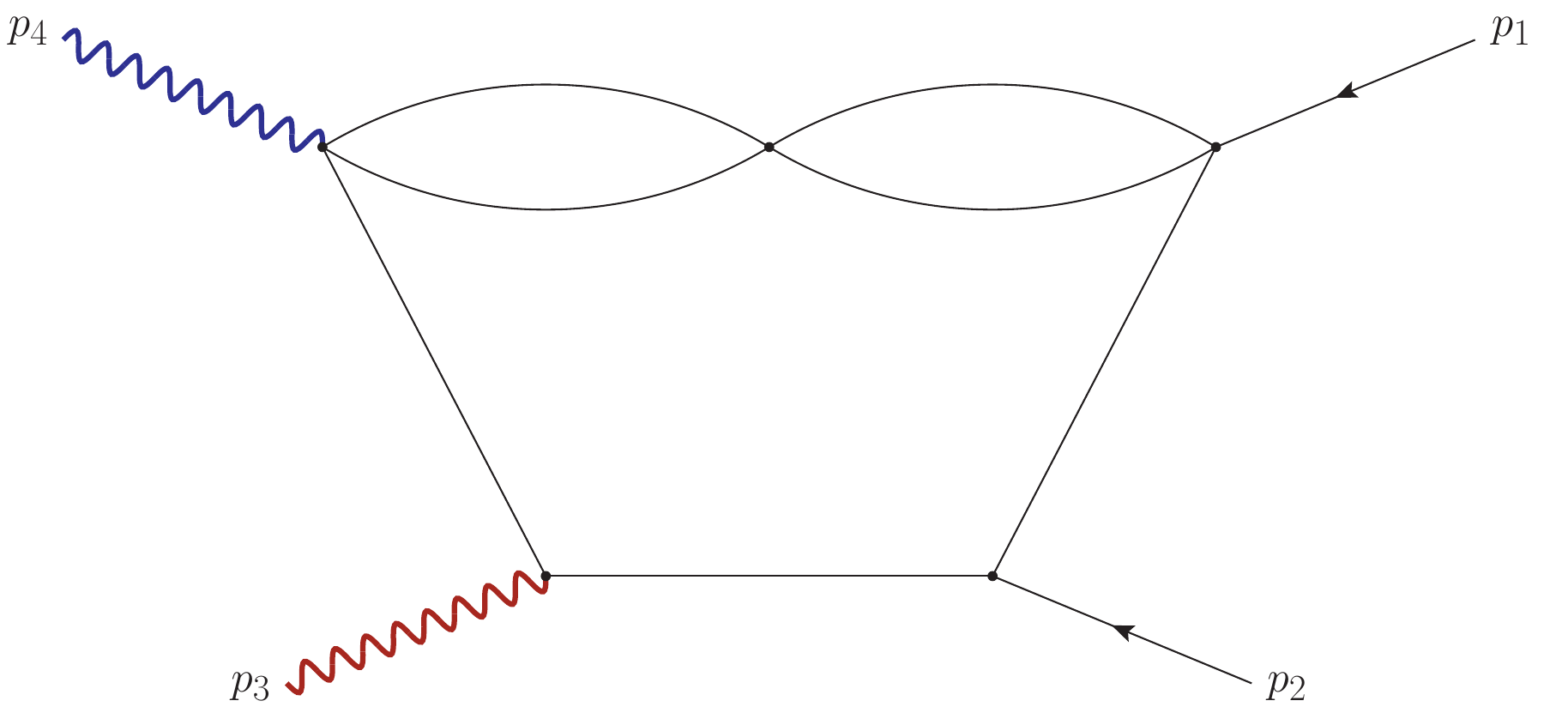}\\
\newline
\begin{equation*}
\{I_{128}^{\text{PT2}}\}
\end{equation*}
\end{multicols}

\textbf{Sector $\mathbf{F_{132}}$[1,1,1,0,0,0,0,1,0,0,0,1,1,1,0]}

\begin{multicols}{2}
\includegraphics[scale=0.20]{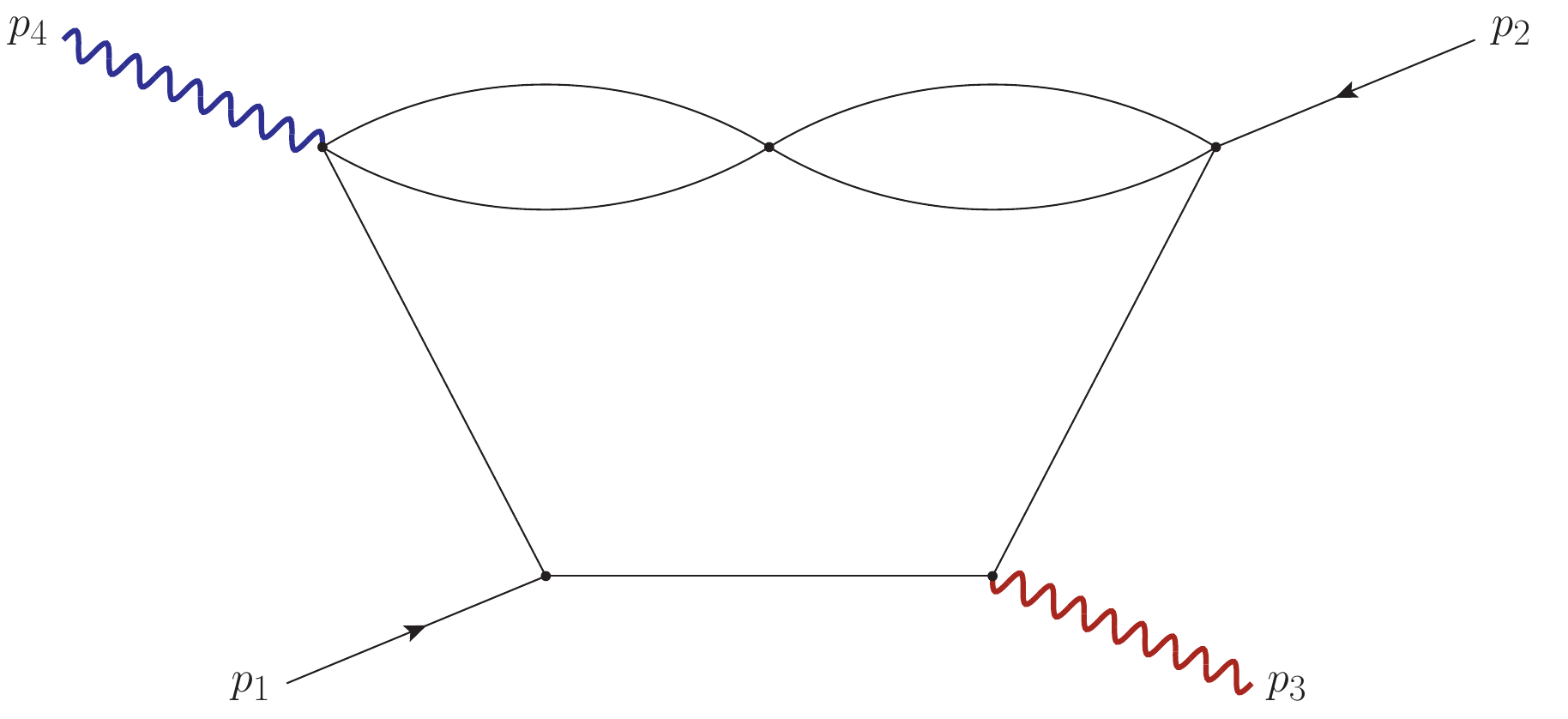}\\
\newline
\begin{equation*}
\{I_{118}^{\text{PT3}}\}
\end{equation*}
\end{multicols}

\textbf{Sector $\mathbf{F_{132}}$[0,1,0,0,0,0,1,1,1,0,0,0,1,1,1]}

\begin{multicols}{2}
\includegraphics[scale=0.20]{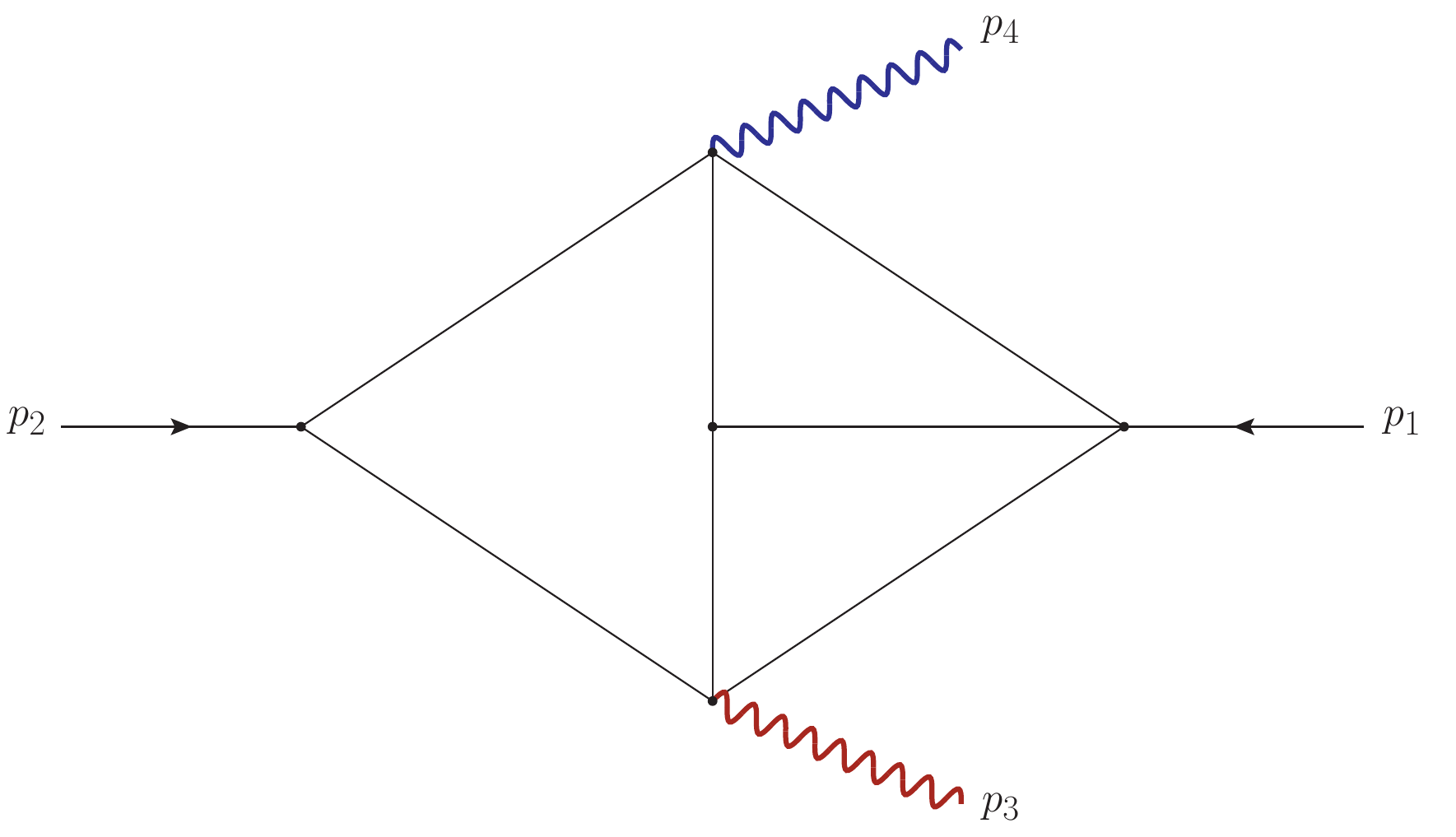}\\
\newline
\begin{equation*}
\{I_{113}^{\text{PT3}},I_{114}^{\text{PT3}}\}
\end{equation*}
\end{multicols}

\textbf{Sector $\mathbf{F_{132}}$[0,1,1,0,0,0,0,1,1,0,0,0,1,1,1]}

\begin{multicols}{2}
\includegraphics[scale=0.20]{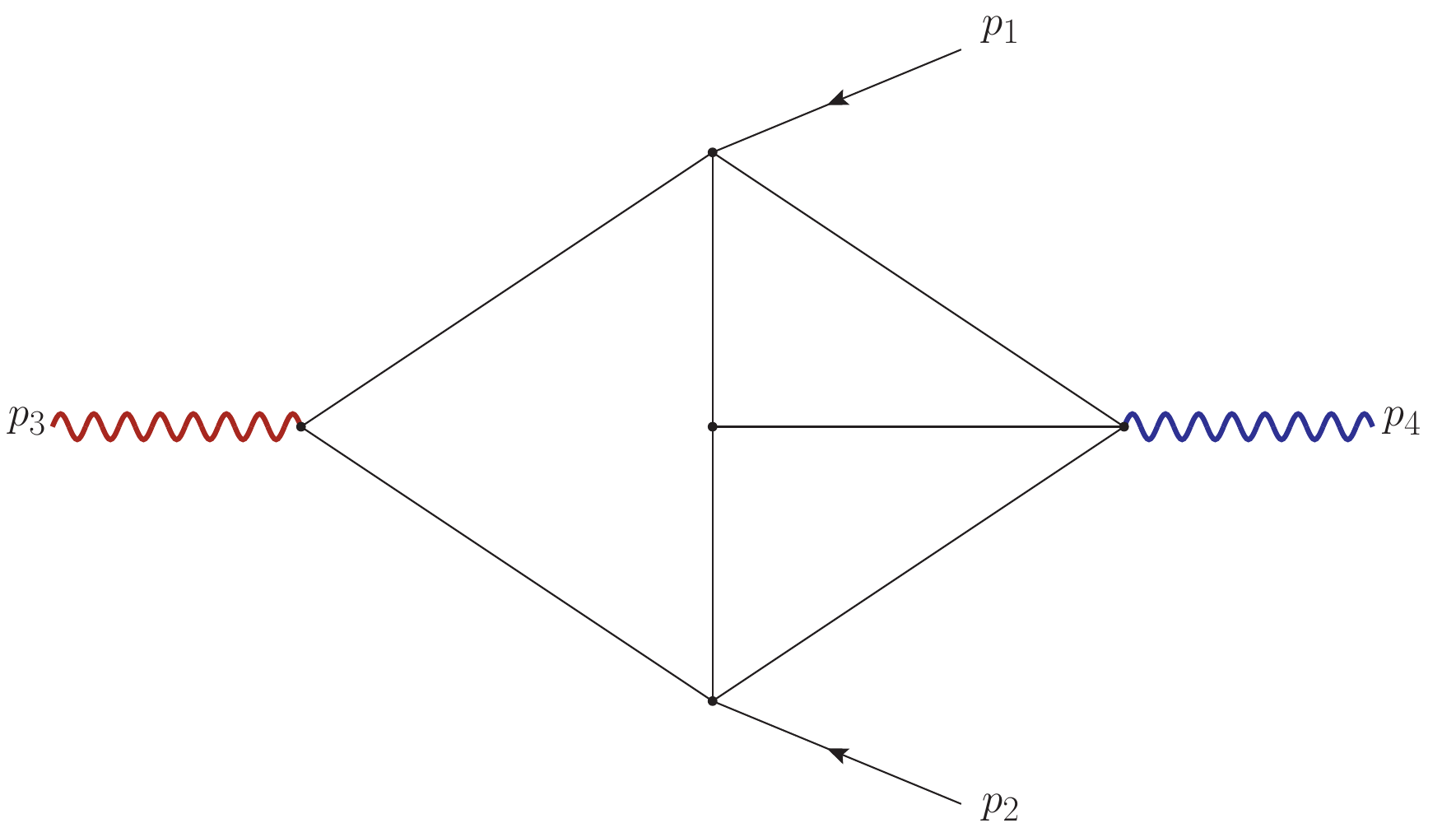}\\
\newline
\begin{equation*}
\{I_{61}^{\text{PT3}},I_{62}^{\text{PT3}},I_{63}^{\text{PT3}},I_{64}^{\text{PT3}},I_{65}^{\text{PT3}},I_{66}^{\text{PT3}},I_{67}^{\text{PT3}}\}
\end{equation*}
\end{multicols}

\textbf{Sector $\mathbf{F_{123}}$[1,0,0,1,0,0,0,1,0,1,1,0,1,0,1]}

\begin{multicols}{2}
\includegraphics[scale=0.20]{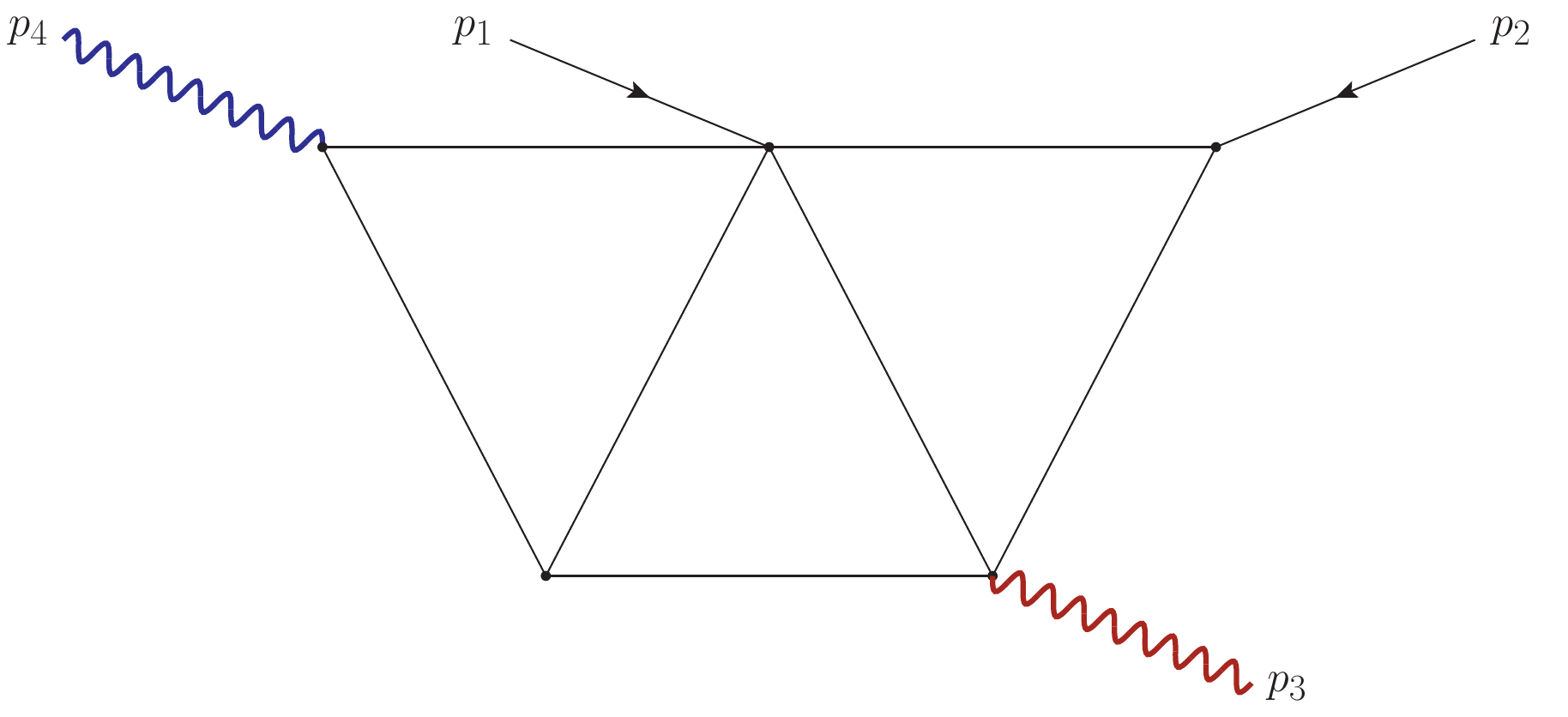}\\
\newline
\begin{equation*}
\{I_{52}^{\text{PL2}},I_{53}^{\text{PL2}}\}
\end{equation*}
\end{multicols}

\textbf{Sector $\mathbf{F_{132}}$[1,0,1,0,0,0,0,1,1,0,0,1,1,0,1]}

\begin{multicols}{2}
\includegraphics[scale=0.20]{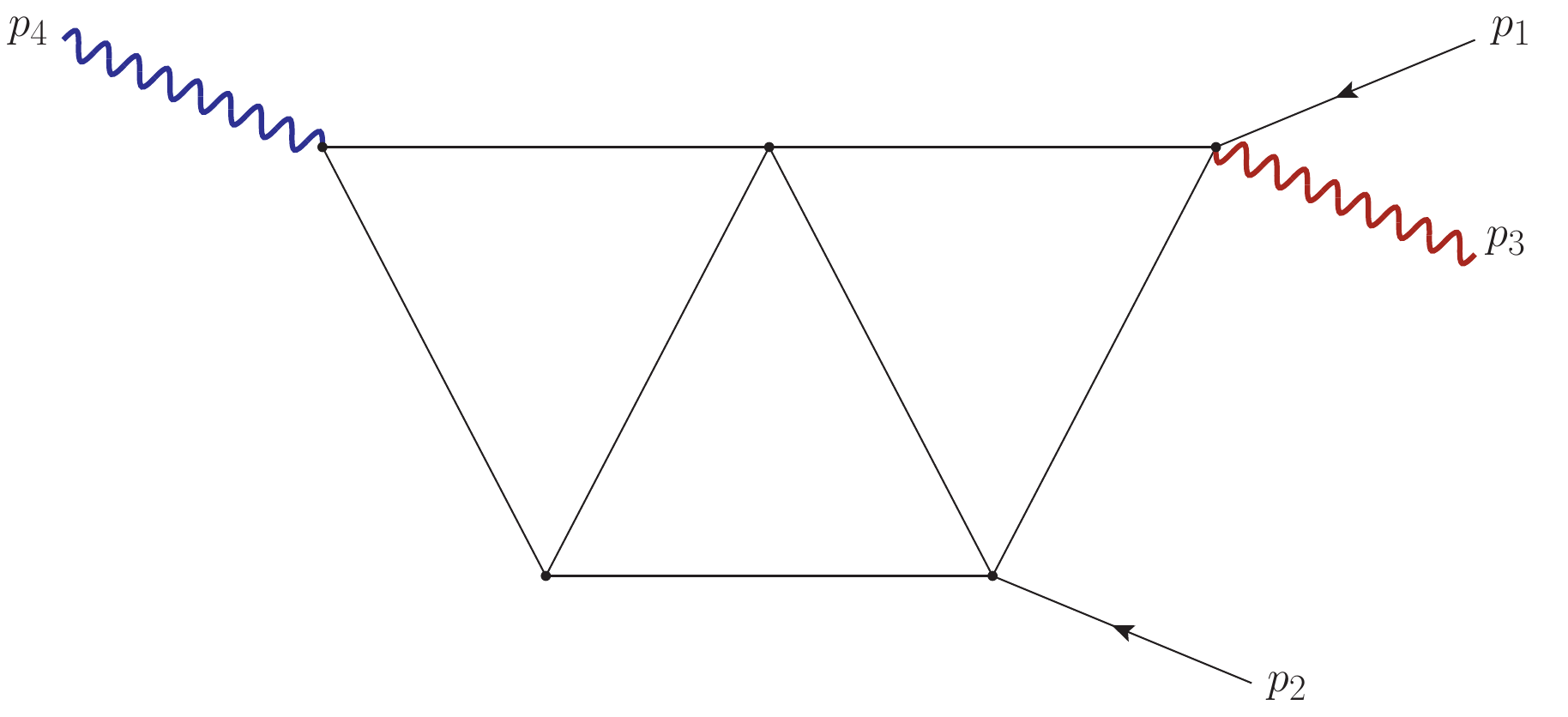}\\
\newline
\begin{equation*}
\{I_{111}^{\text{PT3}},I_{112}^{\text{PT3}}\}
\end{equation*}
\end{multicols}

\textbf{Sector $\mathbf{F_{132}}$[0,1,1,0,0,0,1,0,1,0,0,1,1,0,1]}

\begin{multicols}{2}
\includegraphics[scale=0.20]{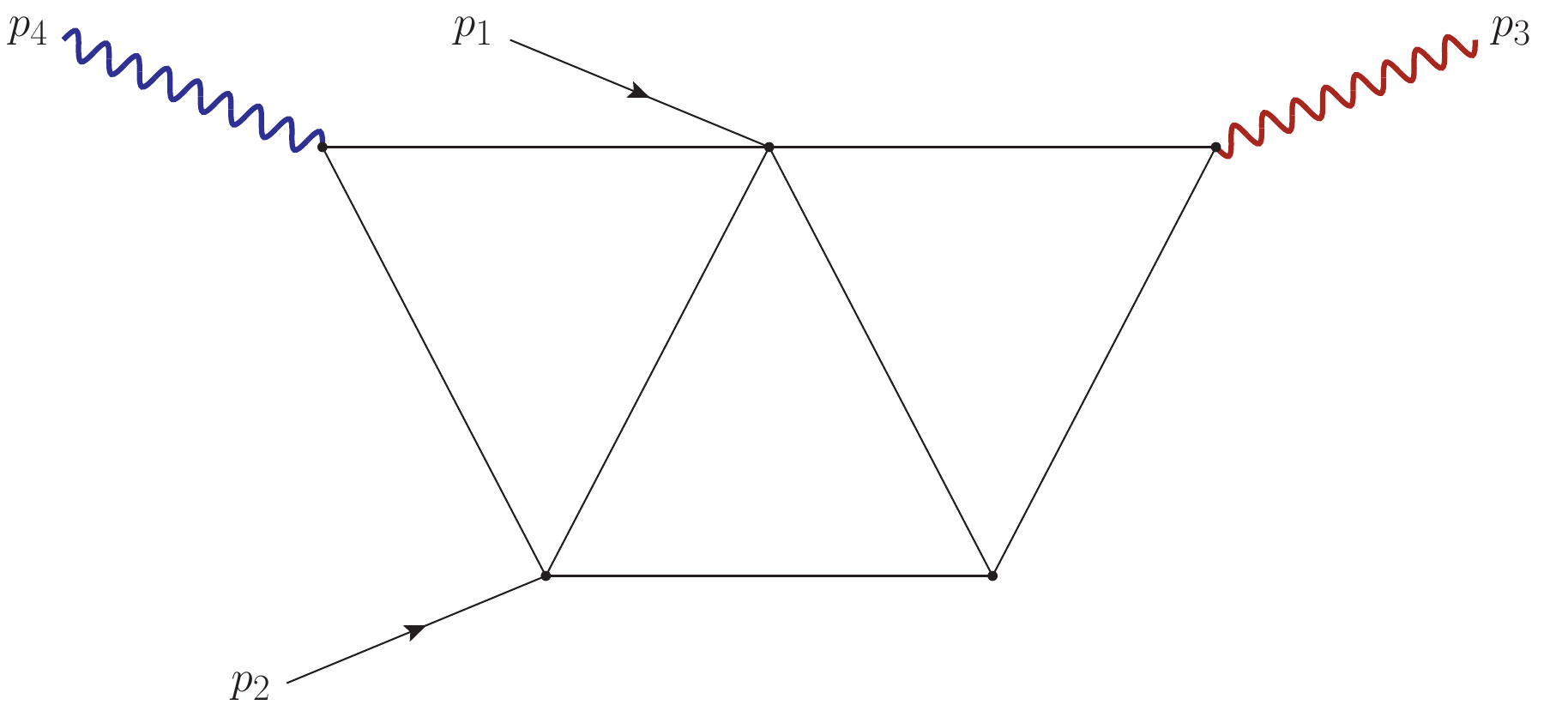}\\
\newline
\begin{equation*}
\{I_{44}^{\text{PL3}},I_{45}^{\text{PL3}},I_{46}^{\text{PL3}},I_{47}^{\text{PL3}},I_{48}^{\text{PL3}}\}
\end{equation*}
\end{multicols}

\textbf{Sector $\mathbf{F_{132}}$[0,1,1,0,1,0,0,0,0,0,1,1,1,0,1]}

\begin{multicols}{2}
\includegraphics[scale=0.20]{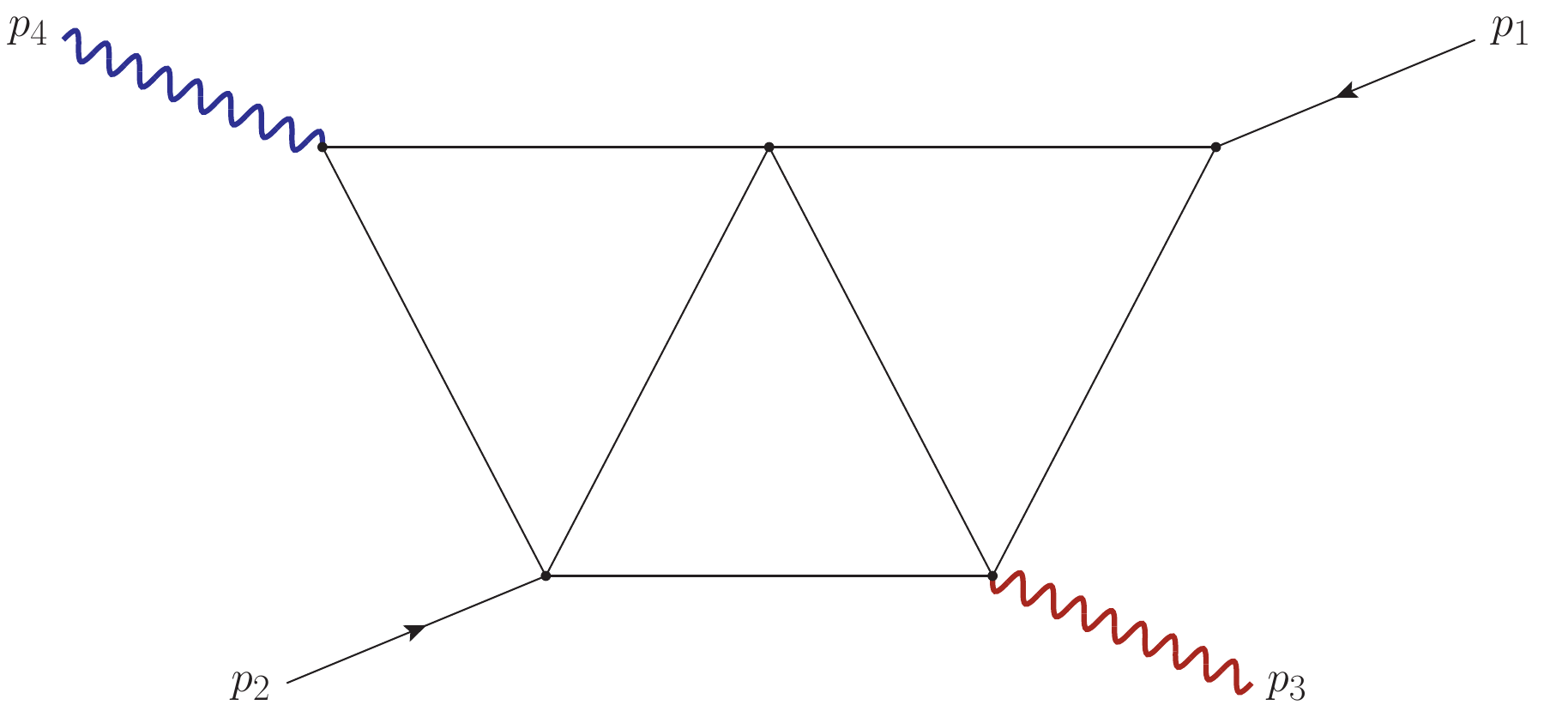}\\
\newline
\begin{equation*}
\{I_{34}^{\text{PL3}},I_{35}^{\text{PL3}},I_{36}^{\text{PL3}},I_{37}^{\text{PL3}},I_{38}^{\text{PL3}}\}
\end{equation*}
\end{multicols}

\textbf{Sector $\mathbf{F_{123}}$[0,1,0,0,0,0,1,1,1,0,0,1,1,1,0]}

\begin{multicols}{2}
\includegraphics[scale=0.20]{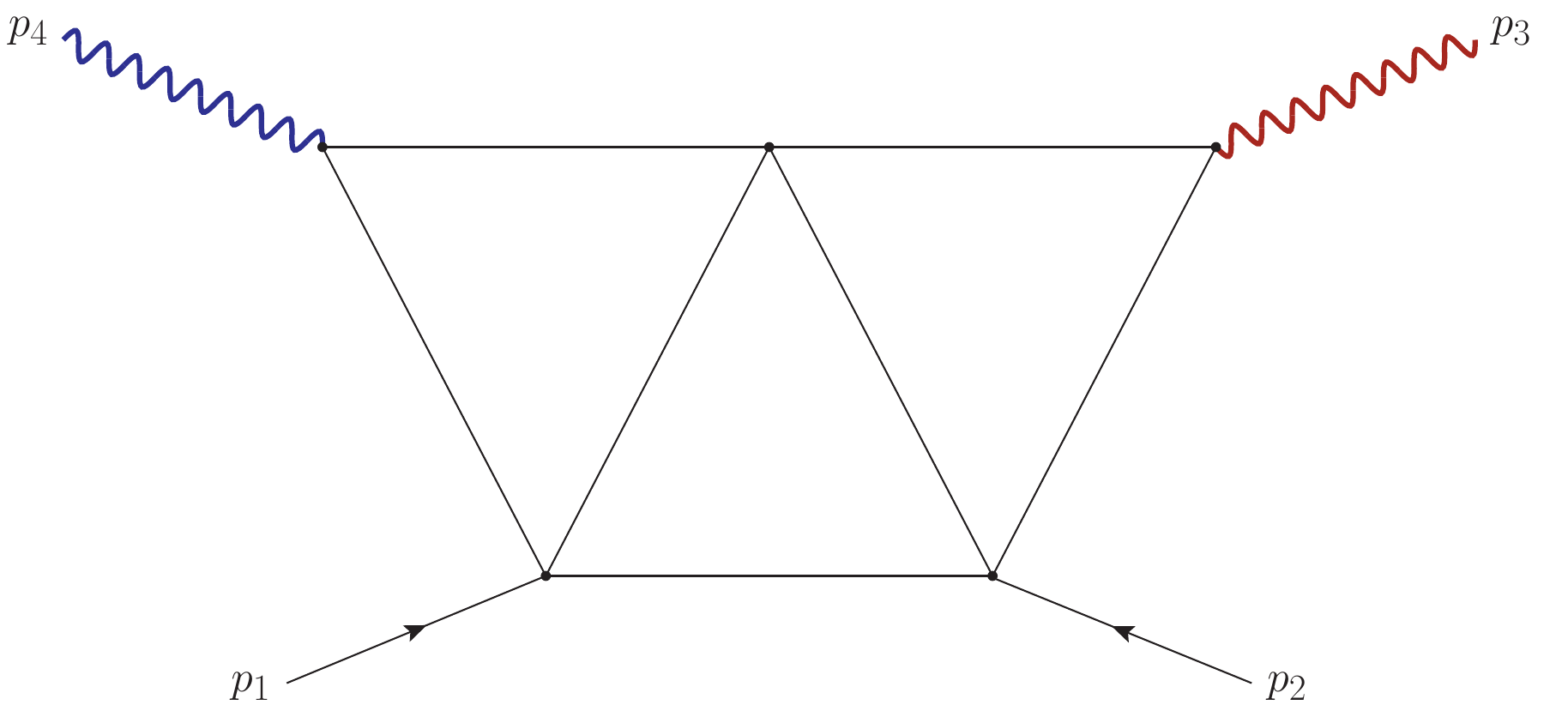}\\
\begin{equation*}
\begin{split}
\{&I_{90}^{\text{PT1}},I_{91}^{\text{PT1}},I_{92}^{\text{PT1}},I_{93}^{\text{PT1}},I_{94}^{\text{PT1}},\\ &I_{95}^{\text{PT1}},I_{96}^{\text{PT1}},I_{97}^{\text{PT1}},I_{98}^{\text{PT1}},I_{99}^{\text{PT1}}, \\ &I_{100}^{\text{PT1}},I_{101}^{\text{PT1}},I_{102}^{\text{PT1}},I_{103}^{\text{PT1}},I_{104}^{\text{PT1}}\}
\end{split}
\end{equation*}
\end{multicols}

\vspace{0.15cm}
\begin{center}
\textbf{\textit{Eight-Propagator Pure Candidates}}\\
\end{center}
\vspace{0.2cm}

\textbf{Sector $\mathbf{F_{123}}$[1,0,0,1,0,1,0,1,0,0,1,1,1,0,1]}

\begin{multicols}{2}
\includegraphics[scale=0.20]{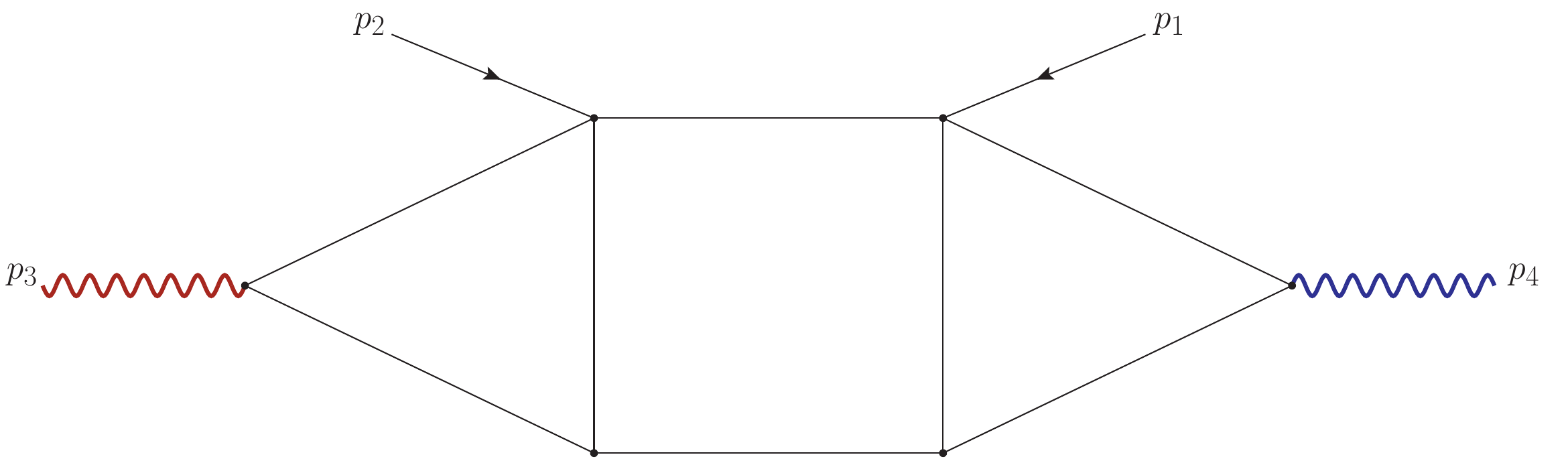}\\
\begin{equation*}
\begin{split}
\{&I_{28}^{\text{PL2}}\}
\end{split}
\end{equation*}
\end{multicols}

\textbf{Sector $\mathbf{F_{132}}$[0,1,1,0,1,0,1,0,1,0,0,1,1,0,1]}

\begin{multicols}{2}
\includegraphics[scale=0.20]{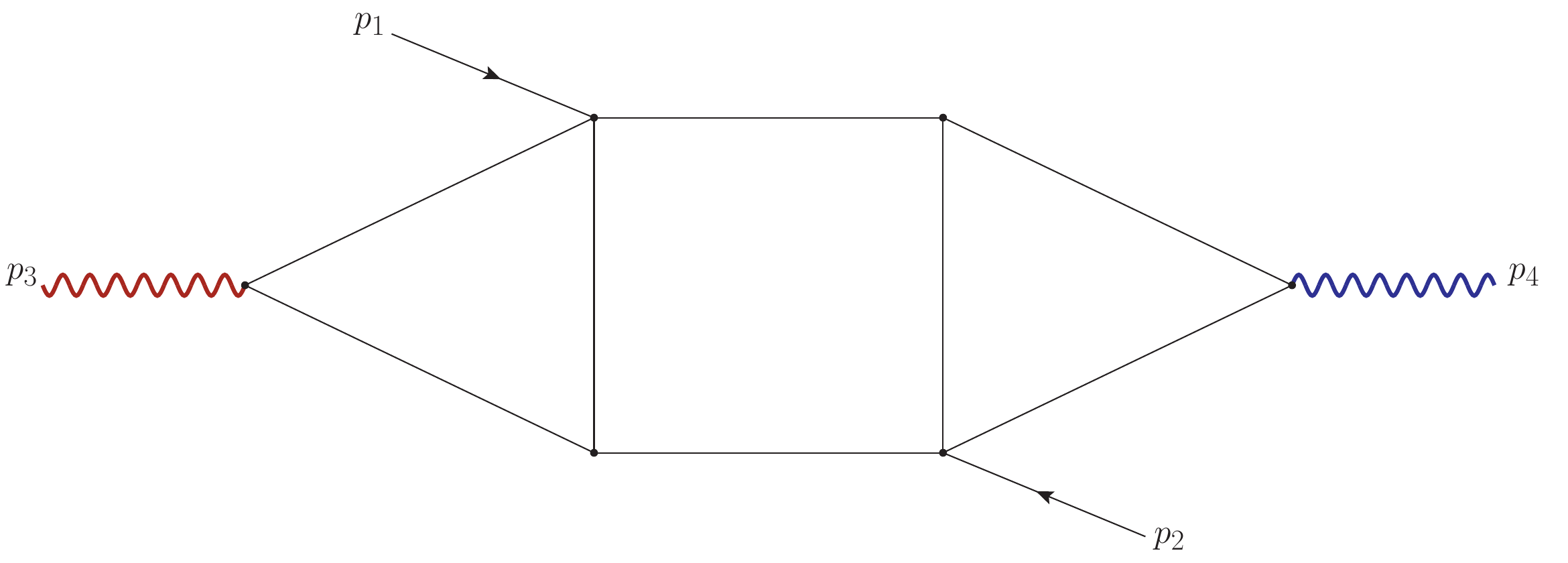}\\
\begin{equation*}
\begin{split}
\{&I_{28}^{\text{PL3}}\}
\end{split}
\end{equation*}
\end{multicols}

\textbf{Sector $\mathbf{F_{123}}$[0,1,1,0,0,0,1,1,1,0,0,1,0,1,1]}

\begin{multicols}{2}
\includegraphics[scale=0.30]{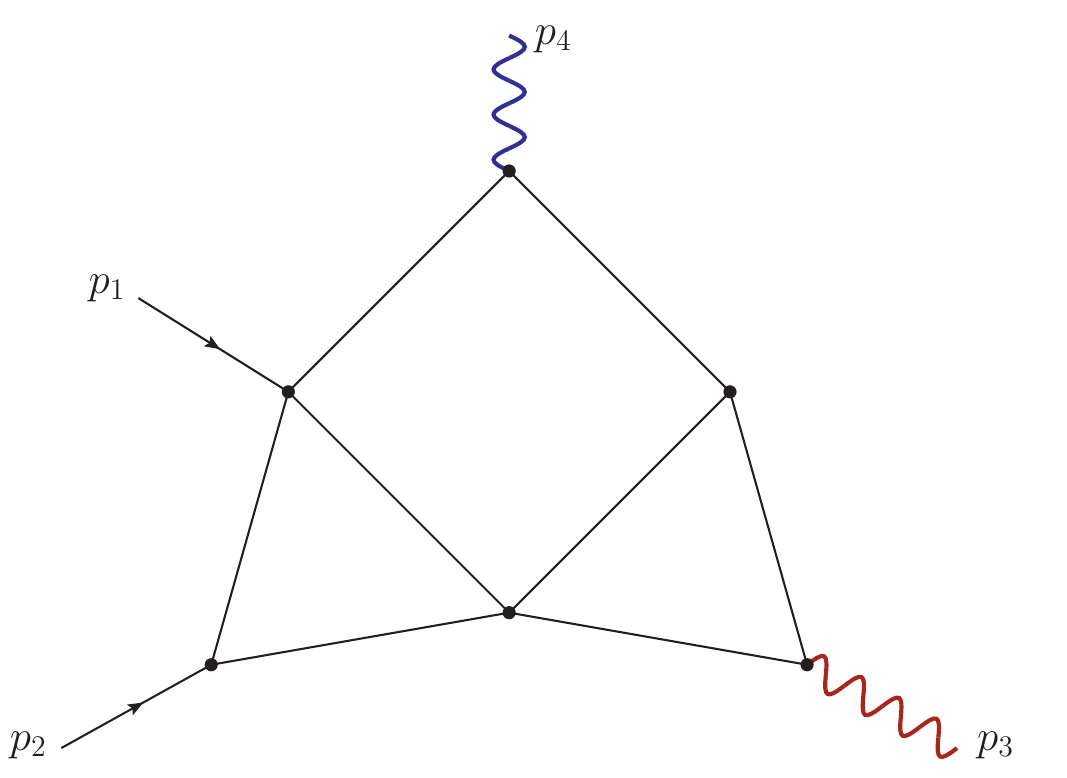}\\
\begin{equation*}
\{I_{89}^{\text{PT1}}\}
\end{equation*}
\end{multicols}

\textbf{Sector $\mathbf{F_{132}}$[0,1,1,0,0,0,1,1,1,0,0,1,1,0,1]}

\begin{multicols}{2}
\includegraphics[scale=0.30]{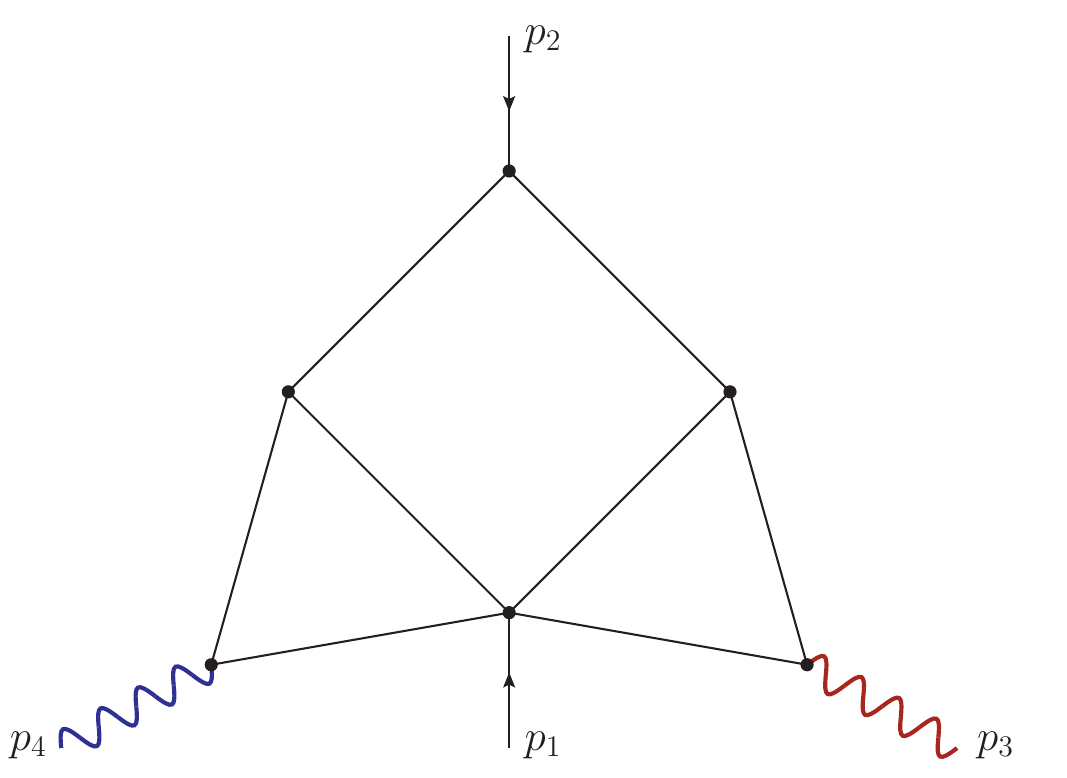}\\
\begin{equation*}
\{I_{53}^{\text{PT3}}\}
\end{equation*}
\end{multicols}

\textbf{Sector $\mathbf{F_{132}}$[1,1,0,0,0,0,1,1,1,0,0,1,1,0,1]}

\begin{multicols}{2}
\includegraphics[scale=0.30]{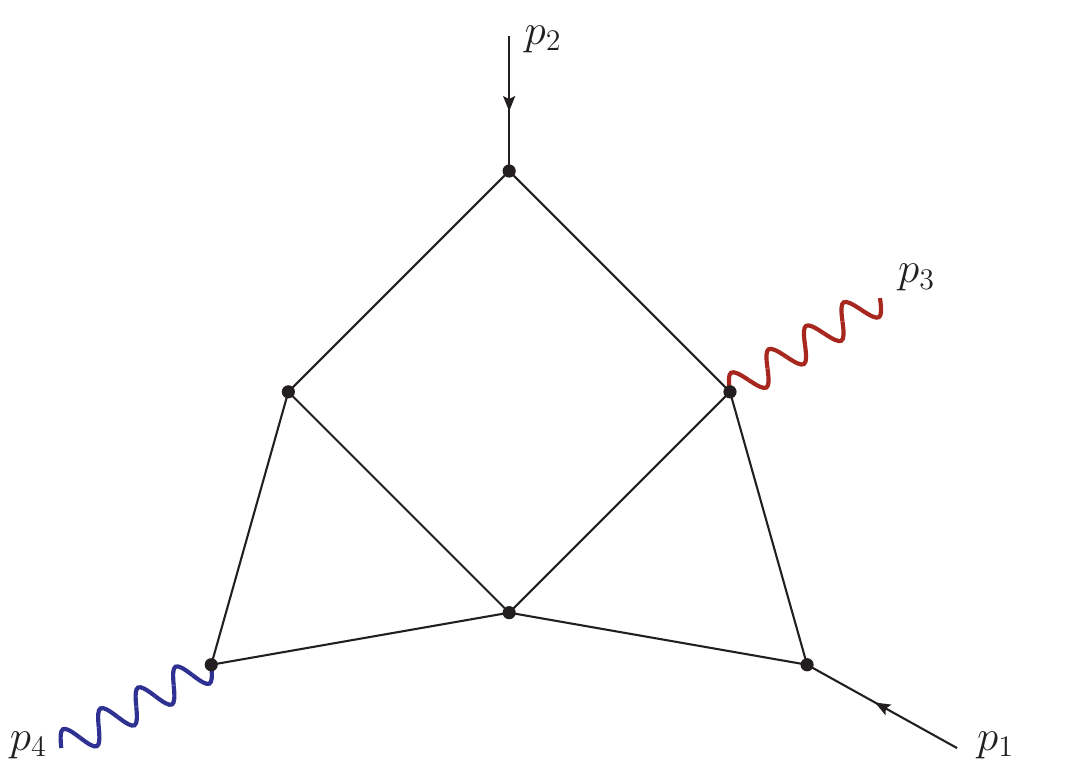}\\
\begin{equation*}
\{I_{51}^{\text{PT3}}\}
\end{equation*}
\end{multicols}

\textbf{Sector $\mathbf{F_{123}}$[0,1,1,0,0,0,1,1,1,0,0,1,1,1,0]}

\begin{multicols}{2}
\includegraphics[scale=0.30]{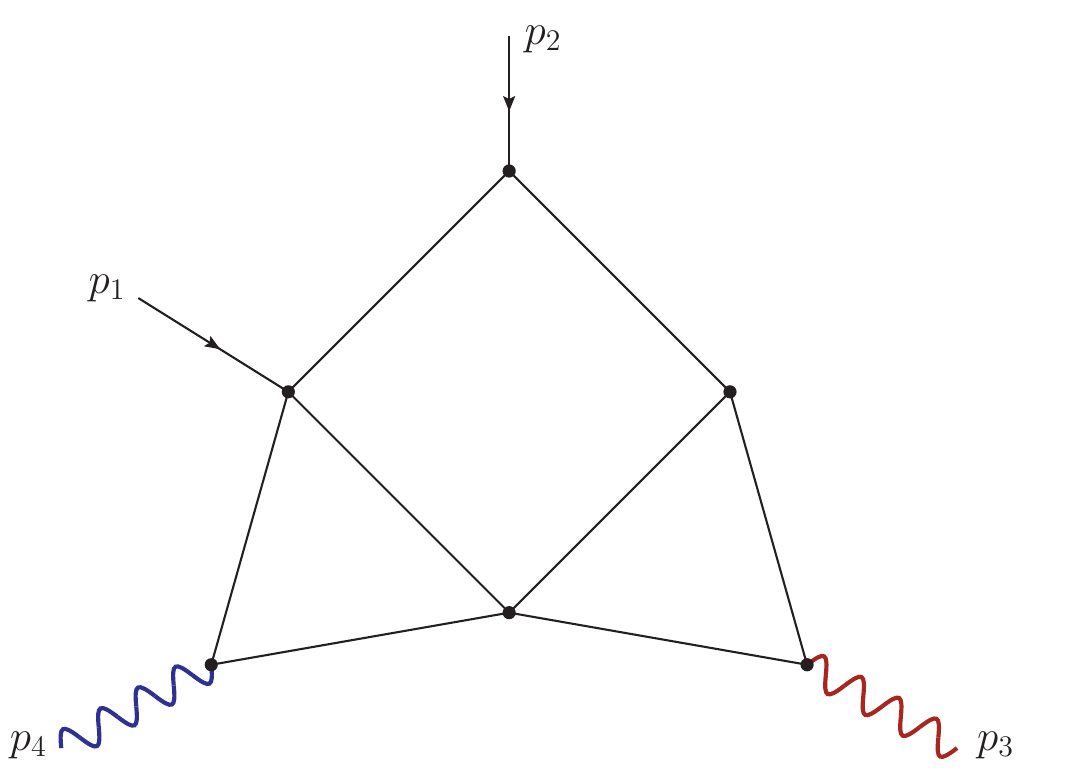}\\
\begin{equation*}
\{I_{77}^{\text{PT1}},I_{78}^{\text{PT1}},I_{79}^{\text{PT1}}\}
\end{equation*}
\end{multicols}

\textbf{Sector $\mathbf{F_{123}}$[1,1,1,0,0,0,0,1,1,0,0,1,1,1,0]}

\begin{multicols}{2}
\includegraphics[scale=0.20]{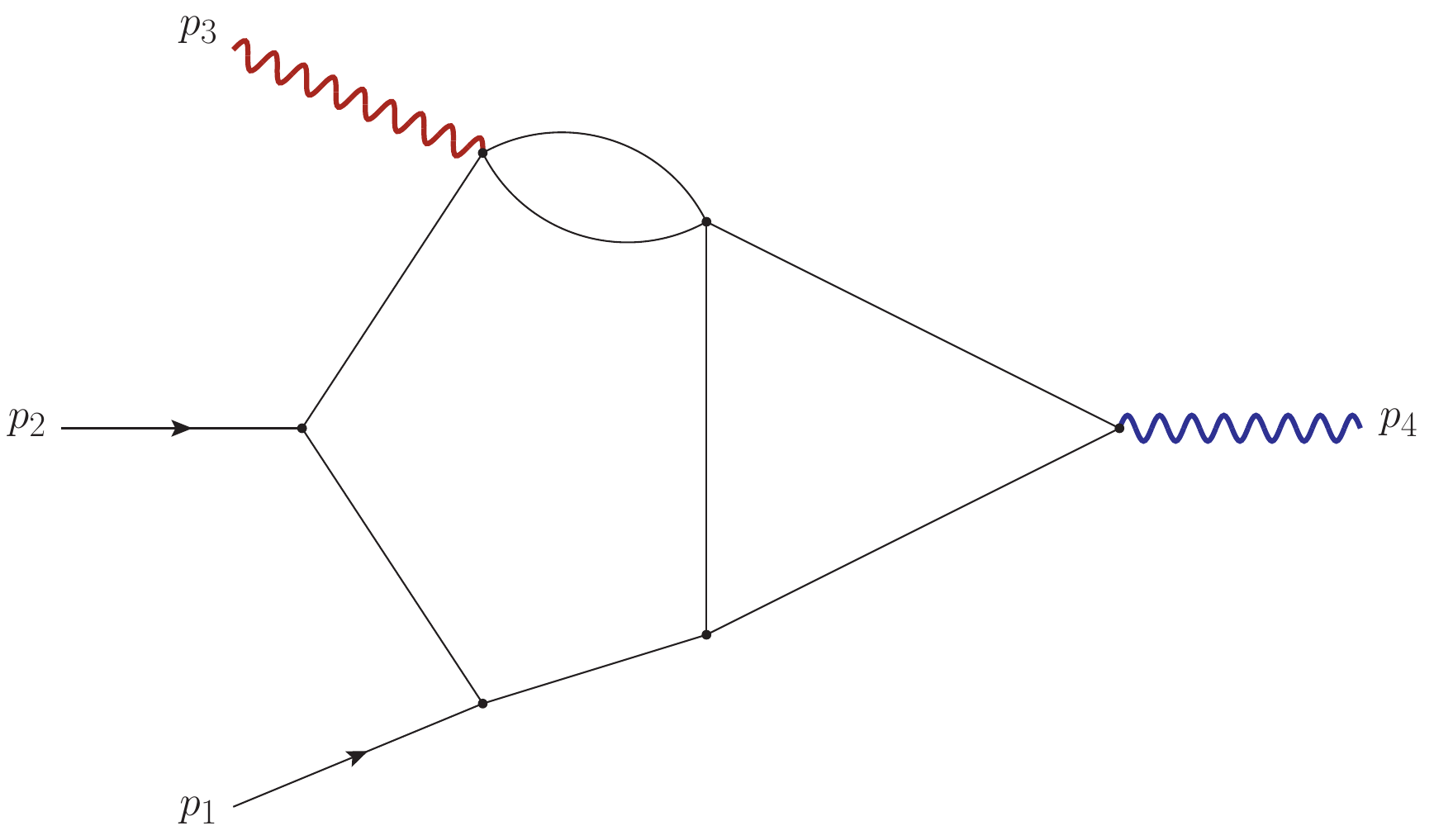}\\
\begin{equation*}
\begin{split}
\{&I_{87}^{\text{PT1}}\}
\end{split}
\end{equation*}
\end{multicols}

\textbf{Sector $\mathbf{F_{123}}$[0,1,1,1,1,0,0,1,1,0,0,0,1,1,0]}

\begin{multicols}{2}
\includegraphics[scale=0.20]{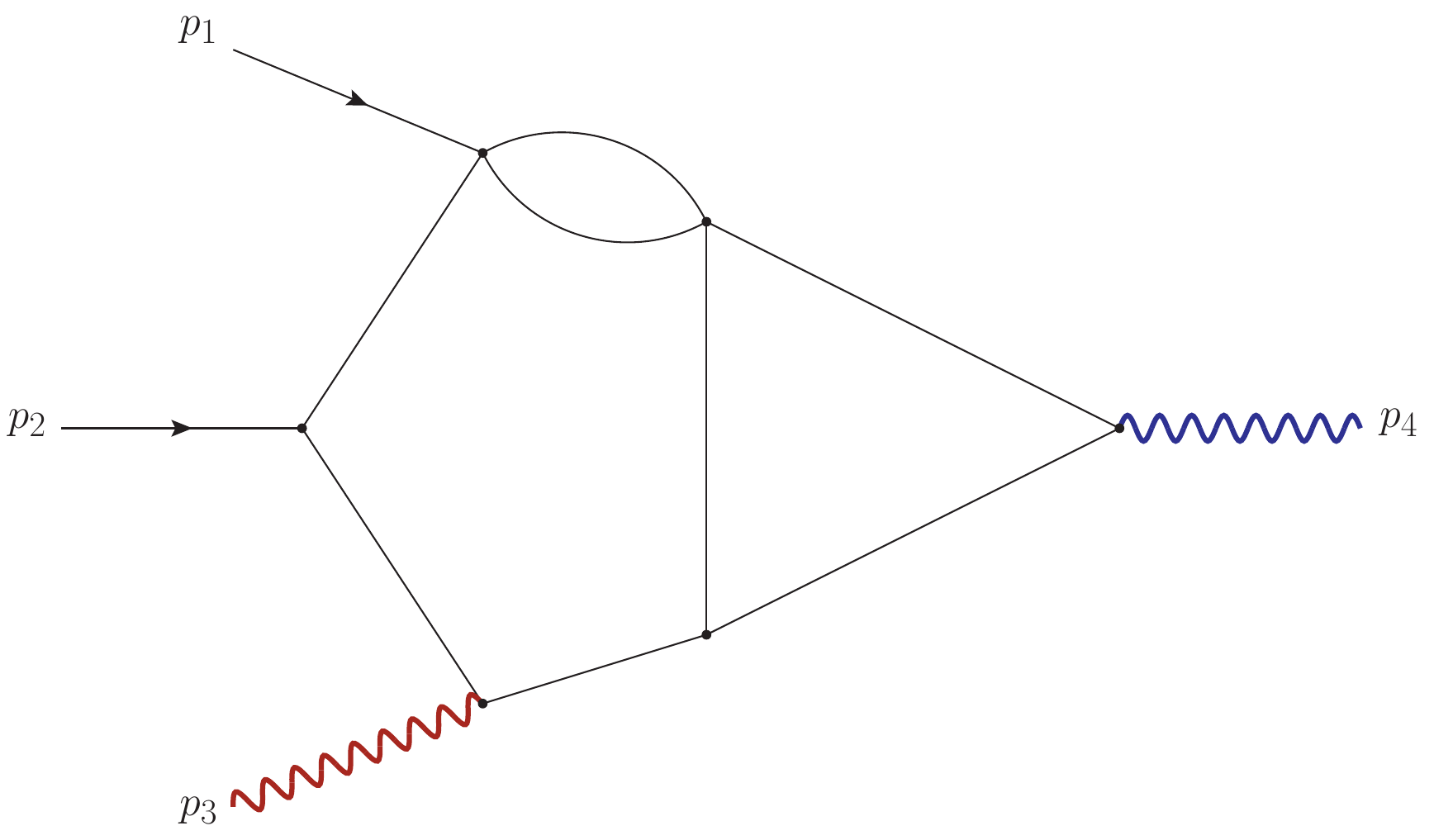}\\
\begin{equation*}
\{I_{54}^{\text{PT2}}\}
\end{equation*}
\end{multicols}

\textbf{Sector $\mathbf{F_{132}}$[1,1,1,0,0,0,0,1,1,0,0,1,1,1,0]}

\begin{multicols}{2}
\includegraphics[scale=0.20]{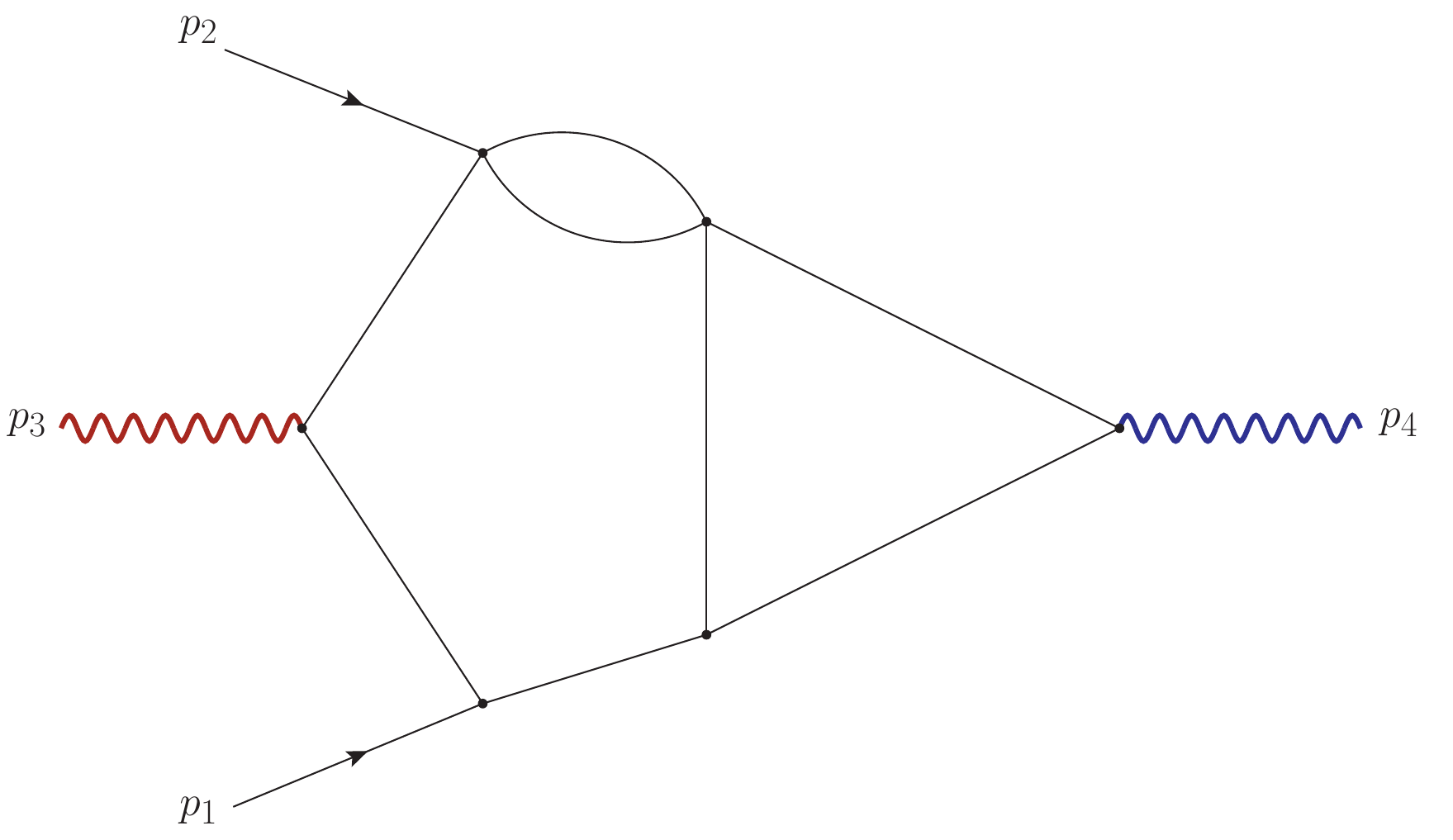}\\
\begin{equation*}
\{I_{50}^{\text{PT3}}\}
\end{equation*}
\end{multicols}

\textbf{Sector $\mathbf{F_{123}}$[1,0,0,0,0,1,0,1,0,1,1,1,1,0,1]}

\begin{multicols}{2}
\includegraphics[scale=0.20]{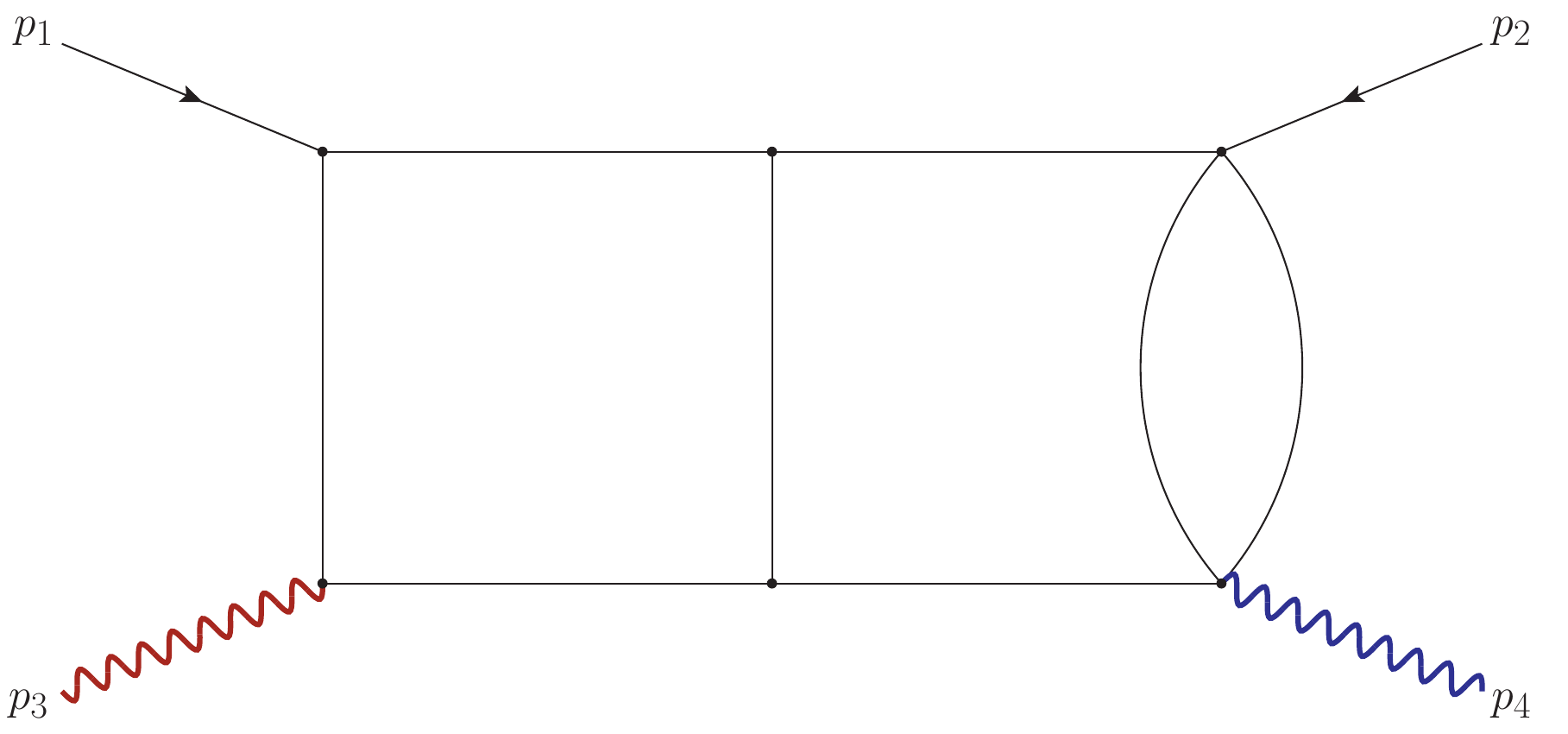}\\
\begin{equation*}
\{I_{26}^{\text{PL2}},I_{27}^{\text{PL2}}\}
\end{equation*}
\end{multicols}

\textbf{Sector $\mathbf{F_{132}}$[0,1,0,0,1,0,1,0,1,0,1,1,1,0,1]}

\begin{multicols}{2}
\includegraphics[scale=0.20]{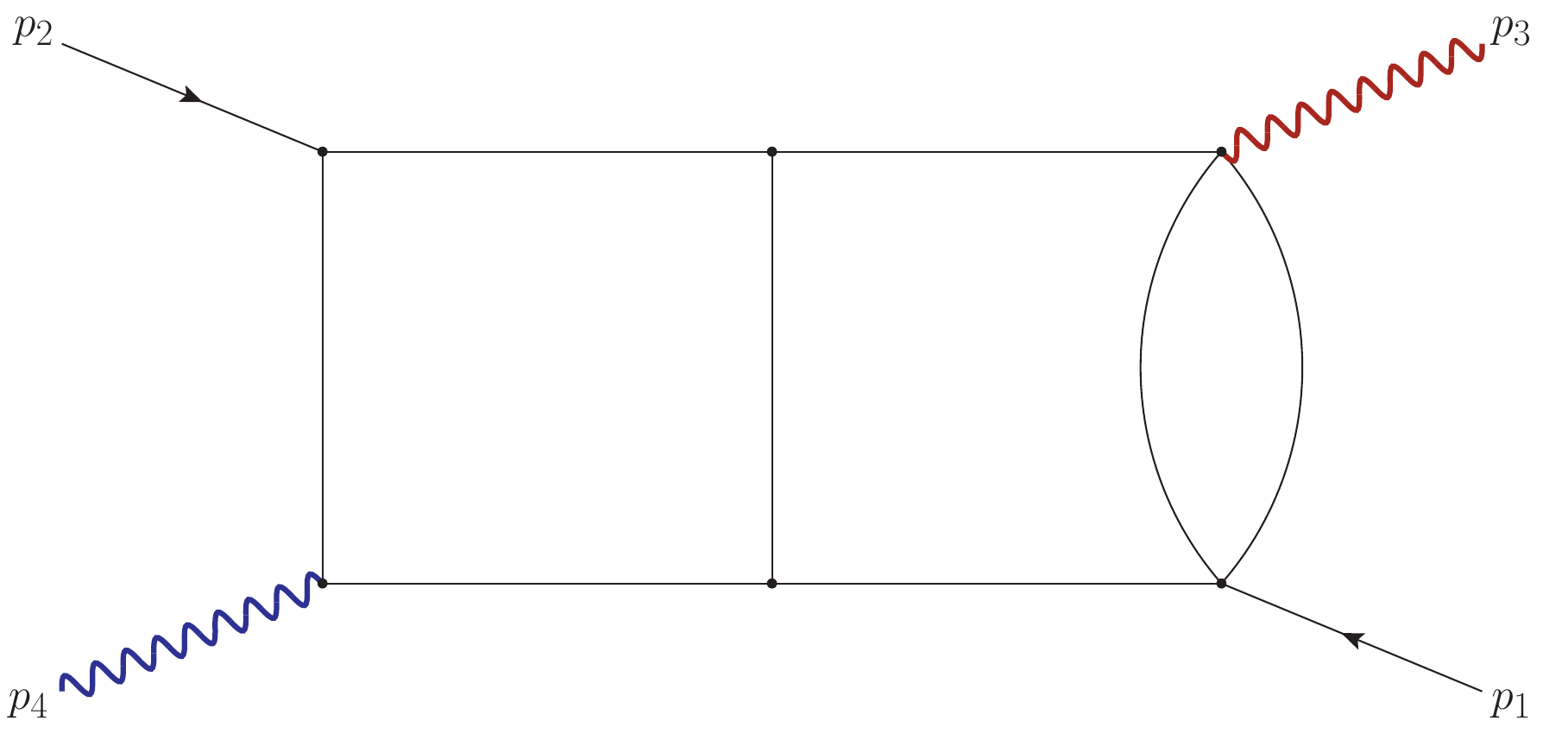}\\
\begin{equation*}
\{I_{26}^{\text{PL3}},I_{27}^{\text{PL3}}\}
\end{equation*}
\end{multicols}

\textbf{Sector $\mathbf{F_{123}}$[0,1,1,1,0,0,0,1,1,1,0,0,0,1,1]}

\begin{multicols}{2}
\includegraphics[scale=0.20]{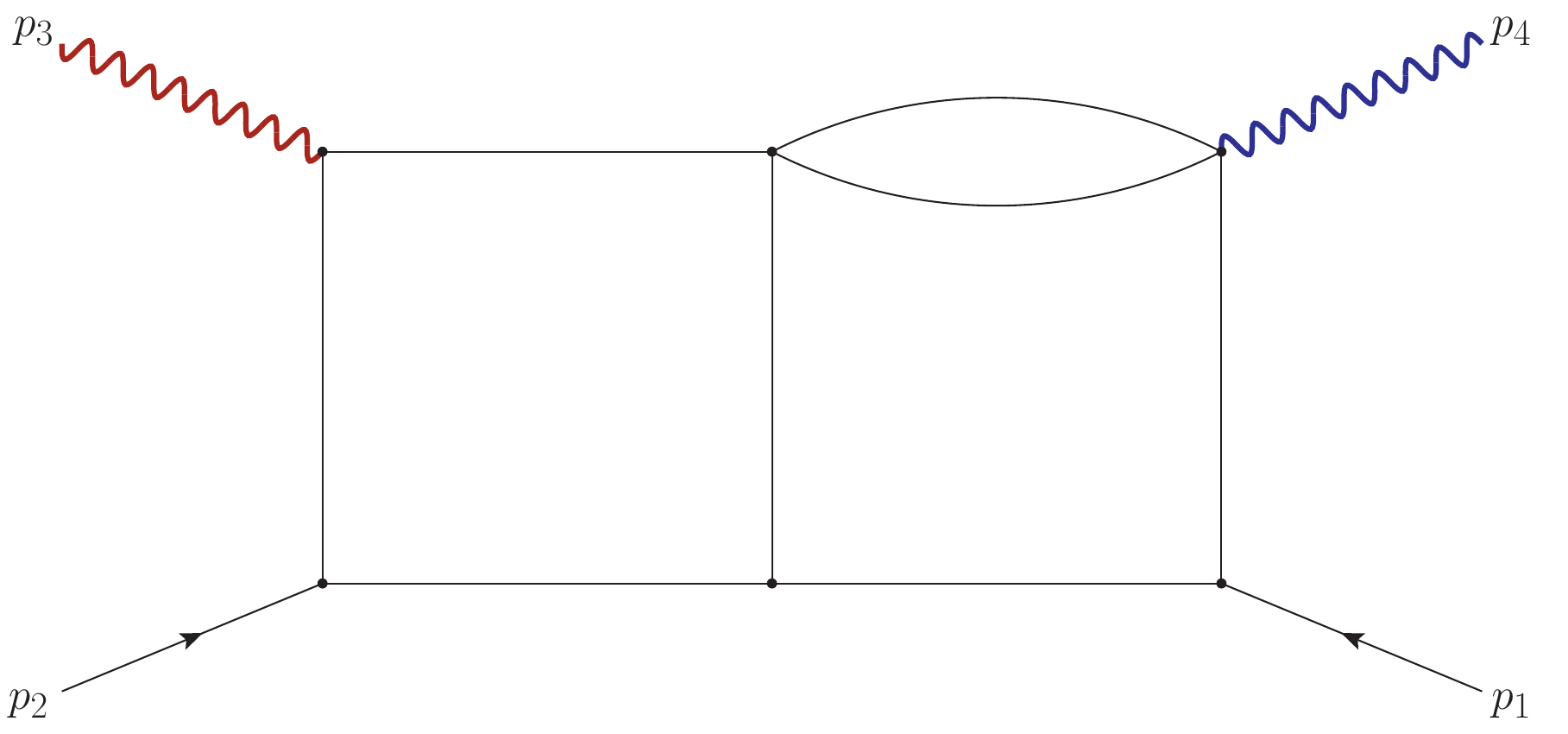}\\
\begin{equation*}
\{I_{50}^{\text{PT2}},I_{51}^{\text{PT2}}\}
\end{equation*}
\end{multicols}

\textbf{Sector $\mathbf{F_{132}}$[1,1,1,0,0,0,1,1,1,0,0,0,1,0,1]}

\begin{multicols}{2}
\includegraphics[scale=0.20]{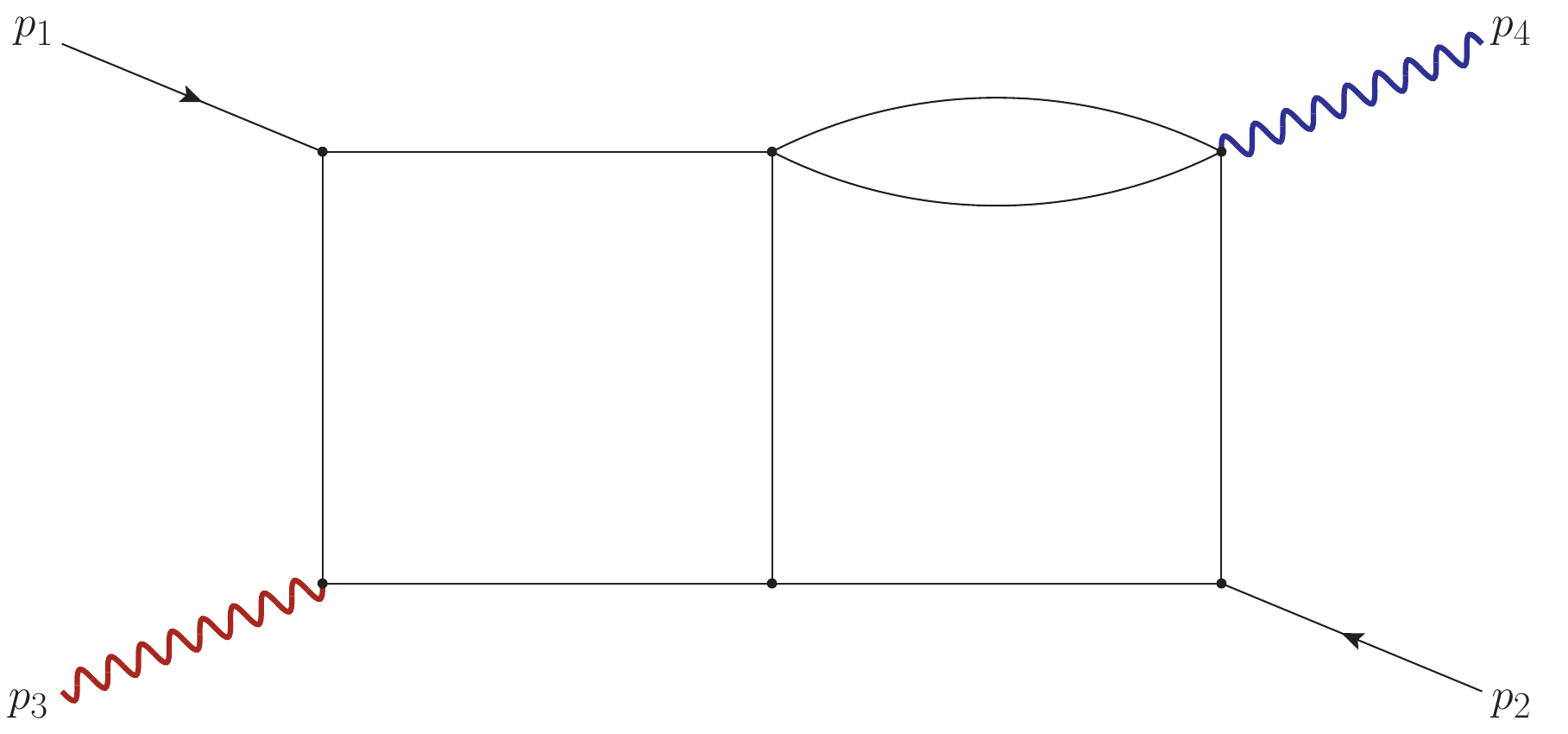}\\
\begin{equation*}
\{I_{46}^{\text{PT3}},I_{47}^{\text{PT3}}\}
\end{equation*}
\end{multicols}

\textbf{Sector $\mathbf{F_{123}}$[0,1,1,1,1,0,0,1,0,1,0,0,1,0,1]}

\begin{multicols}{2}
\includegraphics[scale=0.20]{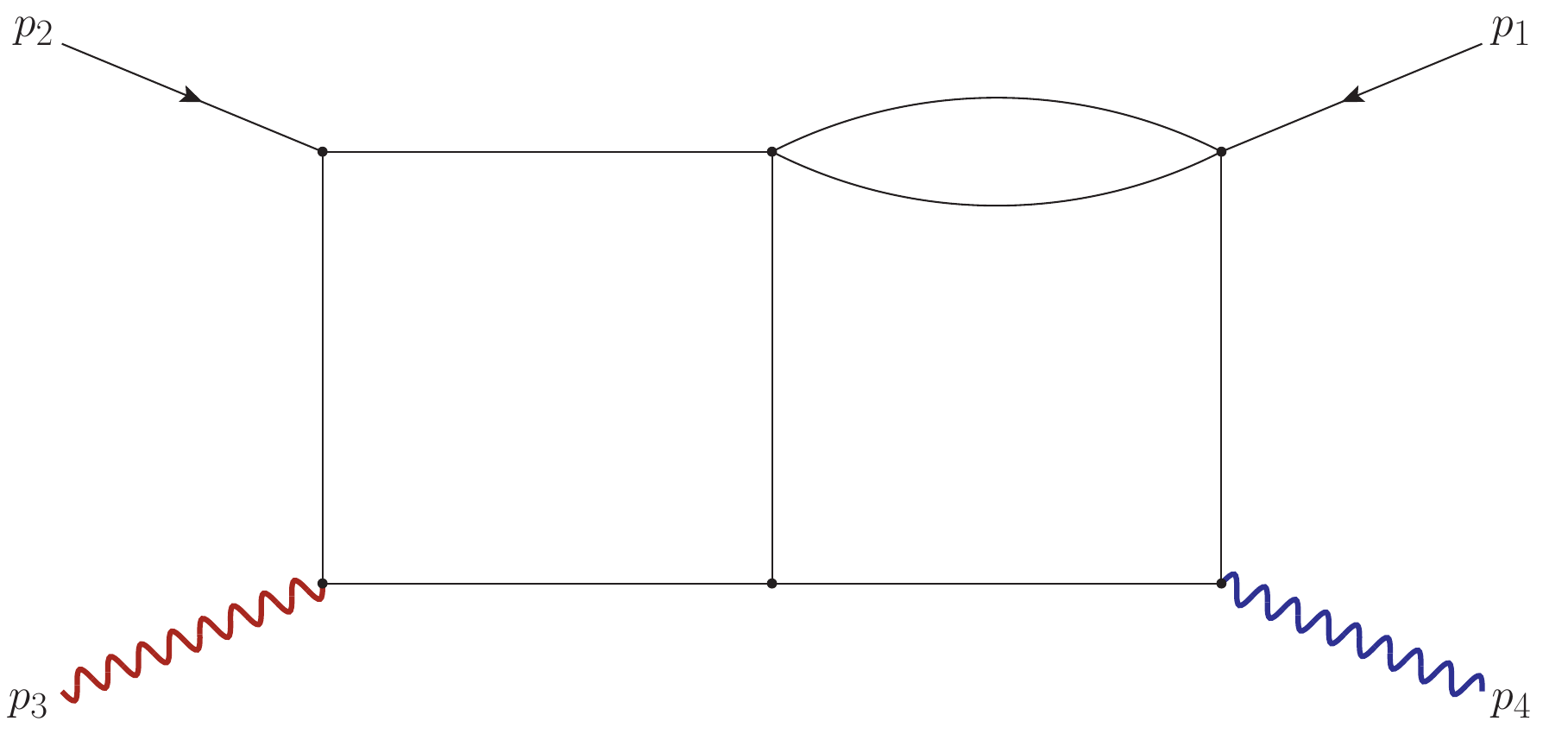}\\
\begin{equation*}
\{I_{40}^{\text{PT2}},I_{41}^{\text{PT2}},I_{42}^{\text{PT2}}\}
\end{equation*}
\end{multicols}

\textbf{Sector $\mathbf{F_{132}}$[1,1,1,0,0,0,1,0,1,0,0,1,0,1,1]}

\begin{multicols}{2}
\includegraphics[scale=0.20]{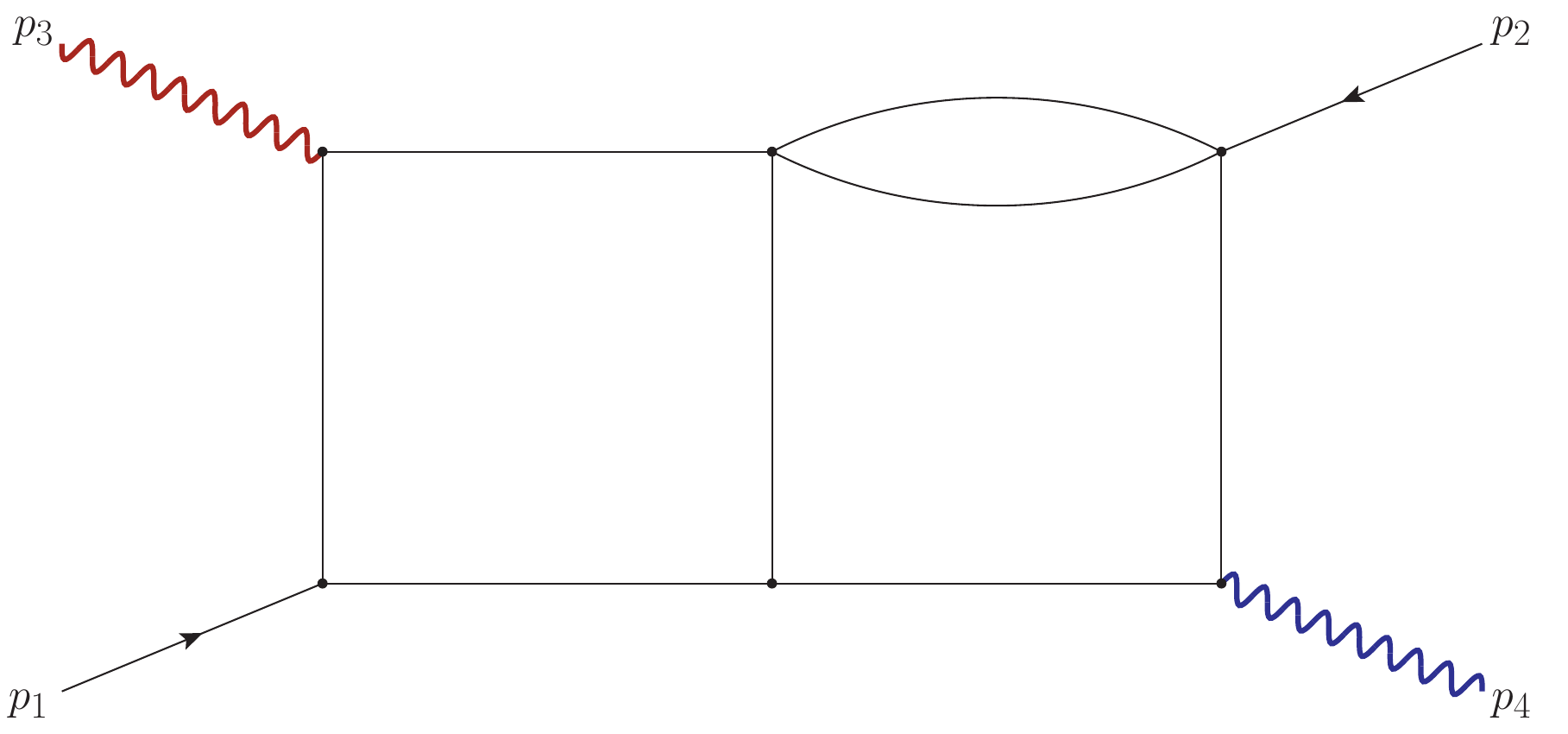}\\
\begin{equation*}
\{I_{41}^{\text{PT3}},I_{42}^{\text{PT3}},I_{43}^{\text{PT3}}\}
\end{equation*}
\end{multicols}

\textbf{Sector $\mathbf{F_{123}}$[1,1,1,0,0,0,1,0,1,0,0,1,0,1,1]}

\begin{multicols}{2}
\includegraphics[scale=0.20]{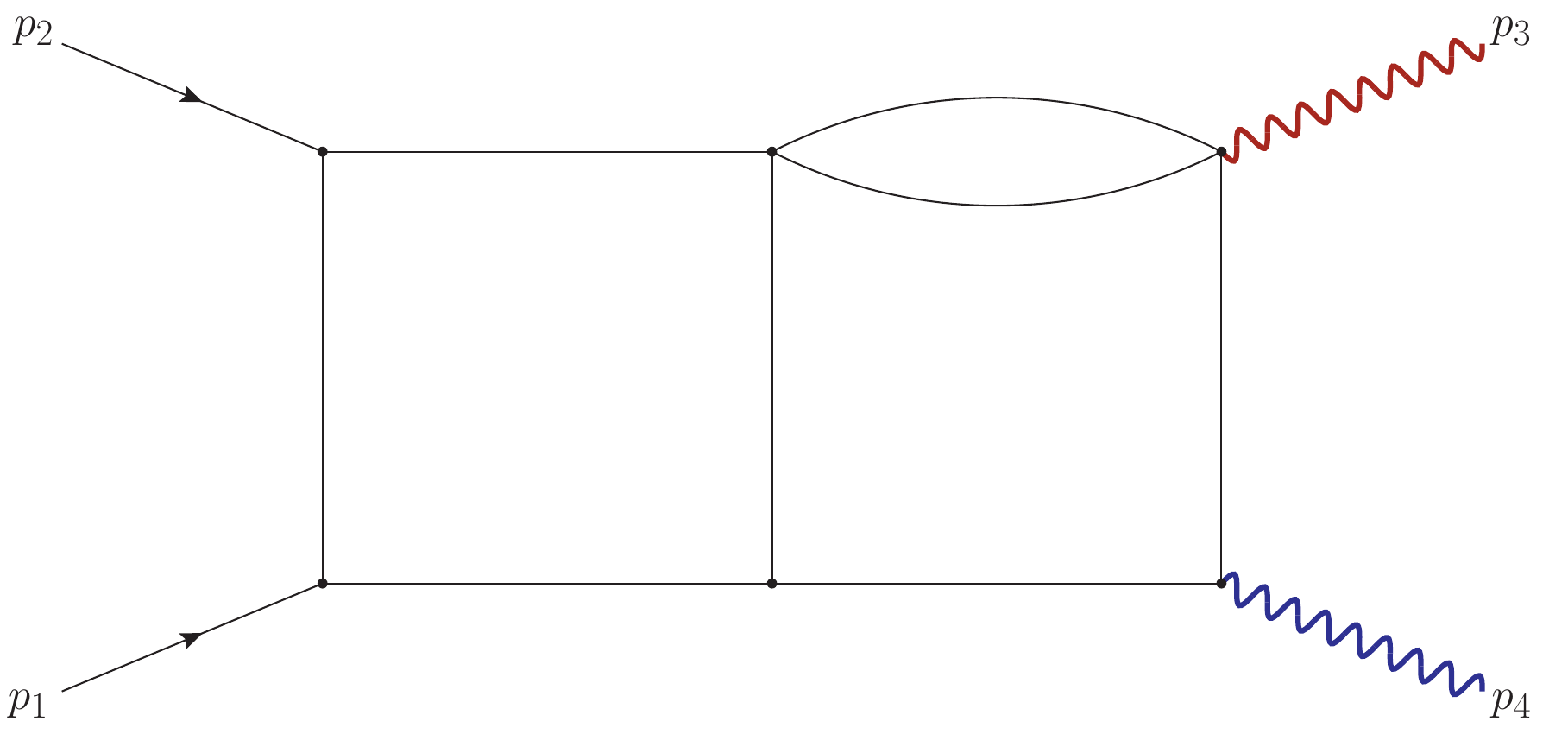}\\
\begin{equation*}
\{I_{70}^{\text{PT1}},I_{71}^{\text{PT1}},I_{72}^{\text{PT1}},I_{73}^{\text{PT1}}\}
\end{equation*}
\end{multicols}

\textbf{Sector $\mathbf{F_{123}}$[1,1,0,1,0,0,0,0,0,1,1,1,1,0,1]}

\begin{multicols}{2}
\includegraphics[scale=0.20]{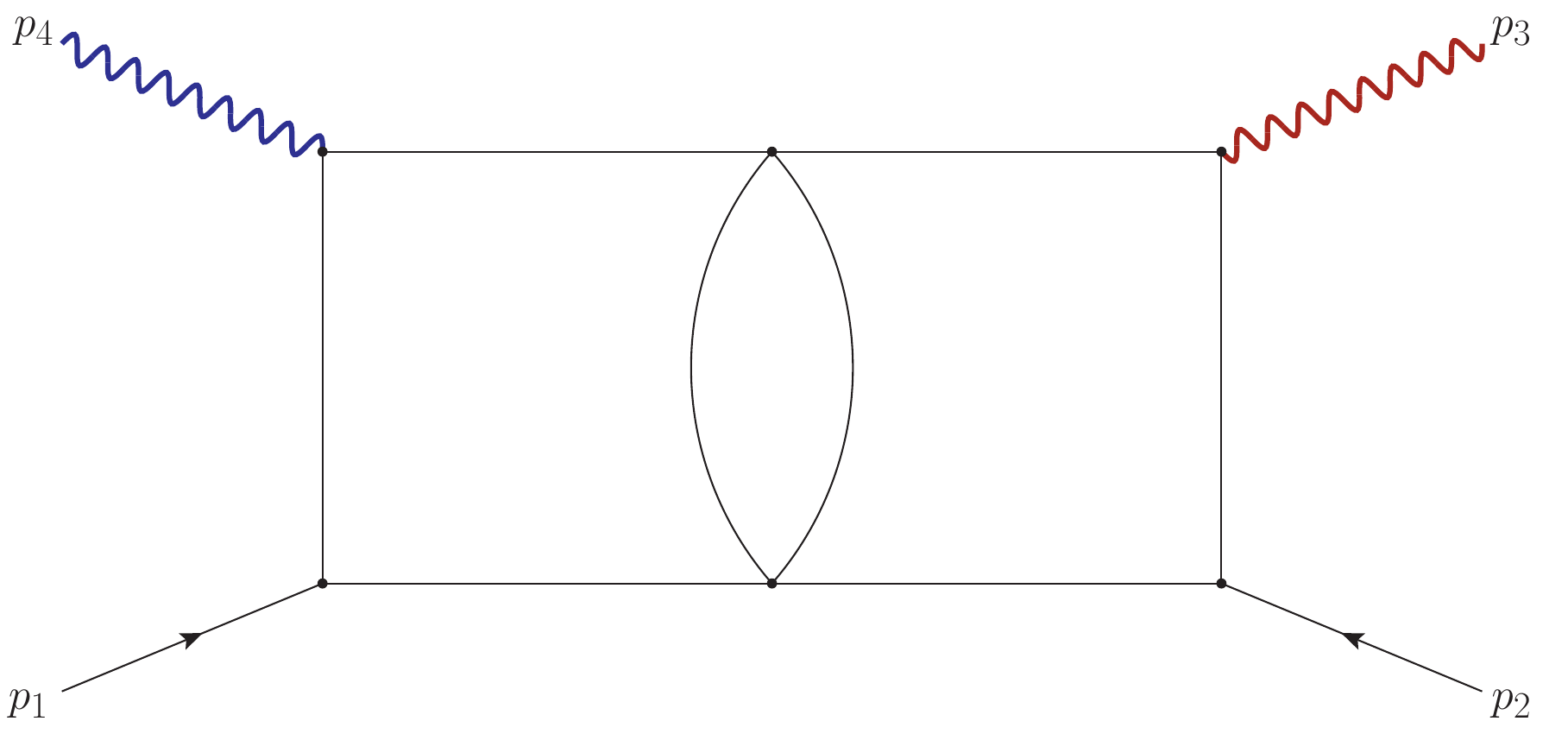}\\
\begin{equation*}
\{I_{22}^{\text{PL2}},I_{23}^{\text{PL2}}\}
\end{equation*}
\end{multicols}

\textbf{Sector $\mathbf{F_{132}}$[1,1,1,0,0,0,0,0,1,0,1,1,1,0,1]}

\begin{multicols}{2}
\includegraphics[scale=0.20]{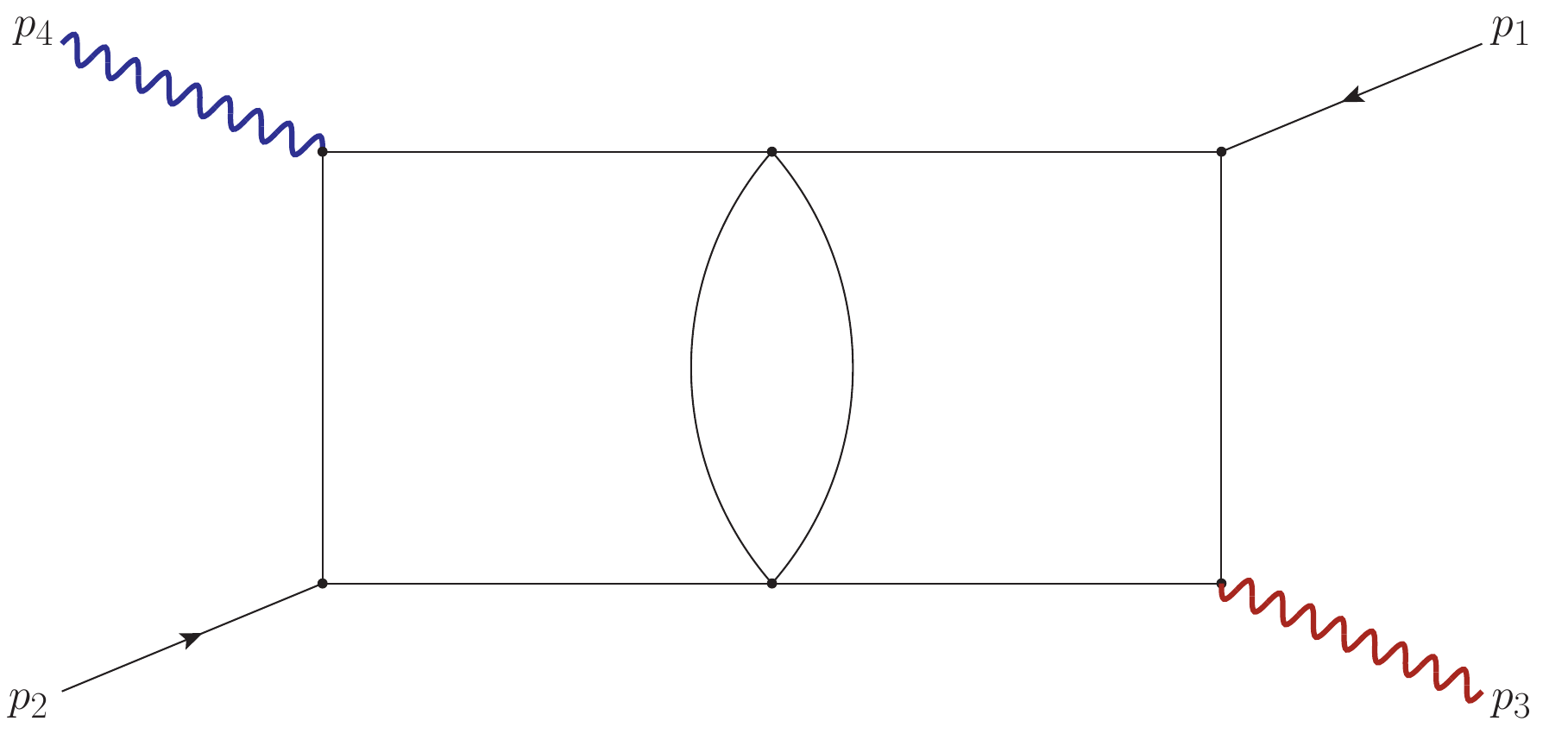}\\
\begin{equation*}
\{I_{22}^{\text{PL3}},I_{23}^{\text{PL3}}\}
\end{equation*}
\end{multicols}

\textbf{Sector $\mathbf{F_{123}}$[1,1,0,0,0,0,0,1,0,1,1,1,1,0,1]}

\begin{multicols}{2}
\includegraphics[scale=0.20]{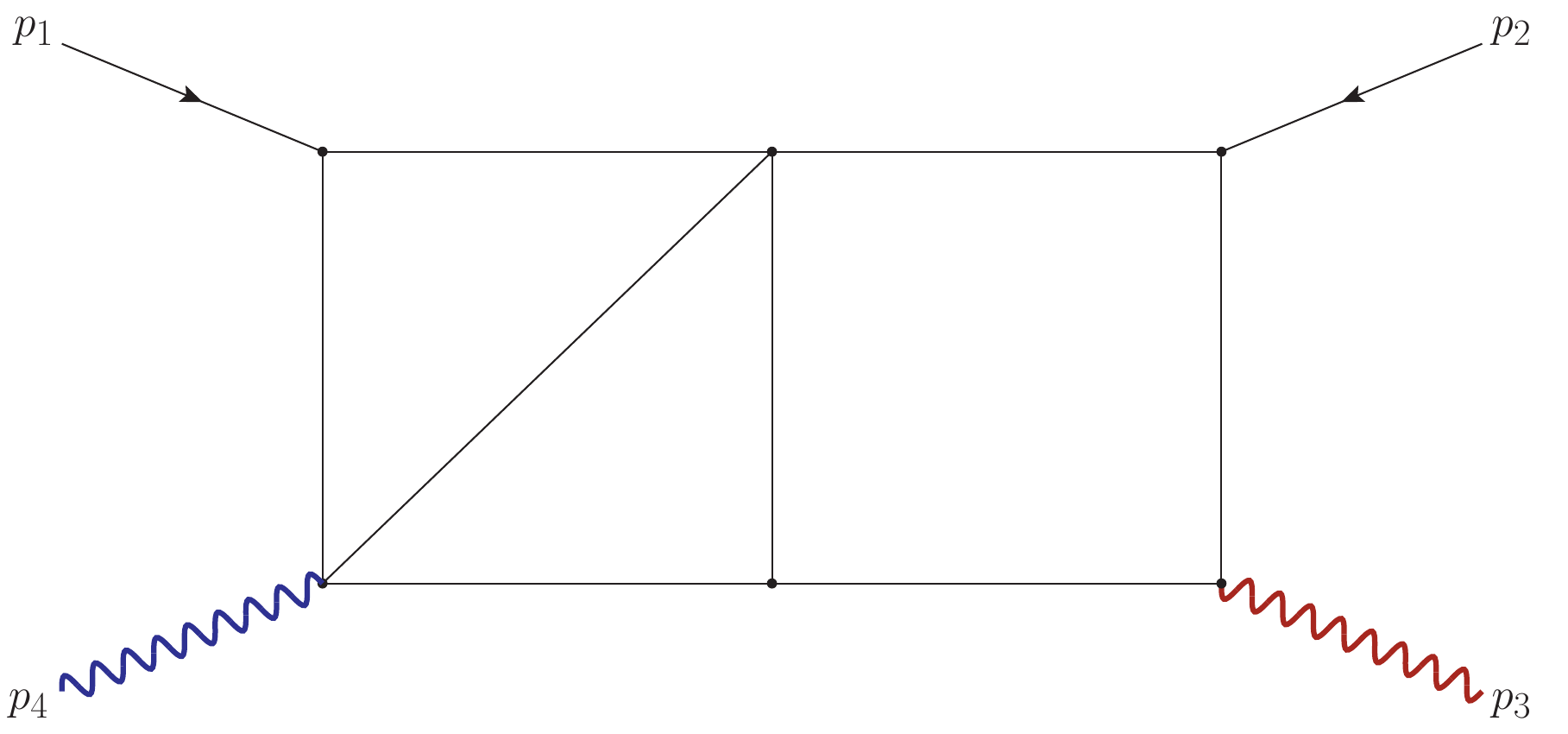}\\
\begin{equation*}
\{I_{24}^{\text{PL2}},I_{25}^{\text{PL2}}\}
\end{equation*}
\end{multicols}

\textbf{Sector $\mathbf{F_{132}}$[1,1,0,0,0,0,1,0,1,0,1,1,1,0,1]}

\begin{multicols}{2}
\includegraphics[scale=0.20]{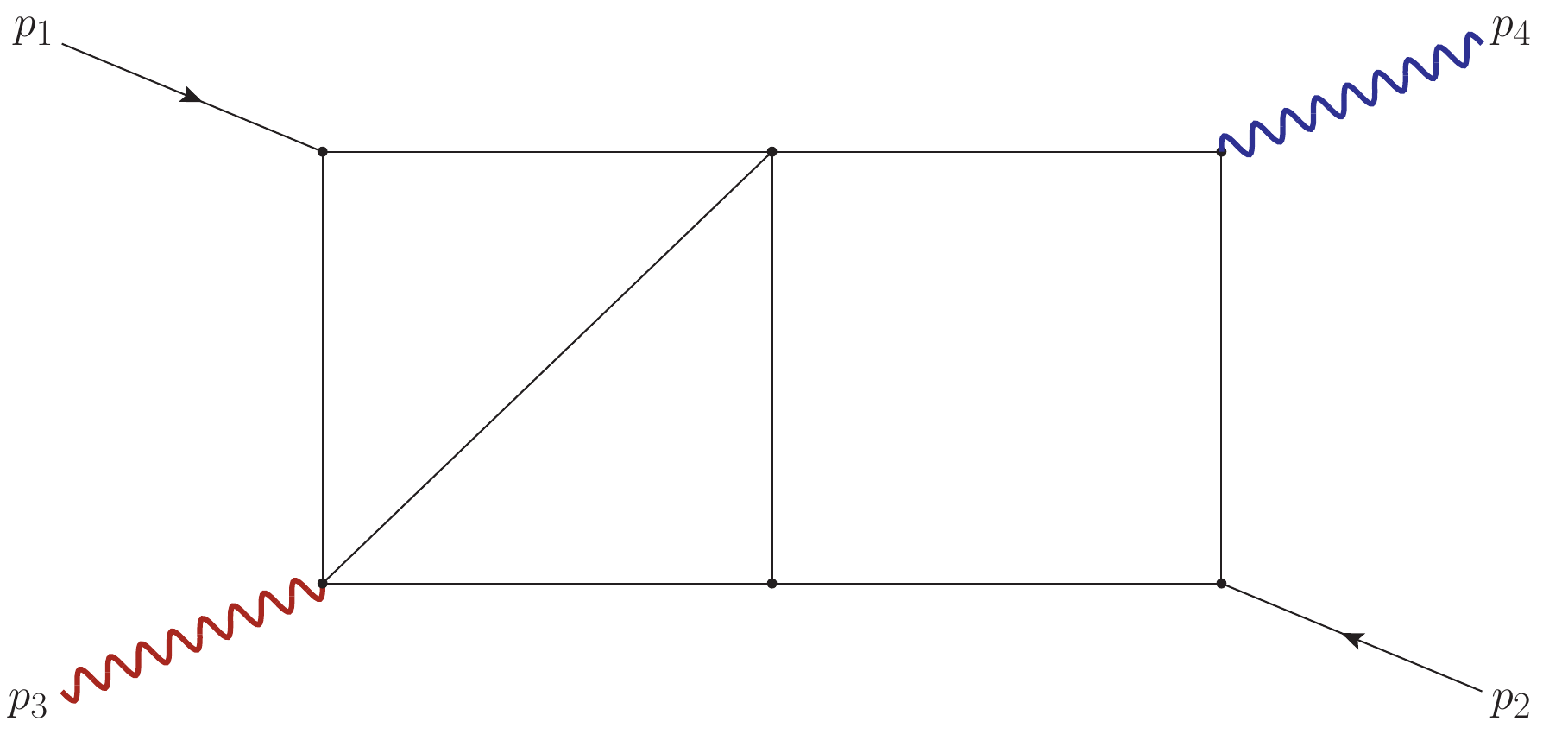}\\
\begin{equation*}
\{I_{24}^{\text{PL3}},I_{25}^{\text{PL3}}\}
\end{equation*}
\end{multicols}

\textbf{Sector $\mathbf{F_{132}}$[0,1,1,0,1,0,0,0,1,0,1,1,1,0,1]}

\begin{multicols}{2}
\includegraphics[scale=0.20]{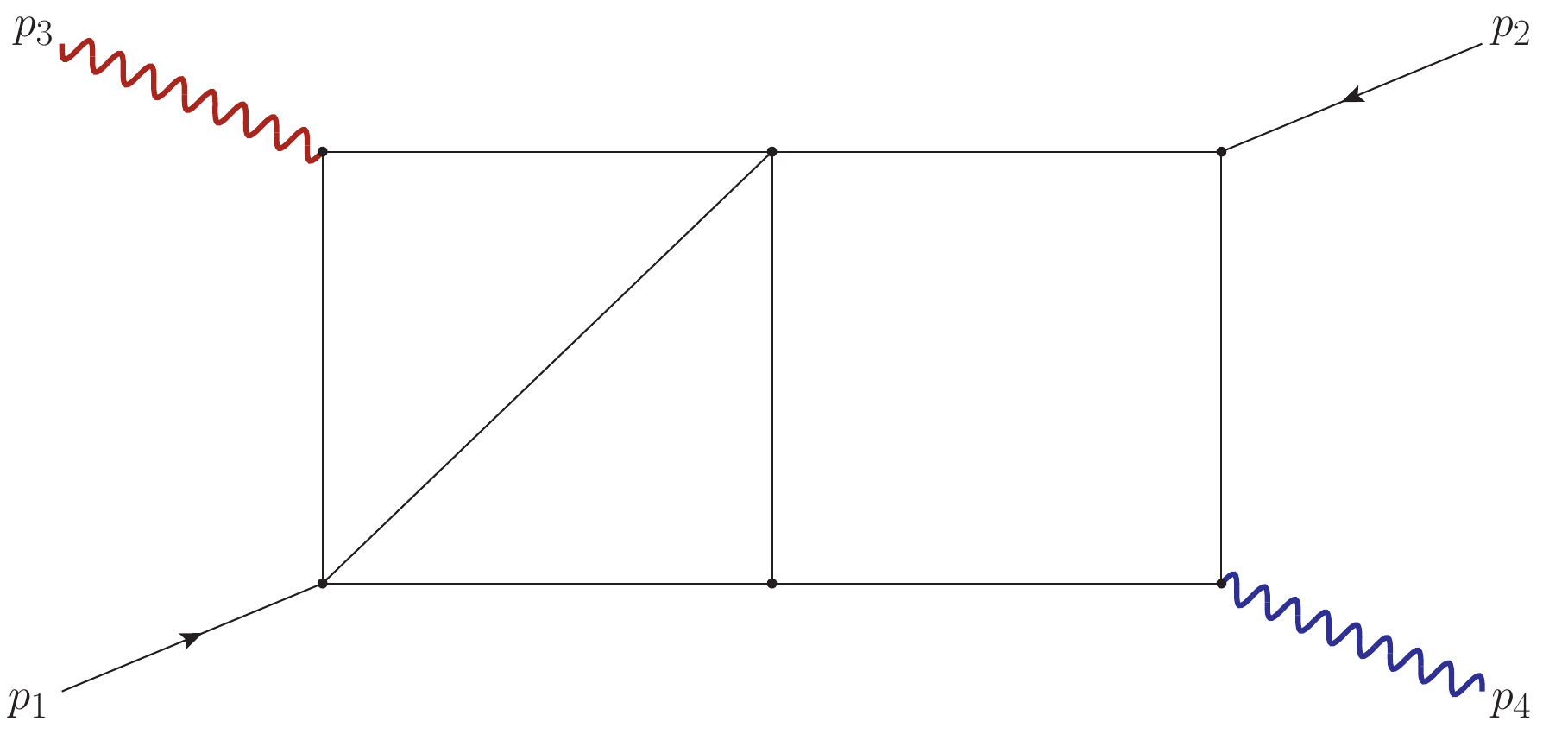}\\
\begin{equation*}
\{I_{14}^{\text{PL3}},I_{15}^{\text{PL3}},I_{16}^{\text{PL3}},I_{17}^{\text{PL3}}\}
\end{equation*}
\end{multicols}

\textbf{Sector $\mathbf{F_{123}}$[1,0,0,1,0,1,0,0,0,1,1,1,1,0,1]}

\begin{multicols}{2}
\includegraphics[scale=0.20]{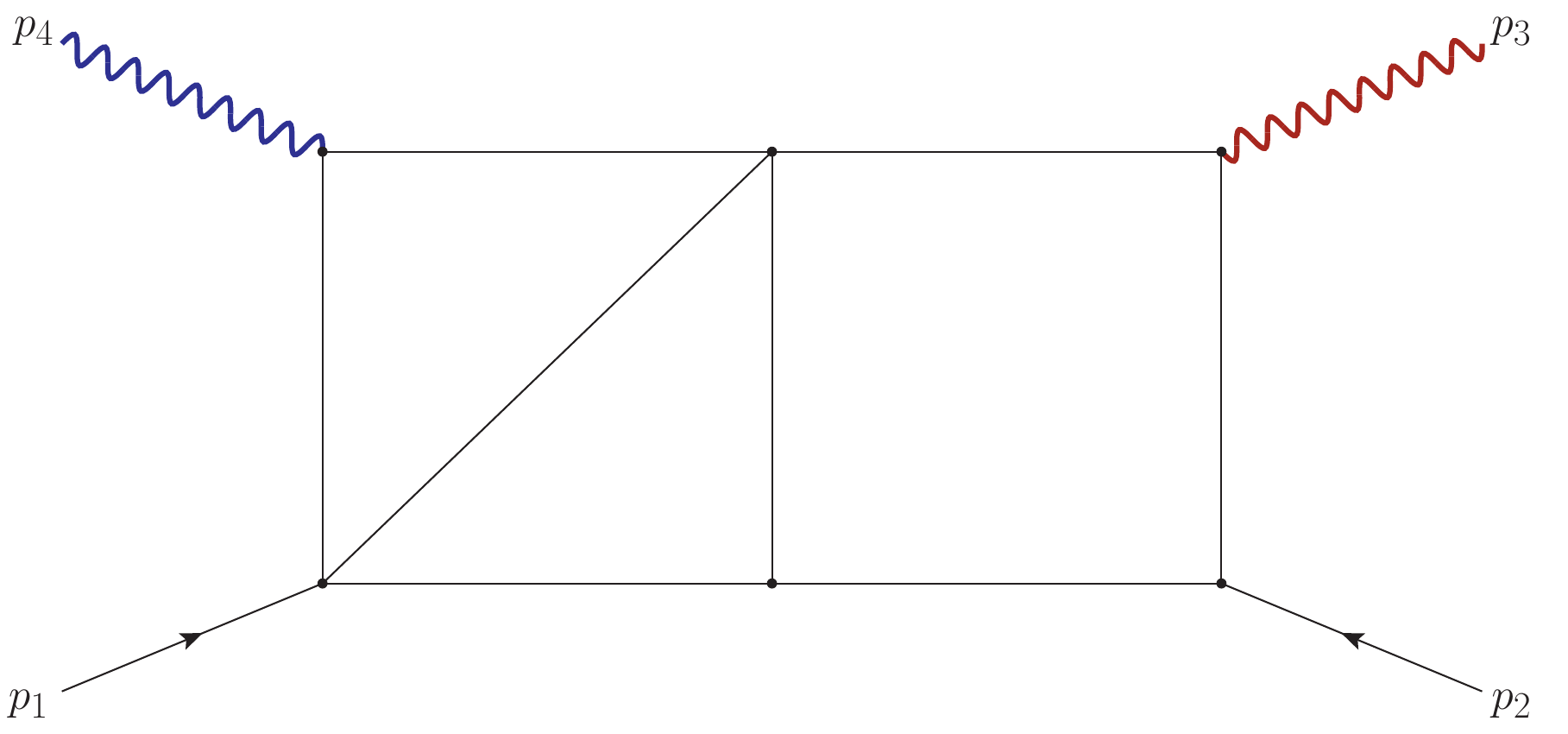}\\
\begin{equation*}
\{I_{14}^{\text{PL2}},I_{15}^{\text{PL2}},I_{16}^{\text{PL2}},I_{17}^{\text{PL2}}\}
\end{equation*}
\end{multicols}

\textbf{Sector $\mathbf{F_{123}}$[1,1,1,0,0,0,0,1,1,0,0,1,1,0,1]}

\begin{multicols}{2}
\includegraphics[scale=0.20]{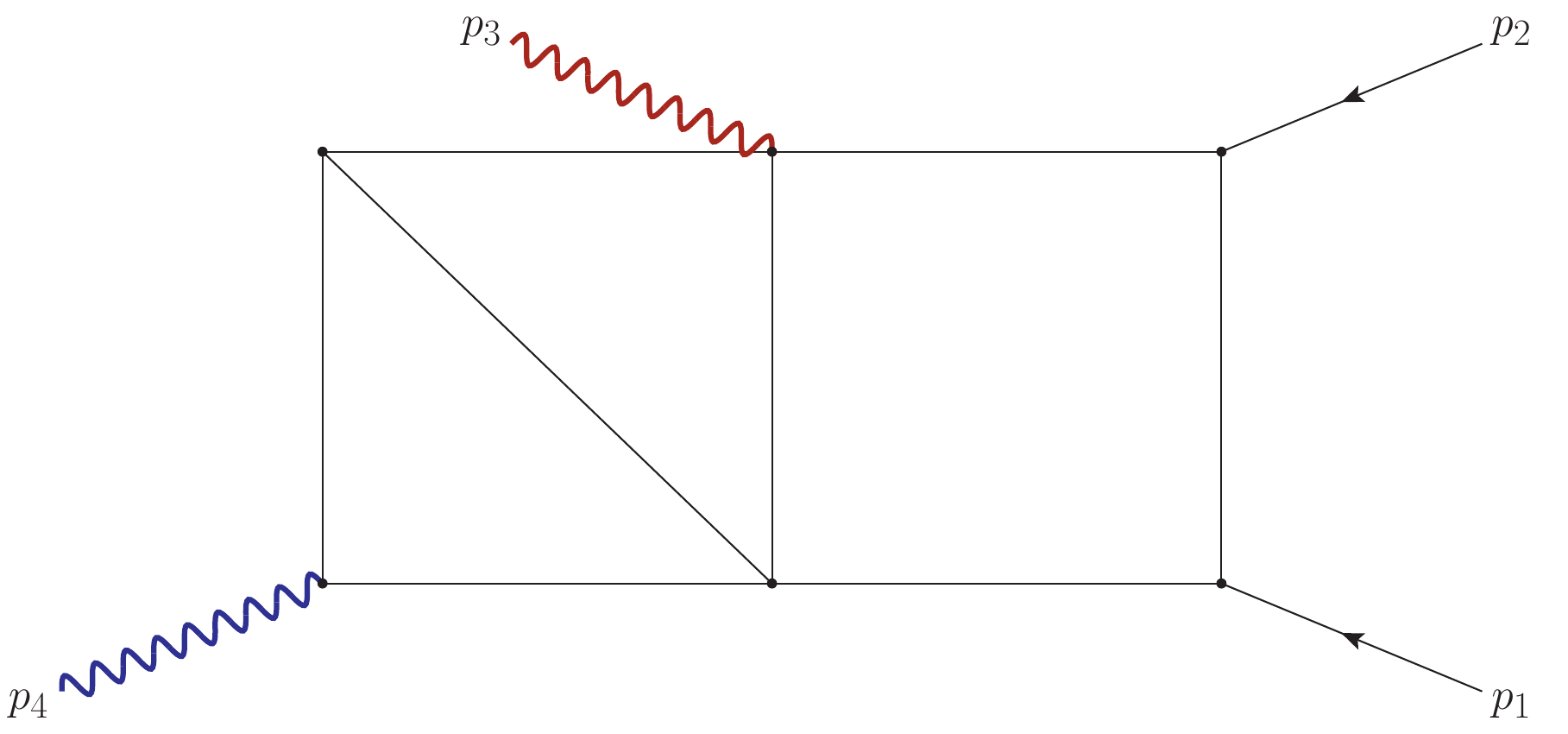}\\
\begin{equation*}
\{I_{82}^{\text{PT1}},I_{83}^{\text{PT1}}\}
\end{equation*}
\end{multicols}

\textbf{Sector $\mathbf{F_{123}}$[0,1,1,1,1,0,0,1,1,0,0,0,0,1,1]}

\begin{multicols}{2}
\includegraphics[scale=0.20]{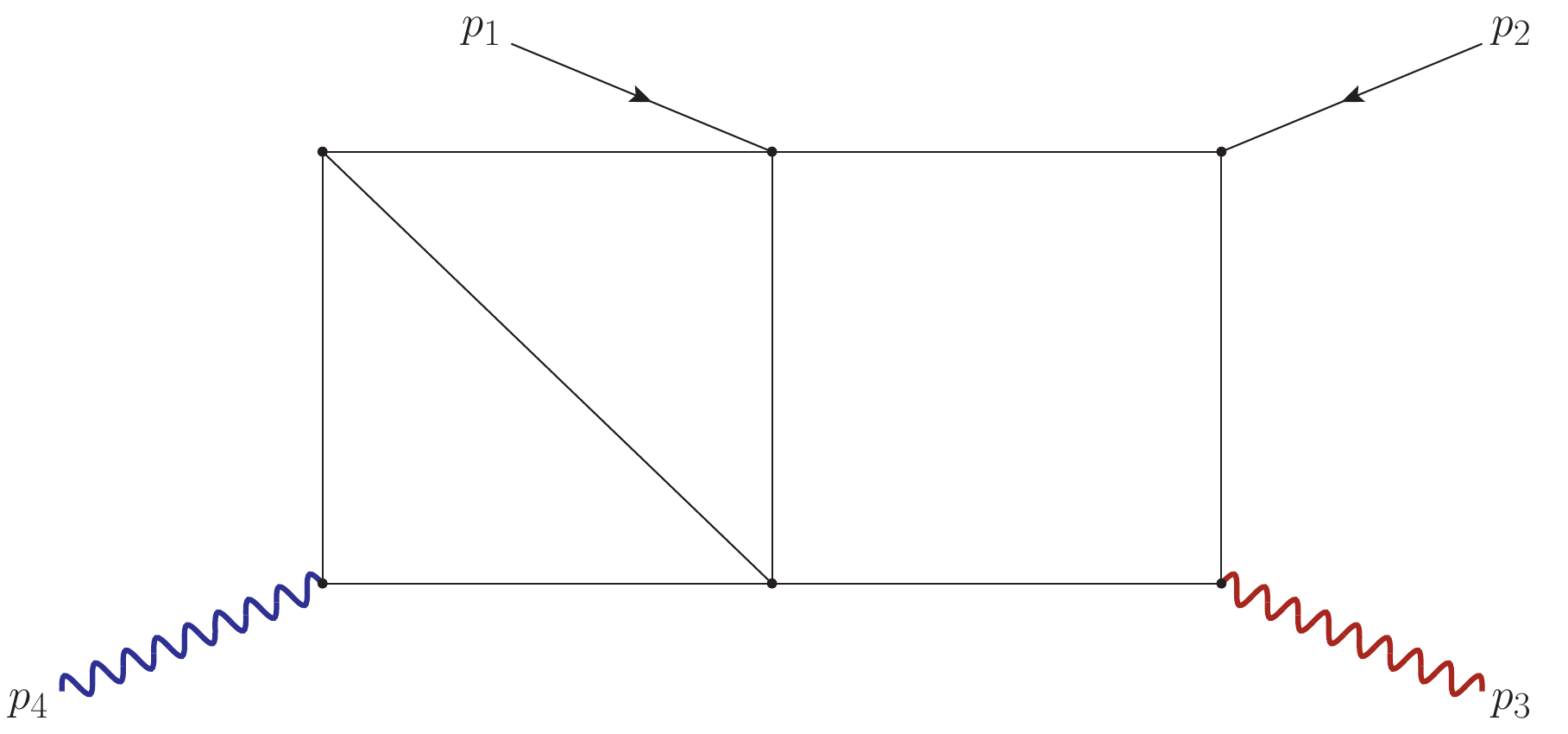}\\
\begin{equation*}
\{I_{46}^{\text{PT2}},I_{47}^{\text{PT2}}\}
\end{equation*}
\end{multicols}

\textbf{Sector $\mathbf{F_{132}}$[1,1,1,0,0,0,0,1,1,0,0,1,1,0,1]}

\begin{multicols}{2}
\includegraphics[scale=0.20]{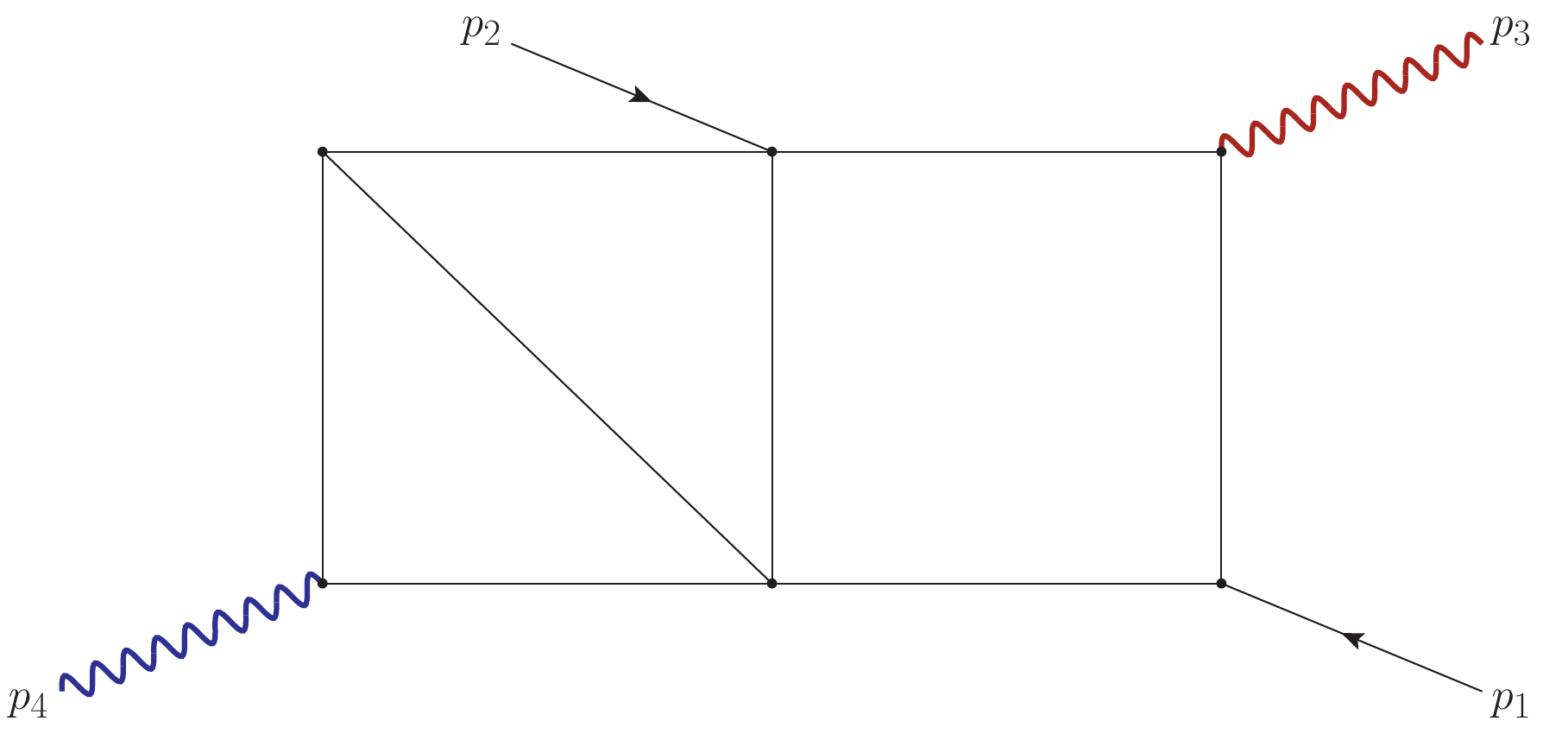}\\
\begin{equation*}
\{I_{48}^{\text{PT3}},I_{49}^{\text{PT3}}\}
\end{equation*}
\end{multicols}

\textbf{Sector $\mathbf{F_{132}}$[0,1,0,0,0,0,1,1,1,0,0,1,1,1,1]}

\begin{multicols}{2}
\includegraphics[scale=0.20]{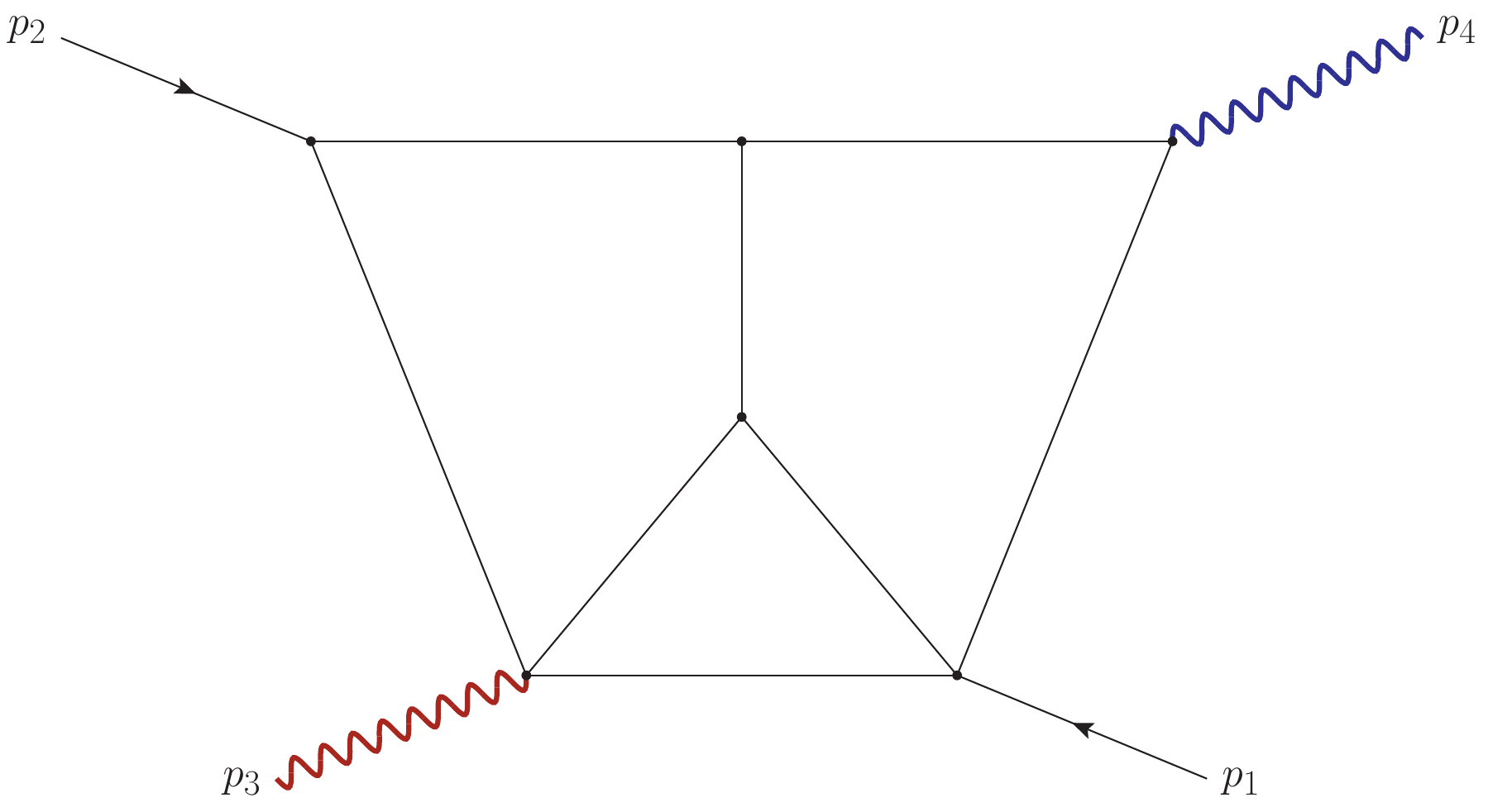}\\
\begin{equation*}
\{I_{32}^{\text{PT3}},I_{33}^{\text{PT3}},I_{34}^{\text{PT3}},I_{35}^{\text{PT3}},I_{36}^{\text{PT3}}\}
\end{equation*}
\end{multicols}

\textbf{Sector $\mathbf{F_{123}}$[0,1,0,0,0,0,1,1,1,0,0,1,1,1,1]}

\begin{multicols}{2}
\includegraphics[scale=0.20]{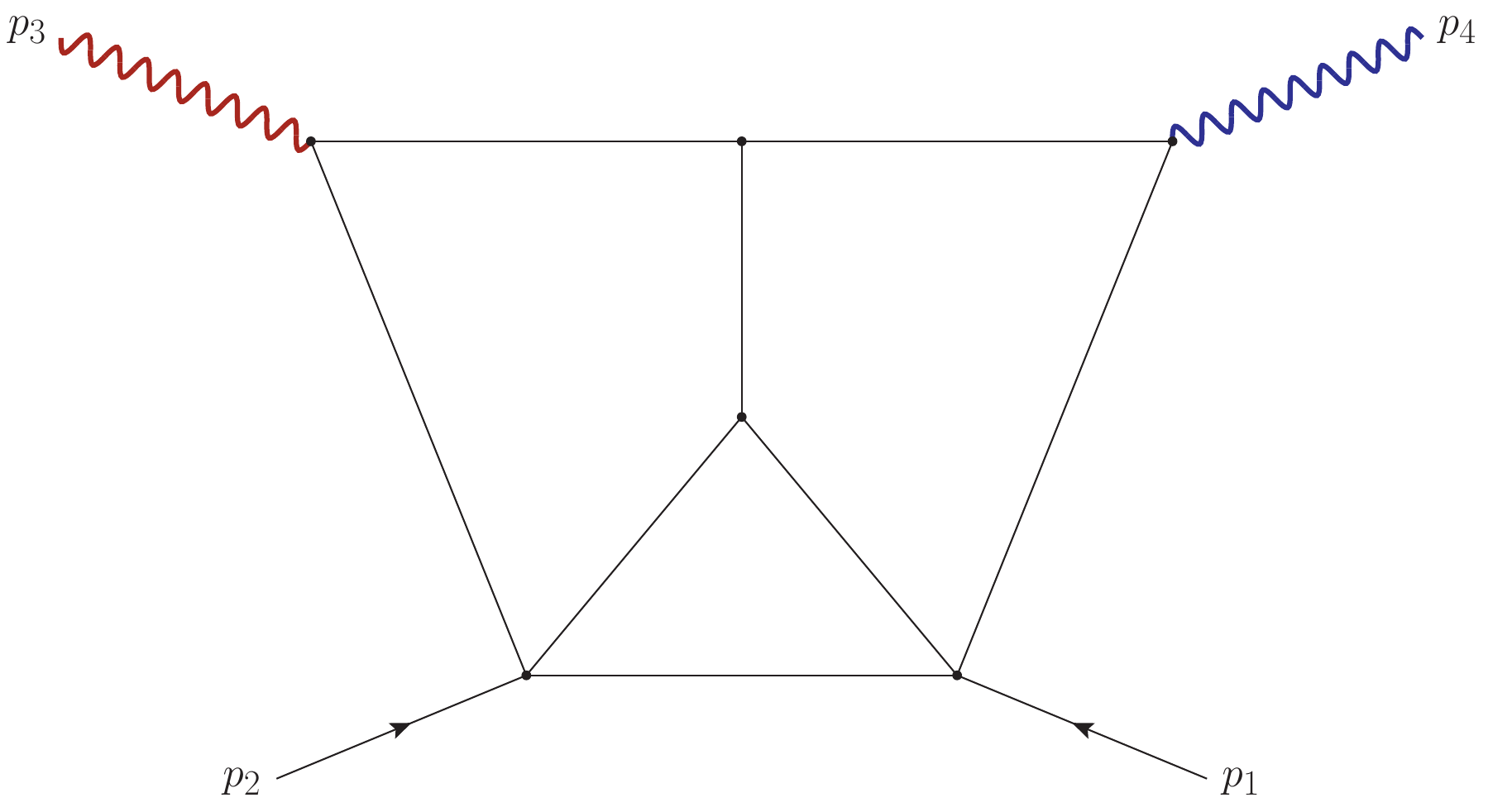}\\
\begin{equation*}
\{I_{37}^{\text{PT1}},I_{38}^{\text{PT1}},I_{39}^{\text{PT1}},I_{40}^{\text{PT1}},I_{41}^{\text{PT1}},I_{42}^{\text{PT1}},I_{43}^{\text{PT1}}\}
\end{equation*}
\end{multicols}

\vspace{0.15cm}
\begin{center}
\textbf{\textit{Nine-Propagator Pure Candidates}}\\
\end{center}
\vspace{0.2cm}

\textbf{Sector $\mathbf{F_{123}}$[0,1,1,1,1,0,0,1,1,1,0,0,0,1,1]}

\begin{multicols}{2}
\includegraphics[scale=0.20]{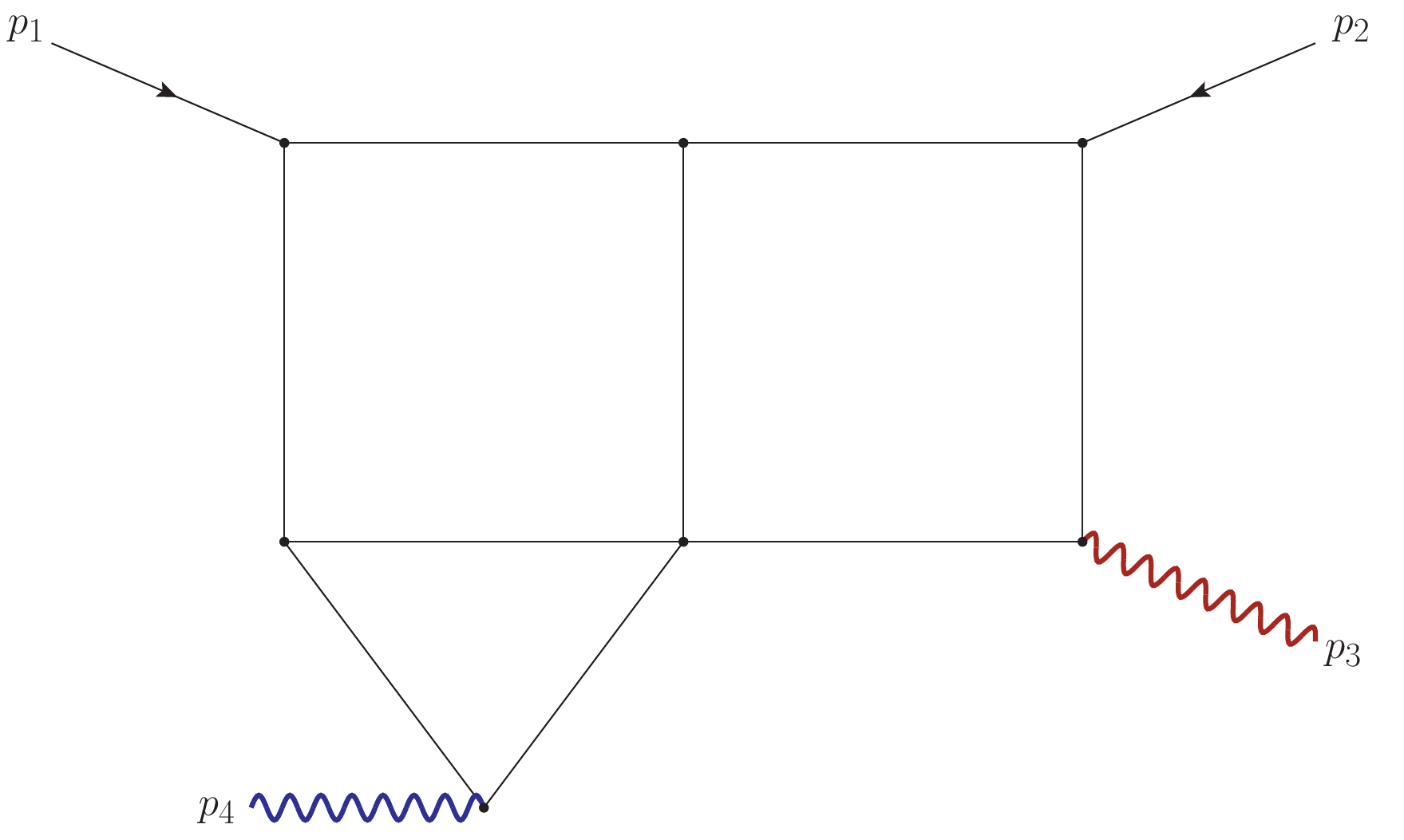}\\
\begin{equation*}
\{I_{19}^{\text{PT2}},I_{20}^{\text{PT2}}\}
\end{equation*}
\end{multicols}

\textbf{Sector $\mathbf{F_{132}}$[1,1,1,0,0,0,1,1,1,0,0,1,1,0,1]}

\begin{multicols}{2}
\includegraphics[scale=0.20]{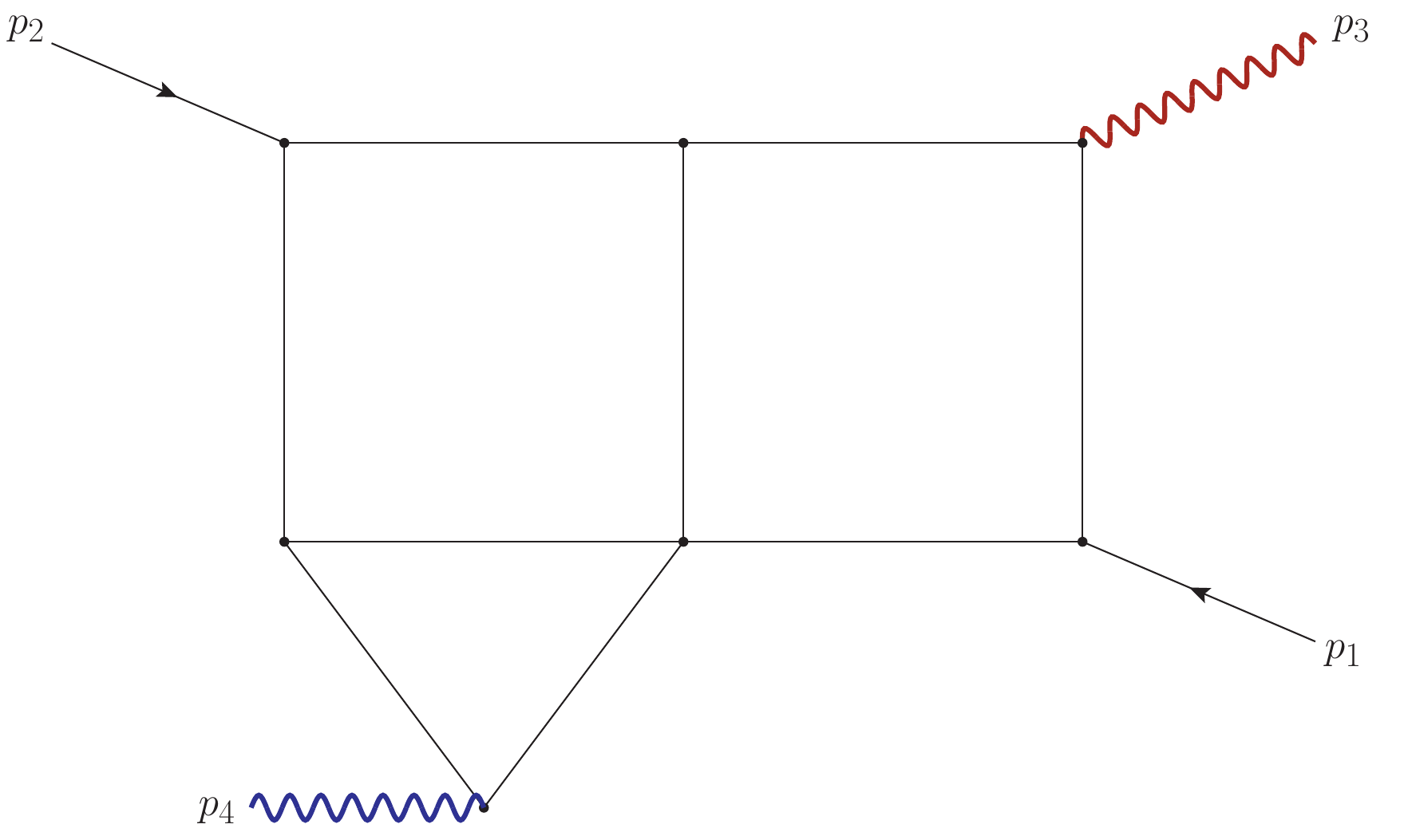}\\
\begin{equation*}
\{I_{18}^{\text{PT3}},I_{19}^{\text{PT3}}\}
\end{equation*}
\end{multicols}

\textbf{Sector $\mathbf{F_{123}}$[1,1,1,0,0,0,1,1,1,0,0,1,0,1,1]}

\begin{multicols}{2}
\includegraphics[scale=0.20]{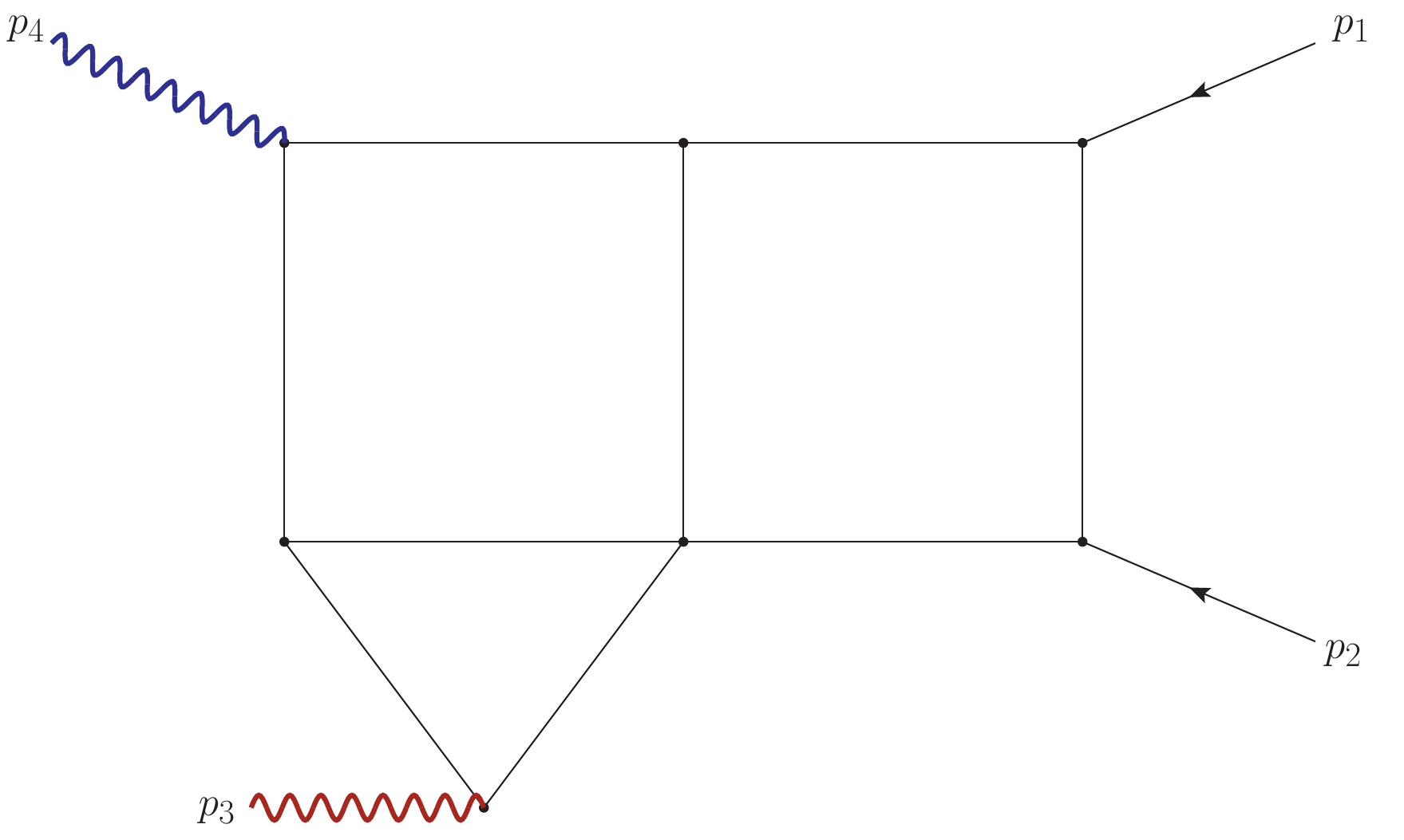}\\
\begin{equation*}
\{I_{34}^{\text{PT1}},I_{35}^{\text{PT1}},I_{36}^{\text{PT1}}\}
\end{equation*}
\end{multicols}

\textbf{Sector $\mathbf{F_{123}}$[1,0,0,1,0,1,0,1,0,1,1,1,1,0,1]}

\begin{multicols}{2}
\includegraphics[scale=0.18]{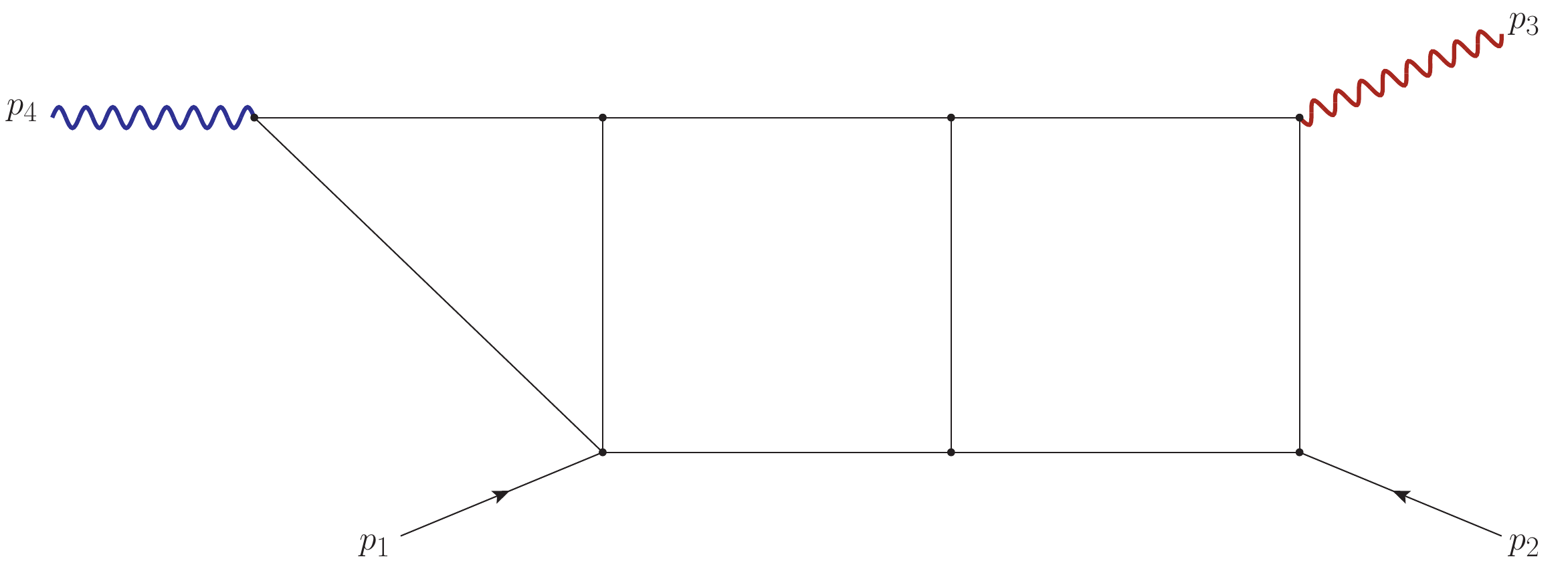}\\
\begin{equation*}
\{I_{7}^{\text{PL2}},I_{8}^{\text{PL2}},I_{9}^{\text{PL2}}\}
\end{equation*}
\end{multicols}

\textbf{Sector $\mathbf{F_{132}}$[0,1,1,0,1,0,1,0,1,0,1,1,1,0,1]}

\begin{multicols}{2}
\includegraphics[scale=0.18]{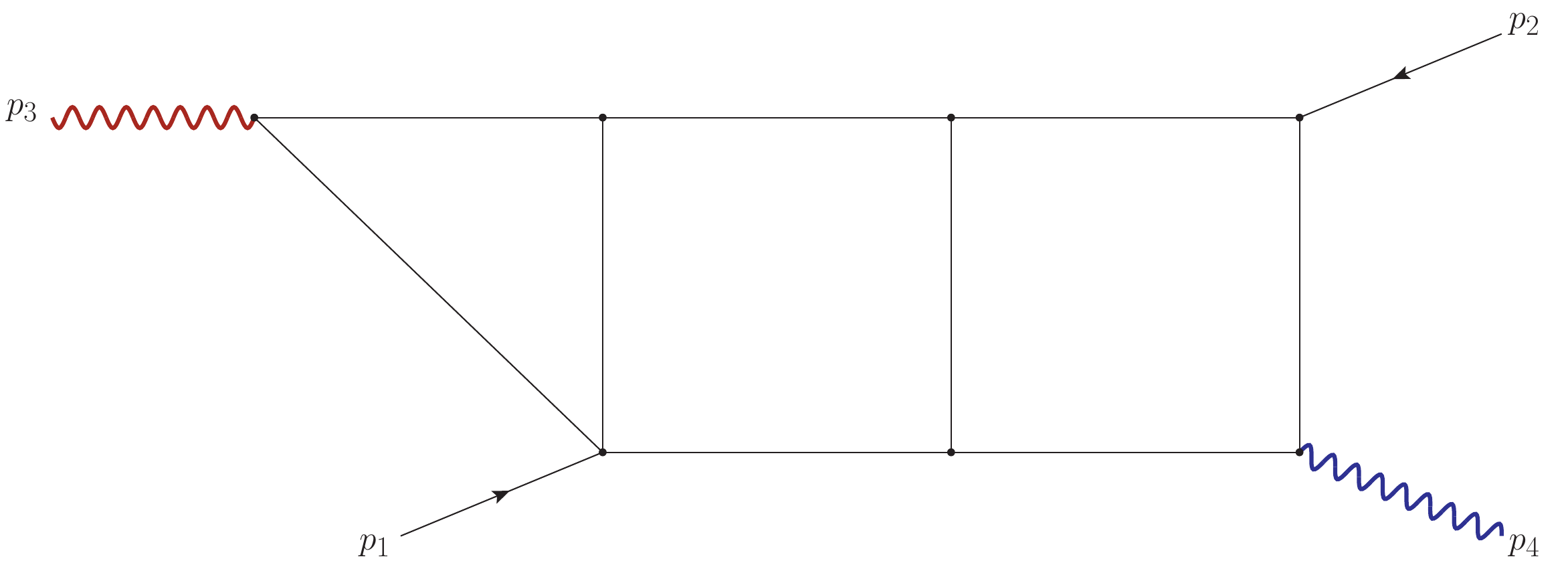}\\
\begin{equation*}
\{I_{7}^{\text{PL3}},I_{8}^{\text{PL3}},I_{9}^{\text{PL3}}\}
\end{equation*}
\end{multicols}

\textbf{Sector $\mathbf{F_{123}}$[1,0,1,0,0,0,1,1,1,0,0,1,1,1,1]}

\begin{multicols}{2}
\includegraphics[scale=0.15]{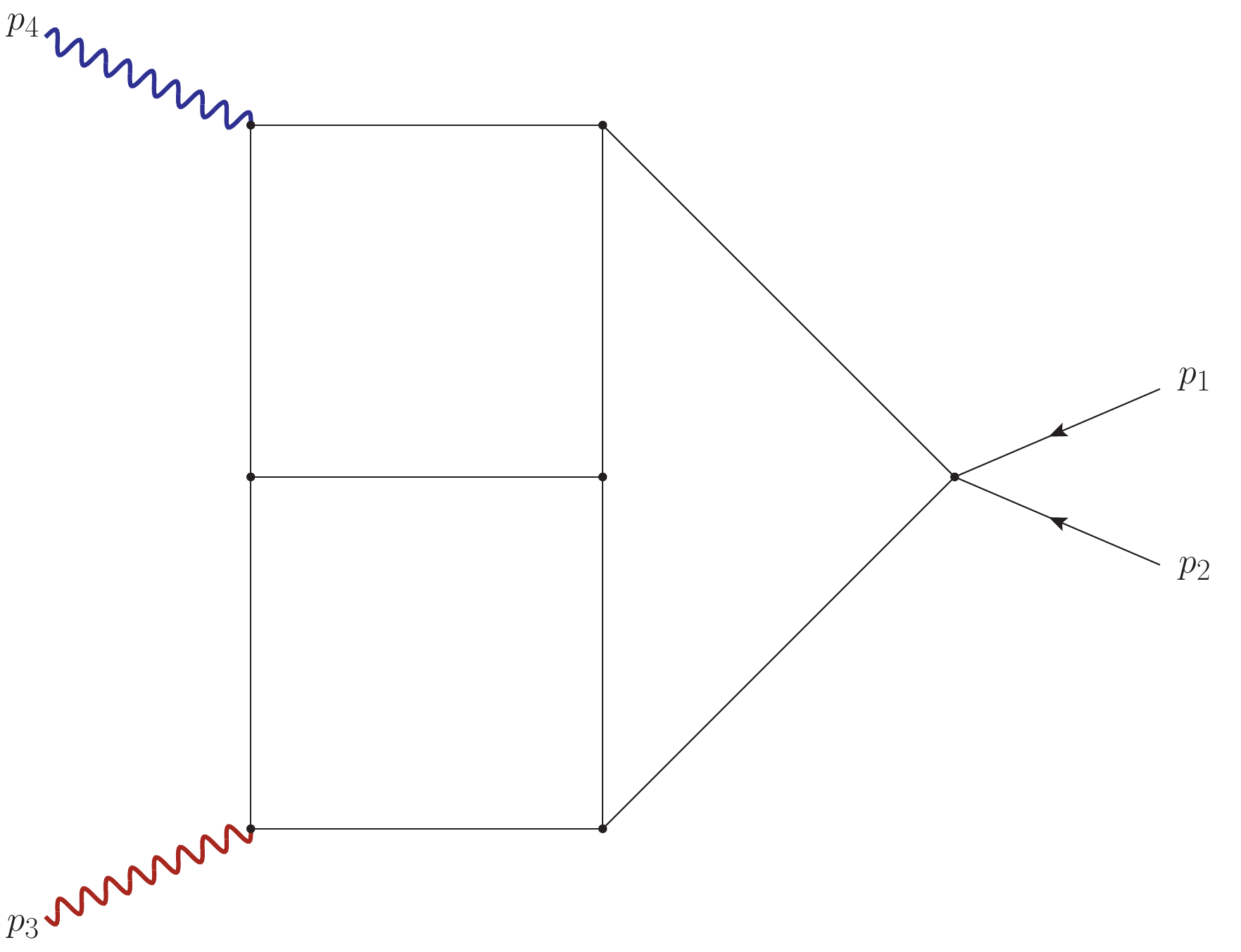}\\
\begin{equation*}
\{I_{27}^{\text{PT1}},I_{28}^{\text{PT1}},I_{29}^{\text{PT1}},I_{30}^{\text{PT1}}\}
\end{equation*}
\end{multicols}

\textbf{Sector $\mathbf{F_{132}}$[1,1,0,0,0,0,1,1,1,0,0,1,1,1,1]}

\begin{multicols}{2}
\includegraphics[scale=0.15]{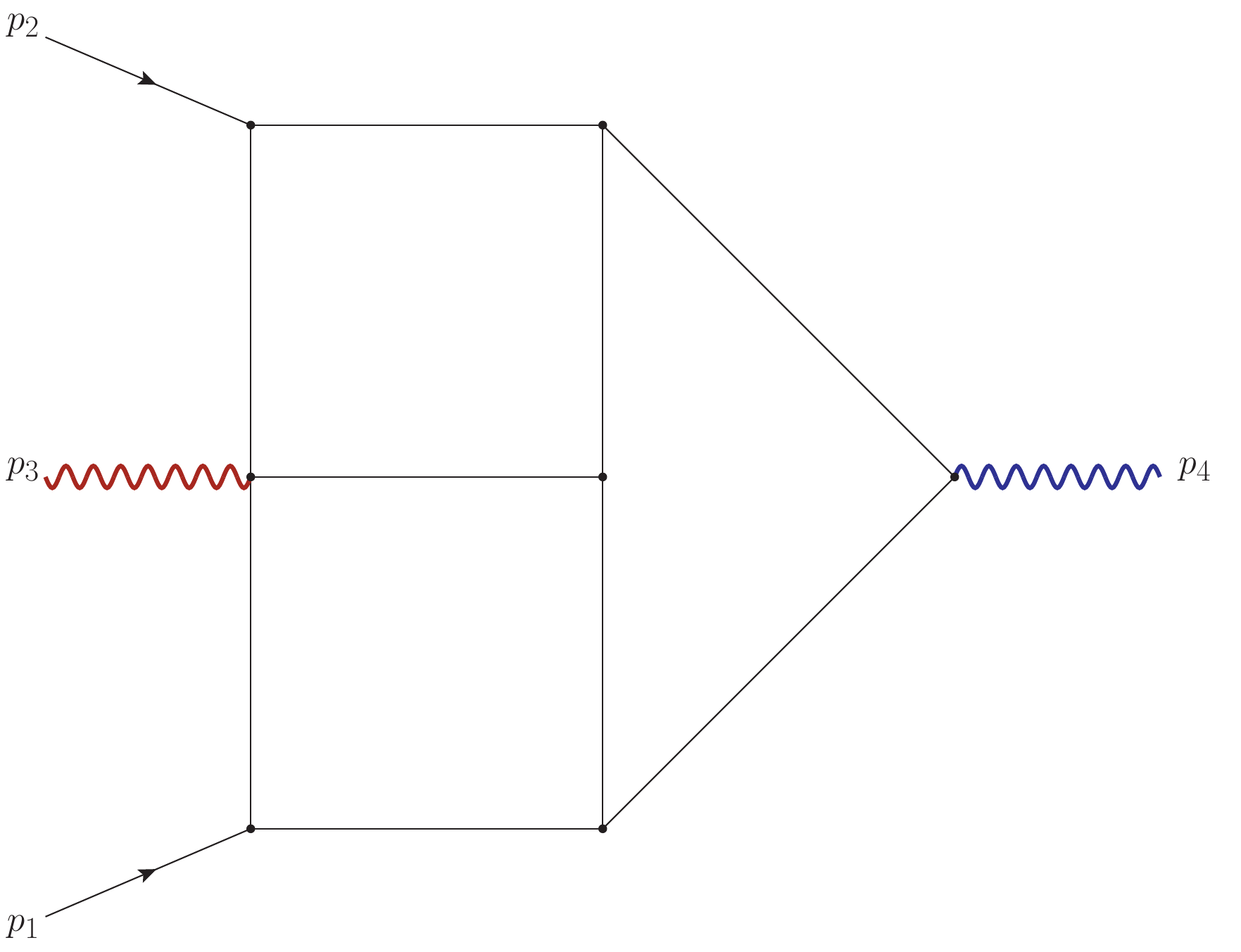}\\
\begin{equation*}
\{I_{14}^{\text{PT3}},I_{15}^{\text{PT3}},I_{16}^{\text{PT3}},I_{17}^{\text{PT3}}\}
\end{equation*}
\end{multicols}

\textbf{Sector $\mathbf{F_{132}}$[0,1,1,0,0,0,1,1,1,0,0,1,1,1,1]}

\begin{multicols}{2}
\includegraphics[scale=0.15]{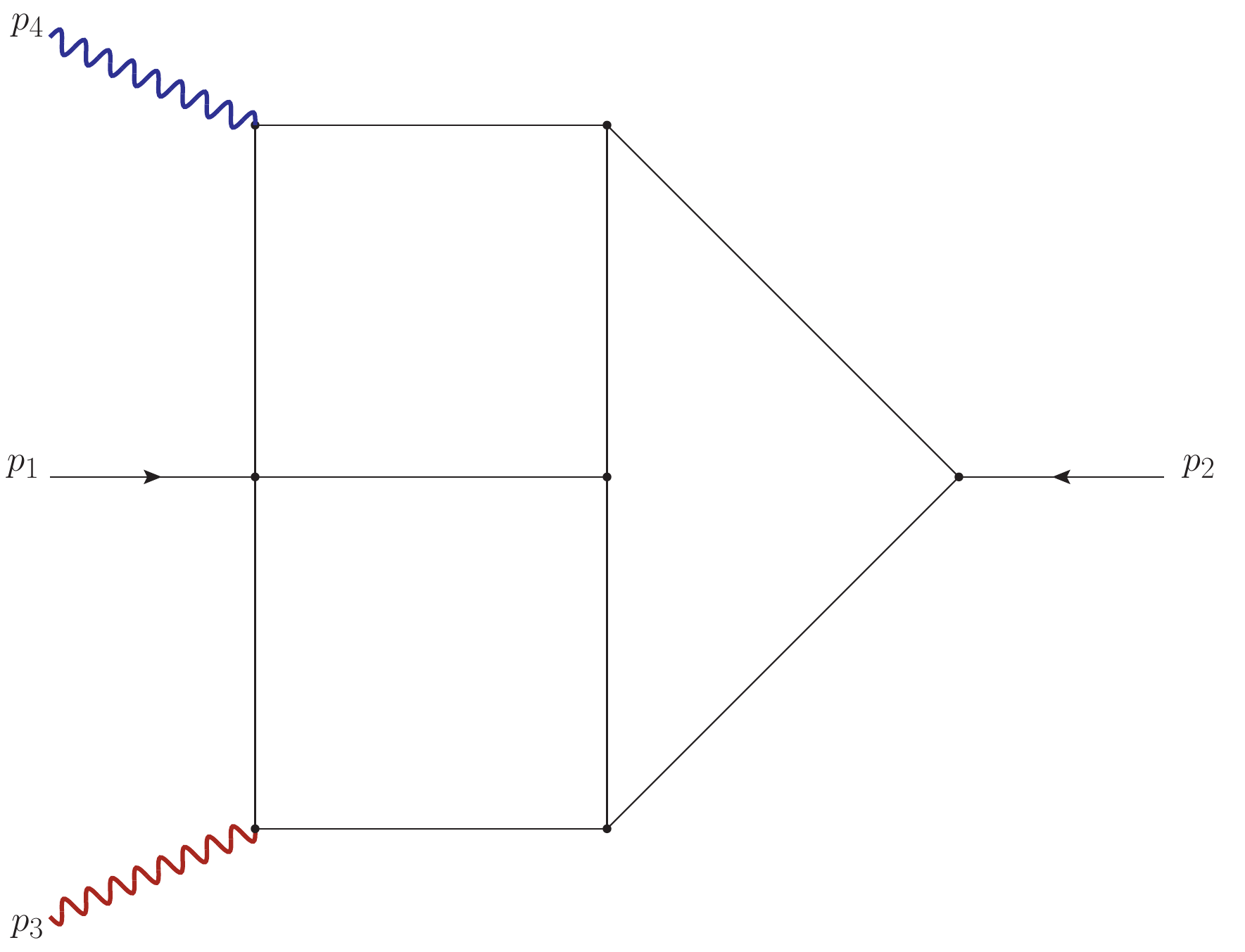}\\
\begin{equation*}
\begin{split}
\{&I_{5}^{\text{PT3}},I_{6}^{\text{PT3}},I_{7}^{\text{PT3}},I_{8}^{\text{PT3}},I_{9}^{\text{PT3}}, \\ &I_{10}^{\text{PT3}},I_{11}^{\text{PT3}},I_{12}^{\text{PT3}},I_{13}^{\text{PT3}}\}
\end{split}
\end{equation*}
\end{multicols}

\textbf{Sector $\mathbf{F_{123}}$[1,1,0,0,0,0,1,1,1,0,0,1,1,1,1]}

\begin{multicols}{2}
\includegraphics[scale=0.15]{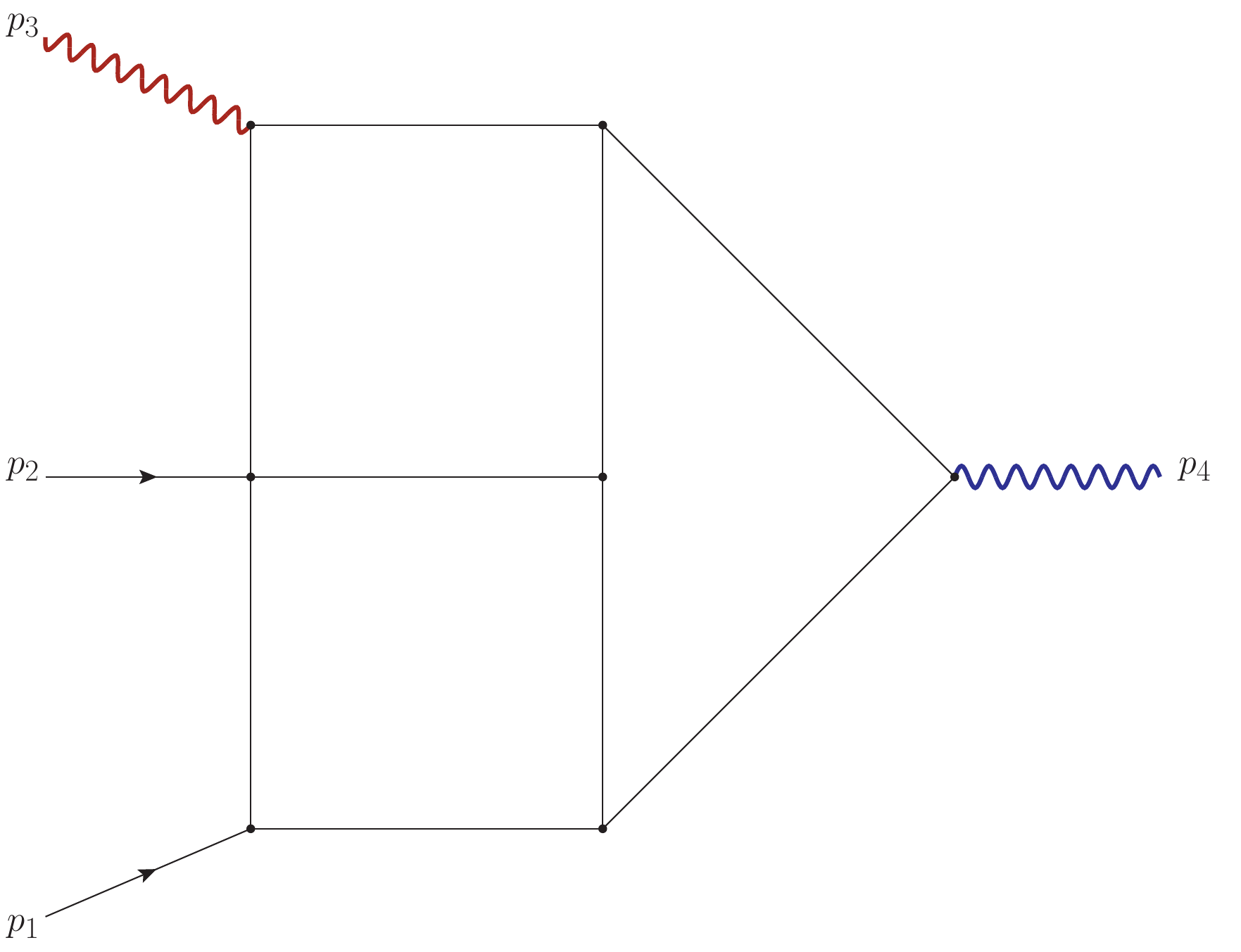}\\
\begin{equation*}
\begin{split}
\{&I_{17}^{\text{PT1}},I_{18}^{\text{PT1}},I_{19}^{\text{PT1}},I_{20}^{\text{PT1}},I_{21}^{\text{PT1}}, \\ &I_{22}^{\text{PT1}},I_{23}^{\text{PT1}},I_{24}^{\text{PT1}},I_{25}^{\text{PT1}},I_{26}^{\text{PT1}}\}
\end{split}
\end{equation*}
\end{multicols}

\vspace{0.15cm}
\begin{center}
\textbf{\textit{Ten-Propagator Pure Candidates}}\\
\end{center}
\vspace{0.2cm}

\textbf{Sector $\mathbf{F_{123}}$[1,1,0,1,0,1,0,1,0,1,1,1,1,0,1]}

\begin{multicols}{2}
\includegraphics[scale=0.39]{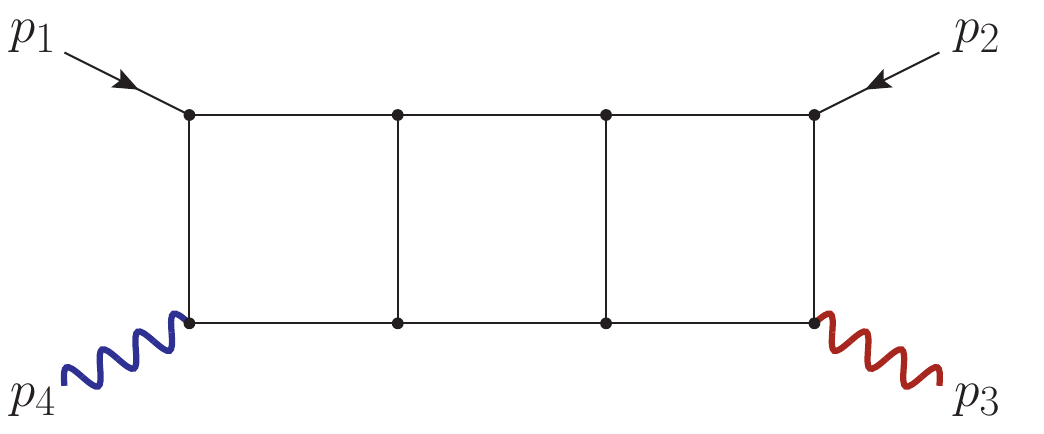}\\
\begin{equation*}
\begin{split}
\{&I_{1}^{\text{PL2}},I_{2}^{\text{PL2}},I_{3}^{\text{PL2}}\}
\end{split}
\end{equation*}
\end{multicols}

\textbf{Sector $\mathbf{F_{132}}$[1,1,1,0,1,0,1,0,1,0,1,1,1,0,1]}

\begin{multicols}{2}
\includegraphics[scale=0.39]{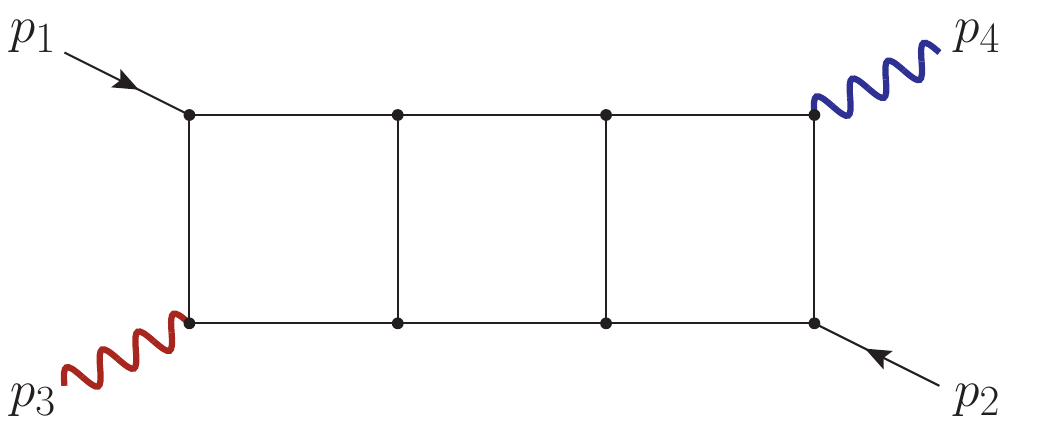}\\
\begin{equation*}
\begin{split}
\{&I_{1}^{\text{PL3}},I_{2}^{\text{PL3}},I_{3}^{\text{PL3}}\}
\end{split}
\end{equation*}
\end{multicols}

\textbf{Sector $\mathbf{F_{123}}$[1,1,1,0,0,0,1,1,1,0,0,1,1,1,1]}

\begin{multicols}{2}
\includegraphics[scale=0.34]{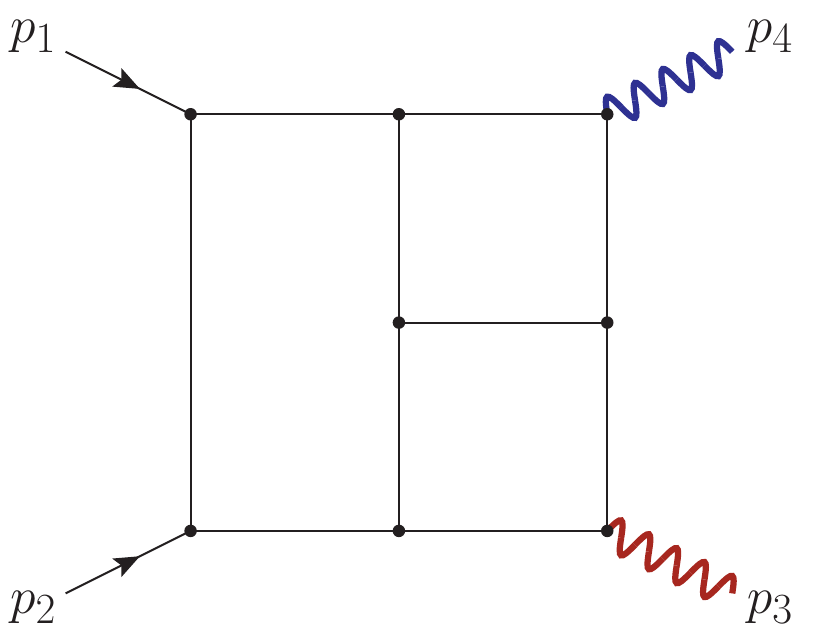}\\
\begin{equation*}
\begin{split}
\{&I_{1}^{\text{PT1}},I_{2}^{\text{PT1}},I_{3}^{\text{PT1}},I_{4}^{\text{PT1}},I_{5}^{\text{PT1}},I_{6}^{\text{PT1}}\}
\end{split}
\end{equation*}
\end{multicols}

\textbf{Sector $\mathbf{F_{123}}$[0,1,1,1,1,0,0,1,1,1,0,0,1,1,1]}

\begin{multicols}{2}
\includegraphics[scale=0.34]{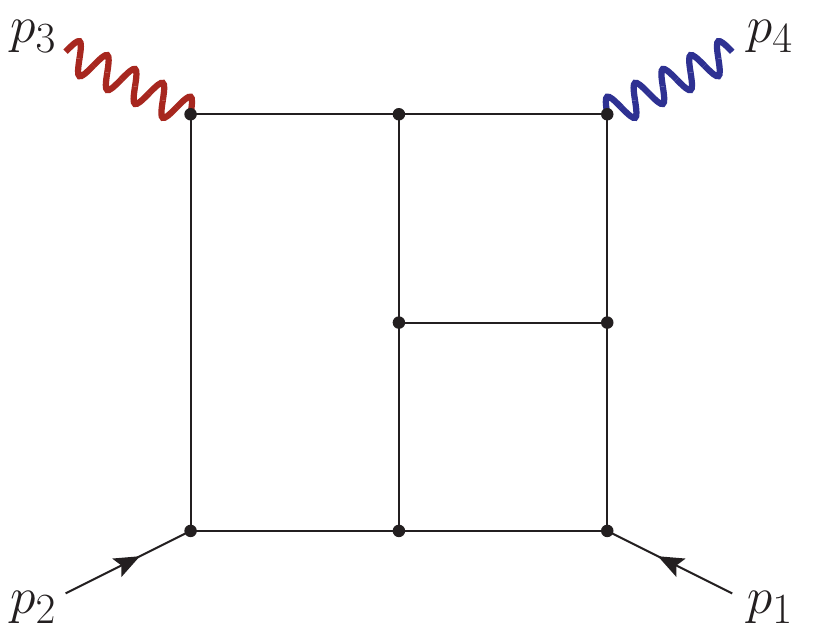}\\
\begin{equation*}
\begin{split}
\{&I_{1}^{\text{PT2}},I_{2}^{\text{PT2}},I_{3}^{\text{PT2}},I_{4}^{\text{PT2}}\}
\end{split}
\end{equation*}
\end{multicols}

\textbf{Sector $\mathbf{F_{132}}$[01,1,1,0,0,0,1,1,1,0,0,1,1,1,1]}

\begin{multicols}{2}
\includegraphics[scale=0.34]{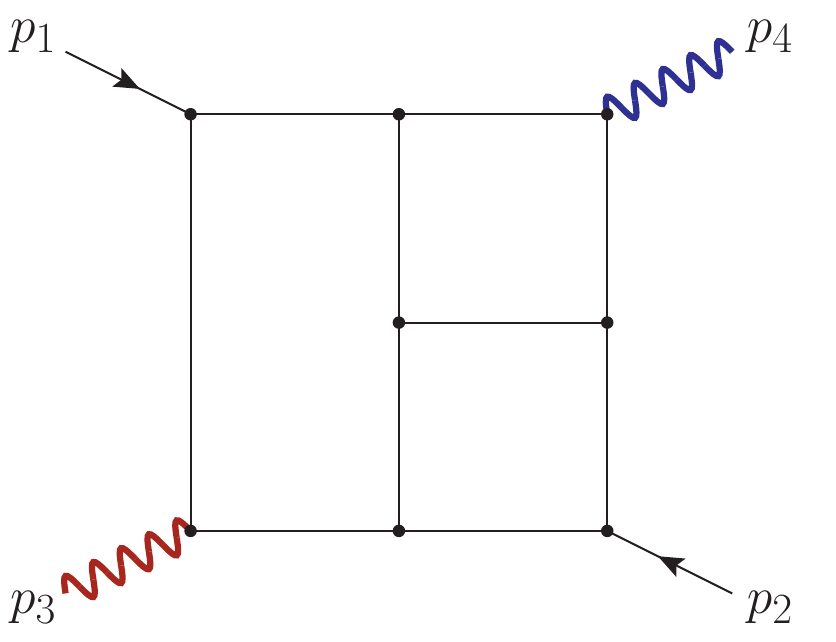}\\
\begin{equation*}
\begin{split}
\{&I_{1}^{\text{PT3}},I_{2}^{\text{PT3}},I_{3}^{\text{PT3}},I_{4}^{\text{PT3}}\}
\end{split}
\end{equation*}
\end{multicols}